\documentclass[unnumsec,namedate,webpdf,contemporary,large]{oup-authoring-template}

\usepackage{bm}
\usepackage{setspace}
\usepackage[expansion=false]{microtype}
\usepackage{float}
\usepackage{threeparttable}
\usepackage{threeparttablex}
\usepackage{longtable}
\usepackage{booktabs}
\theoremstyle{thmstyleone}%
\theoremstyle{thmstyletwo}%
\theoremstyle{thmstylethree}%

\newcommand{\societylogo}{}

\let\OUPtoprule\toprule
\let\OUPmidrule\midrule
\let\OUPbotrule\botrule
\let\toprule\OUPtoprule
\let\midrule\OUPmidrule
\let\botrule\OUPbotrule

\let\bottomrule\botrule

\begin{document}

\journaltitle{Journal Title}
\copyrightyear{2026}
\pubyear{2026}

\firstpage{1}

\title[Dynamic Spatial Bayesian ML Model for Mobility and Inequality]{\textbf{Dynamic Spatial Bayesian Machine Learning Model: Applications to Intergenerational Economic Mobility and Geographic Income Inequality in the United States}}

\author[1,$\ast$]{Hammed A. Olayinka \ORCID{0000-0002-9796-5276}}
\author[2]{Saheed O. Olayinka  \ORCID{0009-0000-4551-5479}}

\address[1]{\orgdiv{Department of Mathematical Sciences}, \orgname{Worcester Polytechnic Institute}, \orgaddress{Worcester, MA, USA}}
\address[2]{\orgdiv{Department of Computer Science and Artificial Intelligence}, \orgname{University of Ibadan}, \orgaddress{Ibadan, Nigeria}}

\corresp[$\ast$]{Corresponding author. \href{mailto:haolayinka@wpi.edu}{haolayinka@wpi.edu}}

\abstract{We develop a Dynamic Spatial Panel Bayesian Additive Regression Trees model with Horseshoe shrinkage (DSP-BART-HS) for high-dimensional spatio-temporal panel data. We jointly evaluate the model against a comprehensive suite of structural spatial econometrics, non-parametric machine learning methods, and small-area estimators across nine data-generating scenarios spanning 200 replicates each, including irregular spatial topologies, dense policy effects, and non-linear individual-level interactions. DSP-BART-HS is the best or statistically indistinguishable from the best estimator in every scenario. Conventional region-time-aggregate comparators suffer severe performance degradation -- trailing by a factor of three or more -- whenever individual-level non-linearity drives outcome variance, a limitation this framework's tree-ensemble design directly addresses. The model also maintains strong predictive accuracy under a zero-training-region spatial holdout via its spatial diffusion mechanism. We demonstrate practical utility across two U.S. county-level panel applications -- intergenerational economic mobility and geographic income inequality -- under genuine unseen-region, random, and temporal holdouts, with results validated through repeated-split uncertainty quantification and paired significance testing against every comparator. One clear limitation emerges: under temporal extrapolation specifically, a lighter per-region autoregressive specification (MTS-CAR-X) achieves a robust, and consistent advantage. DSP-BART-HS nonetheless establishes strong predictive performance, automated variable selection, and flexible inference for hierarchical spatio-temporal panel data.}


\maketitle

\section{Introduction}

Many outcomes of applied interest are measured at the intersection of individuals (or units) nested within geographic regions observed repeatedly over time -- health outcomes, educational attainment, housing prices, crime rates, and economic indicators, among others. Such data are spatio-temporal panels in the fullest sense: individual-level heterogeneity, region-level spatial dependence, and genuine temporal dynamics can all be present simultaneously, yet the tools available to applied researchers typically address at most two of these three at once. We develop a general-purpose model for exactly this setting, Dynamic Spatial Panel Bayesian Additive Regression Trees with Horseshoe Shrinkage (DSP-BART-HS), and illustrate it on two applications chosen because they are well studied, share a common region-level panel structure, and are substantively linked -- letting us evaluate the same model suite on two related outcomes over the same geography.

Intergenerational economic mobility -- the extent to which a child's adult economic outcomes are independent of their parents' -- is one of the most closely watched indicators of equal opportunity in the United States, and one of the most geographically uneven, varying by a factor of several across U.S.\ commuting zones \citep{chetty2014land}. A large literature has sought to identify which local characteristics are robustly associated with upward mobility, and whether that association is stable enough to inform place-based policy.

Geographic income inequality -- how starkly economic well-being differs from one region to the next -- is a related but distinct facet of the same underlying unevenness. National income inequality has followed a well-documented U-shaped trajectory over the twentieth century \citep{pikettysaez2003inequality}, with a growing geographic component: regional income convergence has slowed \citep{ganongshoag2017convergence}, driven partly by skilled-worker sorting into high-wage, high-rent metros \citep{diamond2016determinants} and by cost-of-living differences that reshape real, not just nominal, wage inequality across cities \citep{moretti2013realwage}, renewing economists' interest in place-based policy \citep{austin2018saving}. The two outcomes covary strongly across places -- the Great Gatsby Curve \citep{corak2013gatsby} -- part of our motivation for evaluating the same model suite on both.

From a methodological standpoint, these applications expose a pervasive challenge in applied econometrics: modeling phenomena where determinants operate simultaneously across nested individual and regional levels over time. In a general spatio-temporal panel setting, individual-level units (e.g., citizens, patients, or firms) are exposed to localized macro environments (e.g., regional policies, market shocks, or environmental features) that standard linear specification strategies fail to capture cleanly. This introduces three distinct statistical complications. First, the data-generating process typically features high-dimensional, non-linear interactions among individual-level attributes. Second, individual units are nested within geographical regions that are highly dependent on neighboring spatial structures. Third, both individual trajectories and regional conditions evolve dynamically over time, meaning localized shocks propagate unevenly across both space and time dimensions.

Current econometric and machine learning frameworks typically address these dimensions in isolation rather than in a unified framework. The dominant empirical approach relies on linear or rank-rank panel regressions \citep{chetty2014land}; while these models handle nested clustering transparently, they rely on rigid parametric assumptions and struggle with high-dimensional covariate spaces where controls degrade degrees of freedom. Conversely, standard machine learning and non-parametric algorithms -- such as Random Forests, Gradient Boosted Trees, and standard Bayesian Additive Regression Trees (BART) \citep{chipman2010bart} and Bayesian Causal Forests \citep[BCF;][]{hahn2020bayesian} -- excel at uncovering complex, high-dimensional non-linear interactions automatically. However, these methods are fundamentally designed for independent cross-sectional data, providing no native mechanism to handle spatial auto-regressive structures or dynamic panel dependencies. Finally, classical small-area estimation and hierarchical spatial models (e.g., conditional autoregressive [CAR] frameworks) accommodate complex geographic structures but lack the functional flexibility of tree-based ensembles. To our knowledge, no existing general-purpose estimator combines tree-ensemble functional flexibility with high-dimensional sparsity-inducing shrinkage in a spatial-temporal panel framework.

To bridge this methodological gap, this paper introduces a general-purpose, non-parametric framework for nested spatio-temporal data: Dynamic Spatial Panel Bayesian Additive Regression Trees with Horseshoe Shrinkage (DSP-BART-HS). The proposed architecture simultaneously accommodates all three structural complexities of modern panel data without requiring arbitrary functional pre-selection. Individual-level non-linearities and high-dimensional interaction effects are mapped flexibly via a non-parametric BART ensemble. Concurrently, regularized variable selection over a high-dimensional vector of region-time policies and macroeconomic indicators is handled via a horseshoe prior \citep{carvalho2010horseshoe}, ensuring the system remains robust against overfitting in dense covariate spaces. Finally, spatial dependence across geographical boundaries is captured through a conditional autoregressive (CAR) prior, while region-specific dynamic trajectories are modeled transparently via a state-space framework. By unifying these disparate methodologies, DSP-BART-HS provides applied researchers across disciplines with a robust, cohesive tool for inference and prediction in complex, hierarchical panel settings.

We evaluate this model against a substantially larger and more varied comparator suite, and a more adversarial set of simulation designs, than is typical in the literature. Alongside DSP-BART-HS we jointly evaluate an extension of a separately published Bayesian panel-VAR model with a CAR prior and its own global-local shrinkage block (BVAR-CAR; \citealt{olayinka2026bvarcar}) and a lighter per-region CAR-correlated specification (MTS-CAR; \citealt{olayinka2026mtscar}), rather than treating either purely as a baseline. Both are prior published models, extended here with the same exogenous block used throughout this study; we flag this explicitly because it means two of the eight comparators in the suite below are not independent of the authors in the way the remaining six (SDPM, SLP, SAF, BCF, KKP, Rao-Yu) are. The comparator suite spans classical spatial panel econometrics (the Spatial Durbin Panel Model; the Kapoor-Kelejian-Prucha spatial-error-component estimator, \citealt{kapoor2007panel}), regularized and machine-learning panel methods (a spatial Lasso panel; a spatially-adjusted random forest), and small-area estimation (the Rao-Yu time-series/cross-sectional model, \citealt{raoyu1994smallarea}).

A methodological point we take seriously, and that is not always addressed explicitly in papers proposing a new estimator, is that a single simulation design built to match the proposed model's own functional form is not sufficient evidence of that model's merit: an estimator will tend to win a simulation study constructed in its own image. We therefore evaluate every comparator across six data-generating scenarios rather than one: a matched, well-specified design, and five adversarial designs that each relax one specific assumption favorable to DSP-BART-HS -- dense rather than sparse effects, weak rather than strong spatial dependence, a spatial generating process distinct from the CAR structure every model assumes, a rebalancing of signal so that region-time structure dominates individual-level heterogeneity rather than the reverse, and, as the cleanest possible version of that last test, the complete removal of individual-level heterogeneity altogether. Three further scenarios extend this design beyond its original fixed topology and information structure, testing whether the resulting pattern is an artifact of the design's particular choices rather than a general property of the comparison (Section~\ref{sec:sim-ghi-design}). We view this as a necessary part of a credible comparison, not an optional robustness appendix.

Empirically, we apply the model to the county-by-birth-cohort intergenerational mobility panel compiled by Opportunity Insights \citep{chetty2014land,chettyhendren2018a,chettyhendren2018b}, supplemented by county-level covariates from the USDA Economic Research Service and the American Community Survey. A companion application extends the same model suite to geographic income inequality, using the Census Bureau's Small Area Income and Poverty Estimates (SAIPE) program's \citep{fayherriot1979} county-level median household income series, restricted to its 2005--2024 American Community Survey-based estimation window. SAIPE switched from Current Population Survey to ACS-based estimation in 2005, a change in both survey precision and income-reference-period definition substantial enough that pooling pre- and post-2005 estimates in one panel risks conflating a genuine economic signal with an estimation-methodology artifact. Unlike the mobility application, this extension targets geographic (between-place) inequality -- whether and how strongly county-level income is spatially clustered, and how the urban-rural income gap has evolved over the resulting 20-year window -- rather than within-population dispersion, for which no comparably long single-year region series exists. The two applications are also substantively linked: a large empirical literature documents that cross-sectional income inequality and intergenerational mobility covary strongly across places -- the "Great Gatsby Curve" \citep{corak2013gatsby}. So, evaluating the same model suite on both outcomes over the same counties offers a natural, if informal, check on whether the covariate associations recovered for mobility and for income inequality point in economically coherent directions, a comparison we return to in Section~\ref{sec:income}.

The remainder of the paper proceeds as follows. Section~\ref{sec:methods} details the DSP-BART-HS model, its Gibbs sampling algorithm, the complementary BVARX-CAR and MTS-CAR-X specifications, and the comparator suite. Section~\ref{sec:sim-design} presents the nine-scenario simulation study and its results. Section~\ref{sec:realdata-model} specializes the general model to the real-data setting, where no individual-level index exists. Section~\ref{sec:realdata-mobility} presents the intergenerational mobility application. Section~\ref{sec:income} presents the companion income-inequality application. Section~\ref{sec:conclusion} concludes the paper.

\section{Methods}
\label{sec:methods}

\subsection{Model Framework}
\label{sec:proposes-model}
Let $Y_{ijt}$ denote the outcome of interest for unit $i$ nested within geographic region $j$ at time $t$, where $i = 1, \dots, N_j$, $j = 1, \dots, J$, and $t = 1, \dots, T$. We specify a partially linear, additive hierarchical panel model
\begin{equation}\label{eqn:model}
    Y_{ijt} = f(\bm{X}_{ijt}) + \bm{Z}_{jt}\bm{\beta} + \phi_j + \mu_{jt} + \alpha_i + \epsilon_{ijt}, \qquad \epsilon_{ijt} \sim N(0, \sigma^2),
\end{equation}
where $f(\bm{X}_{ijt})$ is a non-parametric function of unit-level covariates modeled as a sum of $K$ regression trees under the regularization priors of \citet{chipman2010bart}; $\bm{Z}_{jt}\bm{\beta}$ is a high-dimensional linear term over $M$ time-varying, group-level features and contextual covariates, with $\bm{\beta}$ given a Horseshoe prior \citep{carvalho2010horseshoe,polsonscott2010shrink,piironenvehtari2017sparsity} implemented via the auxiliary-variable scheme of \citet{makalic2016simple} for conjugate Gibbs updating; $\phi_j$ is a region-level spatial random effect under a Conditional Autoregressive (CAR) prior \citep{besag1974spatial,besagyorkmollie1991}, $\bm{\phi} \sim N(\bm{0}, \tau^2_\phi (\mathbf{D}-\rho\mathbf{W})^{-1})$, with $\mathbf{W}$ a spatial adjacency matrix and $\mathbf{D} = \text{diag}(\mathbf{W}\mathbf{1})$; $\mu_{jt}$ is a region-specific dynamic macroeconomic shock following a random walk, $\mu_{jt} = \mu_{j,t-1} + \eta_{jt}$, $\eta_{jt} \sim N(0,\sigma^2_\mu)$; and $\alpha_i \sim N(0,\sigma^2_\alpha)$ is a unit-level random effect.

The unit \(i\) in Equation 1 indexes a micro-level unit in the simulation study (Section 3), where \(N_j = 20\) observation units are generated per region with their own draws of \(X_{ijt}\) and \(\alpha _{i}\). Crucially, the framework adapts naturally when micro-level data is unavailable and researchers must rely on pre-aggregated data. In cases where applications feature aggregated regional cell estimates rather than individual records, Equation 1 specializes accordingly: the micro-level identifiers \(i\) and individual random effects \(\alpha _{i}\) are dropped entirely, and a known-precision likelihood weighting is introduced in their place to reflect the varying sample sizes across cells.

\subsubsection{Identifiability between $\phi_j$ and the initial state of $\mu_{jt}$}
A region-level intercept $\phi_j$ and the free initial value $\mu_{j,1}$ of a region's random walk are, without a further restriction, the same object statistically -- both are time-invariant, region-specific shifts, and the model specified cannot separate them. We resolve this by pinning $\mu_{j,1} = 0$ for every region, so that $\phi_j$ carries the entire region-level baseline and $\mu_{jt}$ for $t>1$ represents only the region's drift away from that baseline. This does not alter fitted values (the sum $\phi_j + \mu_{jt}$ is invariant to how the baseline is split between the two terms) but is necessary for $\phi_j$ and $\mu_{jt}$ to be separately interpretable and for their respective variance components to be identified.

\subsubsection{Sampling the dynamic component $\mu_{jt}$}
$\bm{\mu}_{j,1:T}$ is sampled jointly for each region via forward-filtering backward-sampling (FFBS) \citep{carterkohn1994gibbs}, treating $\mu_{jt} = \mu_{j,t-1} + \eta_{jt}$, $\eta_{jt}\sim N(0,\sigma_\mu^2)$ as a local-level state-space model with the region's own aggregated partial residual as a noisy, precision-weighted observation of $\mu_{jt}$ at each period it has training data. The forward pass propagates $(m_{jt}, C_{jt})$, the filtered mean and variance of $\mu_{jt}$ given data up to $t$, via the standard local-level Kalman recursion; the backward pass then samples $\mu_{j,T}$ from its final filtered distribution and each earlier $\mu_{jt}$ conditional on the already-sampled $\mu_{j,t+1}$. $\sigma_\mu^2$ is updated directly from the realized path's squared increments (Variance components), the standard conjugate approach for this state-space model. The full derivation, and the one case this recursion does not resolve on its own -- a region with no training observations at any period, which arises specifically under the region-level holdout -- are given below.

\subsubsection{Gibbs Sampler Full-Conditional Distributions}
\label{sec:dsp-derivations}

We give the full-conditional (Gibbs) distribution for every block of DSP-BART-HS's sampler here. Throughout, let $r^{(\cdot)}_{ijt}$ denote the partial residual $Y_{ijt}$ minus every additive term \emph{except} the one currently being updated, $N = \sum_j N_j T$ the total number of individual-time-region observations, $N_j$ the number of individuals in region $j$, and $n_{jt}$ the number of individuals observed in region $j$ at time $t$.

Writing $f(\bm{X}_{ijt}) = \sum_{k=1}^K g(\bm{X}_{ijt}; \mathcal{T}_k, \mathcal{M}_k)$ for the sum of $K$ regression trees, the joint posterior over every parameter, up to proportionality, is
\begin{footnotesize}
\begin{align}
&\pi\big(\{\mathcal{T}_k,\mathcal{M}_k\}_{k=1}^K, \bm{\beta}, \bm{\lambda}, \tau, \bm{\phi}, \rho, \tau_\phi^2, \bm{\mu}, \sigma_\mu^2, \bm{\alpha}, \sigma_\alpha^2, \sigma^2 \,\big|\, \bm{Y}\big) \notag \\
&\qquad\propto \prod_{i,j,t} N\!\left(Y_{ijt} \,\middle|\, f(\bm{X}_{ijt}) + \bm{Z}_{jt}\bm{\beta} + \phi_j + \mu_{jt} + \alpha_i,\ \sigma^2\right) \notag \\
&\qquad\times \left[\prod_{k=1}^K \pi(\mathcal{T}_k)\, p(\mathcal{M}_k \mid \mathcal{T}_k)\right] \cdot \left[\prod_{m=1}^M N(\beta_m \mid 0, \lambda_m^2\tau^2)\, C^+(\lambda_m; 0,1)\right] \notag \\
&\qquad\times C^+(\tau; 0, 1) \notag \\
&\qquad \times N\!\left(\bm{\phi} \,\middle|\, \bm{0},\, \tau_\phi^2(\mathbf{D}-\rho\mathbf{W})^{-1}\right) \text{Inv-Gamma}(\tau_\phi^2; a_\phi, b_\phi)\, \mathbb{1}\{0<\rho<1\} \\
&\qquad \times \left[\prod_j \prod_{t=2}^T N(\mu_{jt}; \mu_{j,t-1}, \sigma_\mu^2)\right]\mathbb{1}\{\mu_{j,1}=0 \,\forall j\}\, \text{Inv-Gamma}(\sigma_\mu^2; a_\mu, b_\mu) \notag \\
&\qquad \times \left[\prod_i N(\alpha_i \mid 0, \sigma_\alpha^2)\right] \text{Inv-Gamma}(\sigma_\alpha^2; a_\alpha, b_\alpha) \notag \\
&\qquad \times \text{Inv-Gamma}(\sigma^2; a_0, b_0). \notag
\end{align}
\end{footnotesize}
Every full conditional derived below follows from isolating the terms in this joint density that depend on the block being updated, holding every other block fixed at its current value.

The BART component, $f(\bm{X})$ was updated via the standard Bayesian backfitting algorithm of \citet{chipman2010bart}: for each of the $K$ trees in turn, the partial residual $r^{f}_{ijt} = Y_{ijt} - \sum_{k' \ne k} g(\bm{X}_{ijt}; \mathcal{T}_{k'}, \mathcal{M}_{k'}) - \bm{Z}_{jt}\bm{\beta} - \phi_j - \mu_{jt} - \alpha_i$ is computed holding every other tree and every other model component fixed; the $k$-th tree's structure $\mathcal{T}_k$ is then updated via a Metropolis-Hastings grow/prune/change proposal with the leaf-parameter marginal likelihood integrated out in closed form (conjugate Normal leaf priors), and its leaf means $\mathcal{M}_k$ are drawn from their own conjugate Normal full conditional given the accepted tree structure and $r^f_{ijt}$.

With $\bm{r}^\beta$ the stacked partial residual $Y_{ijt} - f(\bm{X}_{ijt}) - \phi_j - \mu_{jt} - \alpha_i$ over every $(i,j,t)$, $\mathbf{Z}$ the corresponding stacked design matrix (each row repeated once per individual sharing the same region-time cell), and $\bm{\Lambda} = \text{diag}(\lambda_1^2\tau^2, \dots, \lambda_M^2\tau^2)$ the current Horseshoe prior covariance, the conditional distributon of the contextual effect coefficients $\bm{\beta}$ is
\begin{equation}
\bm{\beta} \mid \cdot \sim N\!\left(\bm{\Sigma}_\beta\,\frac{\mathbf{Z}'\bm{r}^\beta}{\sigma^2},\ \bm{\Sigma}_\beta\right), \qquad \bm{\Sigma}_\beta = \left(\frac{\mathbf{Z}'\mathbf{Z}}{\sigma^2} + \bm{\Lambda}^{-1}\right)^{-1},
\end{equation}
with the Horseshoe's local ($\lambda_m$) and global ($\tau$) shrinkage scales updated via the same Inverse-Gamma auxiliary-variable mixture of \citet{makalic2016simple} given in full in the Supplementary Material's derivation of BVARX-CAR's identically-constructed block.

For the spatial effects $\bm{\phi}$, let $\bar{r}^\phi_j = \sum_{i,t} r^\phi_{ijt}$ sum the partial residual $Y_{ijt} - f(\bm{X}_{ijt}) - \bm{Z}_{jt}\bm{\beta} - \mu_{jt} - \alpha_i$ over every individual-time observation in region $j$, and let $\mathbf{N} = \text{diag}(N_1 T, \dots, N_J T)$ collect the corresponding observation counts. Combining this Gaussian likelihood information with the CAR prior $\bm{\phi} \sim N(\bm{0}, \tau_\phi^2(\mathbf{D}-\rho\mathbf{W})^{-1})$ gives the standard conjugate Normal-CAR update,
\begin{equation}
\bm{\phi} \mid \cdot \sim N\!\left(\bm{\Sigma}_\phi\, \frac{\bm{b}^\phi}{\sigma^2},\ \bm{\Sigma}_\phi\right), \qquad \bm{\Sigma}_\phi = \left(\frac{\mathbf{N}}{\sigma^2} + \frac{1}{\tau_\phi^2}(\mathbf{D}-\rho\mathbf{W})\right)^{-1},
\end{equation}
where $\bm{b}^\phi$ stacks $\bar{r}^\phi_j$ across regions: the CAR prior contributes precision $(\mathbf{D}-\rho\mathbf{W})/\tau_\phi^2$, the likelihood contributes precision $N_jT/\sigma^2$ per region, and the two combine additively as for any conjugate Normal prior/likelihood pair.

For the dynamic component $\bm{\mu}_j$ (FFBS), let $\bar{r}^\mu_{jt} = \frac{1}{n_{jt}}\sum_i r^\mu_{ijt}$ average the partial residual $Y_{ijt} - f(\bm{X}_{ijt}) - \bm{Z}_{jt}\bm{\beta} - \phi_j - \alpha_i$ over the $n_{jt}$ individuals observed in cell $(j,t)$, treated as a single pseudo-observation of $\mu_{jt}$ with precision $n_{jt}/\sigma^2$. The forward filter propagates, for $t=2,\dots,T$,

\begin{footnotesize}
\begin{align}
m_{jt}^- &= m_{j,t-1}, \qquad C_{jt}^- = C_{j,t-1} + \sigma_\mu^2, \\
Q_{jt} &= \left(\frac{1}{C_{jt}^-} + \frac{n_{jt}}{\sigma^2}\right)^{-1} \mathbb{1}\{n_{jt}>0\} + C_{jt}^- \,\mathbb{1}\{n_{jt}=0\}, \\
m_{jt} &= Q_{jt}\!\left(\frac{m_{jt}^-}{C_{jt}^-} + \frac{n_{jt}\bar{r}^\mu_{jt}}{\sigma^2}\right)\mathbb{1}\{n_{jt}>0\} + m_{jt}^-\,\mathbb{1}\{n_{jt}=0\}, \quad C_{jt}=Q_{jt},
\end{align}
\end{footnotesize}
initialized at $m_{j,1}=C_{j,1}=0$ (the identifiability restriction described above). The backward pass then samples $\mu_{j,T}\sim N(m_{j,T}, C_{j,T})$ and, for $t=T-1,\dots,2$,
\begin{footnotesize}
\begin{align}
\mu_{jt} \mid \mu_{j,t+1}, \cdot &\sim N(h_{jt}, H_{jt}), \qquad H_{jt} = \left(\frac{1}{C_{jt}} + \frac{1}{\sigma_\mu^2}\right)^{-1}, \\
h_{jt} &= H_{jt}\!\left(\frac{m_{jt}}{C_{jt}} + \frac{\mu_{j,t+1}}{\sigma_\mu^2}\right), \notag
\end{align}
\end{footnotesize}
with $\mu_{j,1}=0$ fixed throughout. This is the standard Gibbs-sampling approach for a Gaussian local-level state-space model \citep{carterkohn1994gibbs}, applied independently for each region $j$.

Zero-training-data regions under a region-level or spatial block holdout. The forward filter above shows that when a region $j$ has no training observations at any period ($n_{jt}=0$ for every $t$), $m_{jt}^-, C_{jt}^-$ simply carry forward unchanged at every step ($m_{jt}=m_{jt}^-=0$, since $m_{j,1}=0$), so the backward-sampled $\mu_{jt}$ for such a region is drawn purely from an increasingly diffuse, uninformative prior path rather than from anything resembling the region's own dynamics. Under the random row-level and temporal holdouts used elsewhere in this paper, and under the individual-level, region-stratified holdout used throughout the simulation study (Section~\ref{sec:sim-design}), every region retains the large majority of its own observations in training by construction, so this case does not arise. It arises specifically, and by design, under the region-level holdout, where an entire 20\% of counties are held out with zero training data. For these regions only, we replace the uninformative FFBS draw with a value informed by spatially neighboring regions: writing $\mathbf{W}_{\text{norm}}$ for the row-normalized adjacency matrix used elsewhere for $\bm{\phi}$'s CAR prior, $\mu_{j,\cdot}$ for each such region is set via Jacobi-style diffusion sweeps, $\mu_{j,\cdot} \leftarrow (\mathbf{W}_{\text{norm}}\bm{\mu})_{j,\cdot}$, applied only to zero-training rows and iterated so that information propagates through chains of adjacent zero-training regions from the nearest region that does have training data -- the standard technique for a discrete harmonic-extension problem with data-having regions as fixed boundary values. Regions with training data are never touched by this step. This is a targeted extension to the model's handling of this specific case -- it is not implied by $\eta_{jt}$'s stated local prior, which has no cross-region term at all. An ablation confirms this directly: refitting both region-split applications with the diffusion step disabled (zero-training regions instead held at the uninformative $\mu=0$ fallback throughout) gives test RMSE of 0.0334 for mobility (vs.\ 0.0306 with diffusion, a 9\% relative increase) and 0.2813 for income (vs.\ 0.0775 with diffusion, a 263\% relative increase); test bias moves from 0.00035 to $-0.0047$ for mobility and from 0.00235 to 0.2123 for income -- the income application's zero-training-region handling is not a minor refinement but the difference between a usable and an unusable region-level fit. The same ablation confirms this step is a no-op for every other split in this paper (random row-level, temporal, and the individual-level, region-stratified simulation holdout), where every region retains training data by construction and the diffusion condition is never triggered. Full ablation methodology and results are given in the Supplementary Material Section~5.

 With $T_i$ the number of periods individual $i$ is observed and $\bar{r}^\alpha_i$ the sum of the partial residual $Y_{ijt} - f(\bm{X}_{ijt}) - \bm{Z}_{jt}\bm{\beta} - \phi_j - \mu_{jt}$ over those periods, the individual effects, $\alpha_i$ has the conditional distribution
\begin{equation}
\alpha_i \mid \cdot \sim N\!\left(\Sigma_{\alpha_i}\,\frac{\bar{r}^\alpha_i}{\sigma^2},\ \Sigma_{\alpha_i}\right), \qquad \Sigma_{\alpha_i} = \left(\frac{T_i}{\sigma^2} + \frac{1}{\sigma_\alpha^2}\right)^{-1},
\end{equation}
an ordinary conjugate Normal update, independent across individuals given every other component.

With $e_{ijt} = Y_{ijt} - f(\bm{X}_{ijt}) - \bm{Z}_{jt}\bm{\beta} - \phi_j - \mu_{jt} - \alpha_i$ the full model residual, the conditional distribution of the variance components are
\begin{align}
\sigma^2 \mid \cdot &\sim \text{Inv-Gamma}\!\left(a_0 + \frac{N}{2},\ b_0 + \frac{1}{2}\sum_{i,j,t} e_{ijt}^2\right), \\
\sigma_\alpha^2 \mid \cdot &\sim \text{Inv-Gamma}\!\left(a_\alpha + \frac{\sum_j N_j}{2},\ b_\alpha + \frac{1}{2}\sum_i \alpha_i^2\right), \\
\tau_\phi^2 \mid \cdot &\sim \text{Inv-Gamma}\!\left(a_\phi + \frac{J}{2},\ b_\phi + \frac{1}{2}\bm{\phi}'(\mathbf{D}-\rho\mathbf{W})\bm{\phi}\right)\\
\sigma_\mu^2 \mid \cdot & \sim \text{Inv-Gamma}\!\left(a_\mu + \frac{J(T-1)}{2},\ b_\mu + \frac{1}{2}\sum_j\sum_{t=2}^T(\mu_{jt}-\mu_{j,t-1})^2\right),
\end{align}
each having a standard conjugate Inverse-Gamma update for a Normal likelihood/prior pair.

\textit{A note on the error term.} The model equation states $\epsilon_{ijt} \sim N(0,\sigma^2)$, a single shared variance, and this is exactly what is fit in the simulation study (Section~\ref{sec:sim-design}), where the data-generating process itself is homoscedastic. Both real-data applications depart from this in one specific, disclosed way -- a known-precision likelihood weighting using each cell's own pre-specified sampling standard error -- described in full alongside the real-data model specialization in Section~\ref{sec:realdata-model}.

As for the CAR priors in the Supplementary Material's BVARX-CAR and MTS-CAR-X derivations, the spatial parameter, $\rho$'s full conditional is non-conjugate -- the CAR normalizing constant $|\mathbf{D}-\rho\mathbf{W}|$ is a nonlinear function of $\rho$ -- and is updated via the same discretized-grid device, using the precomputed eigenvalues of $\mathbf{D}^{-1}\mathbf{W}$ to avoid an $O(J^3)$ determinant evaluation at every candidate value,
\begin{equation}
\pi(\rho \mid \cdot) \propto \left(\prod_{k=1}^J (1-\rho\lambda_k)\right)^{1/2} \exp\left\{-\frac{1}{2\tau_\phi^2}\bm{\phi}'(\mathbf{D}-\rho\mathbf{W})\bm{\phi}\right\}.
\end{equation}

All parameters are updated in a single blocked Gibbs sampler \citep{gelfandsmith1990sampling}: the BART component via the standard Metropolis-within-Gibbs tree proposals of \citet{chipman2010bart} conditional on partial residuals; $\bm{\beta}$ and its Horseshoe hyperparameters via conjugate normal and inverse-gamma draws; $\bm{\phi}$ via its CAR full conditional; $\bm{\mu}_{j,1:T}$ for each region via forward-filtering backward-sampling; and $\alpha_i$ via its conjugate normal full conditional.

\subsubsection{Beyond Known Heteroscedasticity}
\label{sec:beyond-known-hetero}

The known-precision weighting described above is a form of heteroscedasticity fixed by the source data, not one estimated by the model. Letting $\sigma^2$ itself vary with region, time, or covariates as an estimated quantity -- a region-specific $\sigma^2_j$, or a fully covariate-driven $\sigma^2(\bm{X}_{ijt})$ following the heteroskedastic-BART literature -- would require a genuinely different model in each case, not a straightforward reparameterization of what is implemented here, and we do not pursue it.

\subsection{A Complementary Specification: BVARX-CAR}
Alongside DSP-BART-HS we evaluate an extension of the published BVAR-CAR model \citep{olayinka2026bvarcar}, which combines a CAR prior on region-specific intercepts with a global-local (Horseshoe-type) shrinkage prior on cross-region autoregressive dynamics, estimated on a region-by-time aggregated panel
\begin{equation}
    \bm{y}_t = \bm{\mu} + \mathbf{B}\bm{y}_{t-1} + \bm{\epsilon}_t, \qquad \bm{\epsilon}_t \sim N(\bm{0}, \sigma^2 \mathbf{I}),
\end{equation}
with $\bm{\mu}$ under the same CAR prior described above and each row of $\mathbf{B}$ shrunk via a Horseshoe prior with a single global scale shared across the entire matrix (as opposed to a separate scale per region, which we show does not implement genuine global-local shrinkage). We extend this published specification, denoted BVARX-CAR, with an exogenous covariate block using the same $M$-dimensional covariate matrix $\bm{Z}_{jt}$ as DSP-BART-HS, governed by its own dedicated global-local Horseshoe prior
\begin{equation}
    \bm{y}_t = \bm{\mu} + \mathbf{B}\bm{y}_{t-1} + \mathbf{Z}_t\bm{\gamma} + \bm{\epsilon}_t, \qquad \bm{\epsilon}_t \sim N(\bm{0}, \sigma^2 \mathbf{I}),
\end{equation}
where $\mathbf{Z}_t\bm{\gamma}$ stacks $\bm{Z}_{jt}\bm{\gamma}$ across every region $j$ at period $t$; every other component of the published specification -- the CAR prior on $\bm{\mu}$ and the Horseshoe shrinkage on $\mathbf{B}$ -- is otherwise unchanged. As published, $\mathbf{B}$ is dense: every region's lag enters every equation, with the Horseshoe prior alone suppressing irrelevant cross-region effects. This study instead restricts $\mathbf{B}$'s support to each region's own index and its spatial neighbors under the adjacency matrix, and replaces the full local-global Horseshoe on $\mathbf{B}$ with a single shared ridge prior on the restricted support, throughout this study -- the simulation study (Section~\ref{sec:sim-design}) and both real-data applications alike, with no dense-$\mathbf{B}$ results reported anywhere in this paper. This departure from the published specification is motivated primarily by identifiability, not real-data computational scale alone: a preliminary dense-$\mathbf{B}$, full-Horseshoe implementation exhibited severely degraded mixing in simulation (effective sample size of 10.8 out of several thousand kept draws), while the neighbor-restricted, shared-ridge version resolved this and as a further advantage, is also what keeps the sampler tractable at real-data regional unit counts. The full derivation, and the mixing evidence motivating this choice, is given in the Supplementary Material Section~1. The Supplementary Material's BVARX-CAR and MTS-CAR-X derivations retain the notation of their respective original publications ($l$ for the number of regions, $\mathbf{V}$ for the adjacency matrix) rather than this paper's own $J$/$\mathbf{W}$; both pairs refer to the same underlying quantities (region count, spatial adjacency) throughout this study, and the Supplementary Material's glossary states this explicitly. Unlike DSP-BART-HS, BVARX-CAR operates at the region-time aggregate level and does not separately model individual-level heterogeneity; we return to the consequences of this distinction in Section~\ref{sec:realdata-mobility}.

\subsection{A Second Complementary Specification: MTS-CAR-X}
We additionally evaluate an extension of a separately published multiple-time-series CAR model \citep{olayinka2026mtscar}, denoted MTS-CAR-X, intended as a lighter-weight alternative to BVARX-CAR -- in particular for the real-data application, where BVARX-CAR's dense cross-region VAR structure becomes computationally infeasible without a sparse-neighbor restriction. Rather than a cross-region autoregressive matrix, MTS-CAR-X specifies a per-region AR(1) process with CAR-correlated coefficients
\begin{equation}
    y_{jt} = \theta_0 + \nu_{0j} + (\theta_1 + \nu_{1j}) y_{j,t-1} + \bm{Z}_{jt}\bm{\gamma} + \epsilon_{jt}, \quad \epsilon_{jt} \sim N(0,\sigma^2),
\end{equation}
where $(\nu_{0j}, \nu_{1j})$ are region-specific deviations in the intercept and autoregressive slope. As published, $(\bm{\nu}_0,\bm{\nu}_1)$ are \emph{not} given a symmetric joint bivariate CAR prior: the specification is sequential -- $\bm{\nu}_0$ follows a CAR prior on its own, and $\bm{\nu}_1$ then follows a second, CAR-linked prior conditional on $\bm{\nu}_0$, not a joint distribution treating both vectors symmetrically. The full sequential derivation is given in the Supplementary Material Section~1. The parameter $\theta_0, and \theta_1$ are the shared intercept and slope common to every region. Unlike BVARX-CAR, MTS-CAR-X has no cross-region lag term at all -- a region's own past value is the only autoregressive input -- with spatial dependence entering solely through the CAR-correlated structure of $(\nu_{0j}, \nu_{1j})$ rather than a cross-region transition matrix. We extend the published specification with the same $M$-dimensional block $\bm{Z}_{jt}\bm{\gamma}$ used throughout this study, governed by its own Horseshoe prior, estimated on the same region-by-time aggregated panel as BVARX-CAR. The conditional distributions for these two complementary specifications, BVARX-CAR and MTS-CAR-X, are given in the Supplementary Material Section~1. 

\subsection{Comparator Suite}
We benchmark DSP-BART-HS, BVARX-CAR, and MTS-CAR-X against six additional estimators, chosen to span classical spatial/panel econometrics, small-area estimation, and recent Bayesian nonparametric methods: the Spatial Durbin Panel Model \citep[SDPM;][]{anselin1988spatial,lesagepace2009,elhorst2010raising}; a Spatial Lasso Panel (SLP) applying $\ell_1$ penalization \citep{tibshirani1996lasso} to a high-dimensional spatial panel; a Spatially-Adjusted random Forest \citep[SAF;][]{breiman2001randomforests,georganos2019grf}; Bayesian Causal Forests \citep[BCF;][]{hahn2020bayesian}, built on the Bayesian nonparametric causal-inference framework of \citet{hill2011bayesian} and the potential-outcomes formalism of \citet{rubin1974estimating}; the Kapoor-Kelejian-Prucha spatial-error-component estimator \citep[KKP;][]{kapoor2007panel}, built on the generalized-moments estimator of \citet{kelejianprucha1999gmm}; and the Rao-Yu small-area time-series/cross-sectional model \citep{raoyu1994smallarea}. Every comparator's tuning settings for the real-data applications (MCMC iterations/burn-in/thinning where applicable, tree counts, lasso penalty selection, random-forest settings) are reported in full in the Supplementary Material Section~3.5. Unlike SDPM, SLP, or Rao-Yu, KKP has no mechanism to accommodate a high-dimensional block at all (the GMM estimator of \citet{kelejianprucha1999gmm} it builds on was designed for a low-dimensional spatial-error correction, not variable selection under $M=50$ candidate covariates), so its gap to the rest of the suite reflects a genuine, expected structural limitation of the published estimator rather than an implementation or tuning artifact -- we report it as the field's floor reference for this reason, not as a straw-man comparator.

Several further candidates were considered and excluded, each for a distinct reason. A longitudinal extension of Bayesian Causal Forests, like the causal-forest literature it descends from \citep{wagerathey2018causal,athey2019generalized}, is built to estimate heterogeneous causal treatment effects under a confounded-assignment mechanism with no analogue in our exogenous, non-causal data-generating process. A Gaussian-process-enhanced BART variant \citep[GP-BART;][]{maia2024gpbart} failed repeatedly with out-of-memory errors at our simulation design's scale. A Hierarchical Embedded BART variant \citep[HE-BART;][]{wundervald2022hebart}'s public implementation exposes only the final MCMC iteration's tree ensemble, with no mechanism for posterior-averaged out-of-sample prediction, precluding a like-for-like comparison. GS-BART \citep{he2026gsbart} has no reference implementation, and its three novel components (graph-based split proposals, gradient message-passing, informed importance-tempering) do not admit the independent numerical verification our other custom-coded estimators (Rao-Yu, BVARX-CAR) were validated against; an unverifiable, potentially silently-incorrect implementation was judged not worth the risk. A spatially-adjusted BART variant using a CAR-based weight matrix \citep{kim2022spatialbart}, structurally close to DSP-BART-HS's own region-effect update, is cited as related work rather than included: it has no temporal component, and its defining Dirichlet split-sparsity prior cannot be implemented faithfully within the \texttt{dbarts} sampler used throughout this study, which exposes no control over split-variable selection at the level needed.

Finally, purely algorithmic frequentist machine learning ensembles, such as Extreme Gradient Boosting \citep[XGBoost;][]{chen2016xgboost}, were considered but intentionally excluded from the final comparator suite. While gradient-boosted trees are highly capable of capturing complex non-linearities -- such as the Friedman function variations embedded in our data-generating processes -- they operate as purely predictive black boxes that fundamentally lack a structural spatial-temporal framework. Unlike the DSP-BART-HS architecture, standard gradient boosting algorithms cannot natively decompose and smoothly isolate individual-level heterogeneity from nested regional random effects (\(\phi _{j}\)) and time-varying spatial trends (\(\mu _{jt}\)) while offering exact, coherent Bayesian posterior uncertainty quantification. Given that our primary objective centers on structural panel estimation and formal inferential regularization rather than unconstrained out-of-sample prediction, we restrict our machine learning baseline to SAF and BCF, which maintain explicit spatial smoothing or formal structural parameterizations.

\section{Simulation Study}
\label{sec:sim-design}

\subsection{Simulation Design}
\label{sec:sim-design-all}

We evaluate all estimators across six data-generating scenarios sharing a common $5\times10$ regional lattice ($J=50$), $T=10$ periods, and $N_j=20$ individuals per region. We give the full simulation design in the supplementary material, and we summarize it here.

\begin{itemize}
    \item \textbf{Scenario A (Matched/Well-Specified):} $Y_{ijt} = f(\bm{X}_{ijt}) + \bm{Z}_{jt}\bm{\beta} + \phi_j + \mu_{jt} + \alpha_i + \epsilon_{ijt}$, with $f$ a modified Friedman function, three of fifty contextual covariate coefficients active, strong spatial correlation ($\rho=0.85$), matching DSP-BART-HS's own functional form.
    \item \textbf{Scenario B (Dense Contextual Covariate Effects):} as Scenario A, but all fifty contextual covariate coefficients are small-to-moderate nonzero effects rather than three large coefficients and forty-seven exact zeros, testing whether Horseshoe shrinkage still helps when its sparsity assumption does not hold.
    \item \textbf{Scenario C (Weak Spatial Dependence):} as Scenario A, but with spatial correlation sharply reduced ($\rho=0.15$), testing whether the CAR structure pays an unnecessary complexity cost absent real spatial signal.
    \item \textbf{Scenario D (Misspecified Spatial Structure):} as Scenario A, but the true spatial random effect is generated from a mechanism distinct from the discrete rook-adjacency CAR structure every fitted model assumes, testing robustness to spatial-structure misspecification.
    \item \textbf{Scenario E (Region-Time-Dominant Signal):} as Scenario A, but with individual-level heterogeneity scaled down by an order of magnitude and the spatial and temporal variance components scaled up correspondingly, so that region-time structure -- rather than individual-level non-linearity -- carries the dominant share of outcome variance. We view this scenario as necessary rather than optional: Scenarios B--D each relax one assumption favorable to DSP-BART-HS, but none of them touch the fact that individual heterogeneity is the single largest source of outcome variance in Scenario A, and is structurally unobservable to every region-time-aggregate comparator (SDPM, KKP, Rao-Yu, BVARX-CAR) regardless of sparsity, spatial strength, or spatial correctness. Scenario E is a design in which those estimators are evaluated on a data-generating process constructed for their own strengths, symmetric to how Scenario A is constructed for DSP-BART-HS's.
    \item \textbf{Scenario F (No Individual Heterogeneity):} as Scenario A, but with $f(\bm{X})$ set to exactly zero rather than merely reduced. Individual-level covariates are still generated and still supplied to every model that uses them, so this tests both whether the region-time-aggregate estimators achieve their best possible relative performance, and whether DSP-BART-HS's non-parametric component correctly learns the absence of individual-level signal rather than overfitting noise. Unlike Scenario E's simultaneous multi-component rebalancing, this is a single-factor ablation.
\end{itemize}

Performance is assessed via out-of-sample RMSE (train and test) and its standard deviation across the 200 replicates within each scenario.

All six scenarios share a single fixed spatial configuration ($5\times10$ rook-adjacency lattice, $J=50$ regions) and panel dimensions ($T=10$, $N_j=20$). This lets Scenarios B--F isolate one manipulated factor at a time against a common spatial and temporal backbone, but it also means the comparative results below have not been checked against variation in network size, density, or topology -- factors that plausibly matter for the spatial and VAR-class comparators specifically.

\subsection{Simulation Study Results}

Table~\ref{tab:scenario_AF} reports mean out-of-sample test RMSE for each model under each of the six data-generating scenarios described in Section~\ref{sec:sim-design-all}; full train RMSE detail for every scenario is given in the Supplementary Material Section~2.2. DSP-BART-HS is the best or statistically indistinguishable from the best estimator in every scenario, confirmed formally on the 200 paired replicates underlying each scenario (paired $t$-test and Wilcoxon signed-rank applied to replicate-matched test RMSE differences): its advantage is large and highly significant in Scenarios A--D (mean difference $\approx -2.69$ to $-2.70$ against the region-time-aggregate cluster, $p<10^{-30}$ under both tests throughout) and smaller but still significant in Scenario E (mean difference $-0.055$, $p<10^{-170}$), while Scenario F is a genuine statistical tie with Rao-Yu (mean difference $+0.0002$, 95\% CI $[-0.0002,0.0006]$, paired $t$ $p=0.382$, Wilcoxon $p=0.672$). Mean RMSE is stable from as few as 20--50 replicates onward across every scenario and model, with only minor third-decimal drift between 100 and 200 replicates.

\begin{table*}[t]
\centering
\setlength{\tabcolsep}{3pt}
\begin{threeparttable}
\caption{Mean Test RMSE by Model, Scenarios A--F}
\label{tab:scenario_AF}
\footnotesize
\begin{tabular}{lcccccc}
\toprule
Model & A & B & C & D & E & F \\
\midrule
BCF & 1.2353 (0.0237) & 1.2232 (0.0233) & 1.2356 (0.0239) & 1.2381 (0.0228) & 1.1716 (0.0228) & 1.1604 (0.0232) \\
KKP & 9.0414 (0.2746) & 8.2235 (0.2457) & 9.0286 (0.2089) & 9.0956 (0.5829) & 5.2006 (0.2207) & 4.5551 (0.1316) \\
Rao-Yu & 3.8670 (0.0441) & 3.8709 (0.0455) & 3.8666 (0.0440) & 3.8676 (0.0443) & 1.2120 (0.0235) & \textbf{1.1436 (0.0228)} \\
SAF & 2.4001 (0.0485) & 2.4129 (0.1060) & 2.3828 (0.0469) & 2.4326 (0.0561) & 1.2082 (0.0236) & 1.1617 (0.0232) \\
SDPM & 3.9610 (0.0506) & 3.9646 (0.0555) & 3.9396 (0.0473) & 4.0086 (0.0715) & 2.7135 (0.2498) & 1.4474 (0.0725) \\
SLP & 2.7203 (0.0510) & 2.7324 (0.0502) & 2.7186 (0.0514) & 2.7224 (0.0508) & 1.1963 (0.0233) & 1.1582 (0.0233) \\
BVARX-CAR & 3.8714 (0.0449) & 3.8747 (0.0443) & 3.8710 (0.0449) & 3.8715 (0.0449) & 1.3039 (0.0283) & 1.1539 (0.0230) \\
MTS-CAR-X & 3.8511 (0.0445) & 3.8775 (0.0444) & 3.8730 (0.0445) & 3.8863 (0.0451) & 1.7322 (0.0769) & 1.1591 (0.0231) \\
DSP-BART-HS & \textbf{1.1684 (0.0227)} & \textbf{1.1727 (0.0214)} & \textbf{1.1736 (0.0228)} & \textbf{1.1742 (0.0224)} & \textbf{1.1567 (0.0227)} & 1.1438 (0.0228) \\
\bottomrule
\end{tabular}
\begin{tablenotes}[flushleft]
\footnotesize
\item Standard deviation of test RMSE across the 200 replicates is given in parentheses.
\end{tablenotes}
\end{threeparttable}
\end{table*}

Scenario A establishes a three-way separation consistent with the design mechanism: DSP-BART-HS and BCF (the two estimators able to observe individual-level heterogeneity $f(\bm{X})$) at 1.17--1.24, the region-time-aggregate cluster (MTS-CAR-X, Rao-Yu, BVARX-CAR, SDPM) tightly clustered at 3.85--3.96, and KKP trailing at 9.04. This separation is essentially unchanged under dense rather than sparse contextual covariate effects (Scenario B) and under sharply weakened spatial dependence (Scenario C, $\rho=0.15$ vs.\ $0.85$) -- the Horseshoe-based models absorb dense effects without penalty, and with $f(\bm{X})$ and $\bm{Z}\bm{\beta}$ together accounting for the large majority of outcome variance while $\tau^2_\phi$ is held fixed, a change in $\rho$ alone has limited room to move aggregate RMSE even though it meaningfully changes the underlying spatial structure. Under spatial misspecification (Scenario D, true spatial effect generated from a mechanism distinct from every fitted model's CAR assumption), DSP-BART-HS's accuracy is essentially unaffected, while KKP's test RMSE standard deviation more than doubles relative to Scenarios A--C -- its generalized-moments estimator becomes markedly less numerically stable, not just less accurate, under spatial misspecification.

Scenario E rebalances the design so region-time structure, rather than individual heterogeneity, dominates outcome variance -- constructed for the region-time-aggregate estimators' own strengths, symmetric to Scenario A's construction for DSP-BART-HS. The field compresses substantially (BCF, SLP, SAF, and Rao-Yu all close to within roughly 0.05 of DSP-BART-HS), but DSP-BART-HS still records the lowest test RMSE, and BVARX-CAR, MTS-CAR-X, SDPM, and KKP still lag by a wide margin even in a design built for them. Scenario F removes individual heterogeneity entirely, the cleanest possible test: Rao-Yu ties DSP-BART-HS outright, and BVARX-CAR, SLP, MTS-CAR-X, BCF, and SAF all sit within the same tight band -- a substantial change for BVARX-CAR and MTS-CAR-X specifically, whose test RMSE falls roughly 70\% relative to Scenario A, moving them from the high-RMSE cluster into DSP-BART-HS's competitive band. SDPM and KKP, by contrast, do not close nearly as much of their gap even under this most favorable possible scenario -- reading Scenarios E and F together, the same two estimators (SDPM, KKP) lag disproportionately under two different manipulations of the same underlying mechanism (rebalancing individual heterogeneity in E, removing it in F), consistent with SDPM's spatial-lag reduced form and KKP's generalized-moments estimator each carrying an estimator-specific cost rather than a single shared "region-time-aggregate" disadvantage applying uniformly across all five non-individual-level comparators.

\begin{figure*}[t]
\centering
\includegraphics[width=0.70\textwidth]{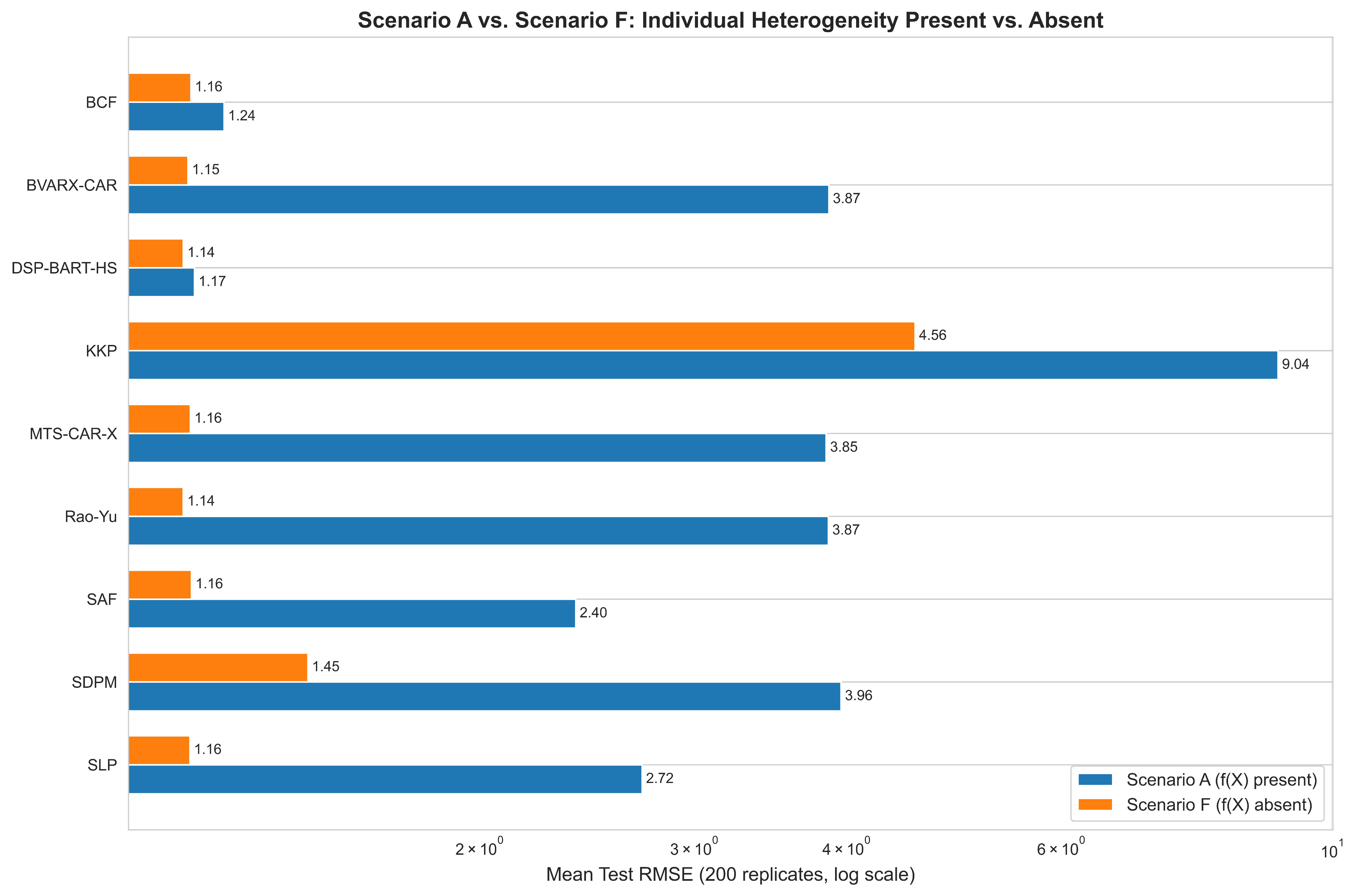}
\caption{Mean Test RMSE by Model, Scenario A vs.\ Scenario F (log scale)}
\label{fig:a-vs-f}
\end{figure*}

\begin{figure*}[t]
\centering
\includegraphics[width=0.80\textwidth]{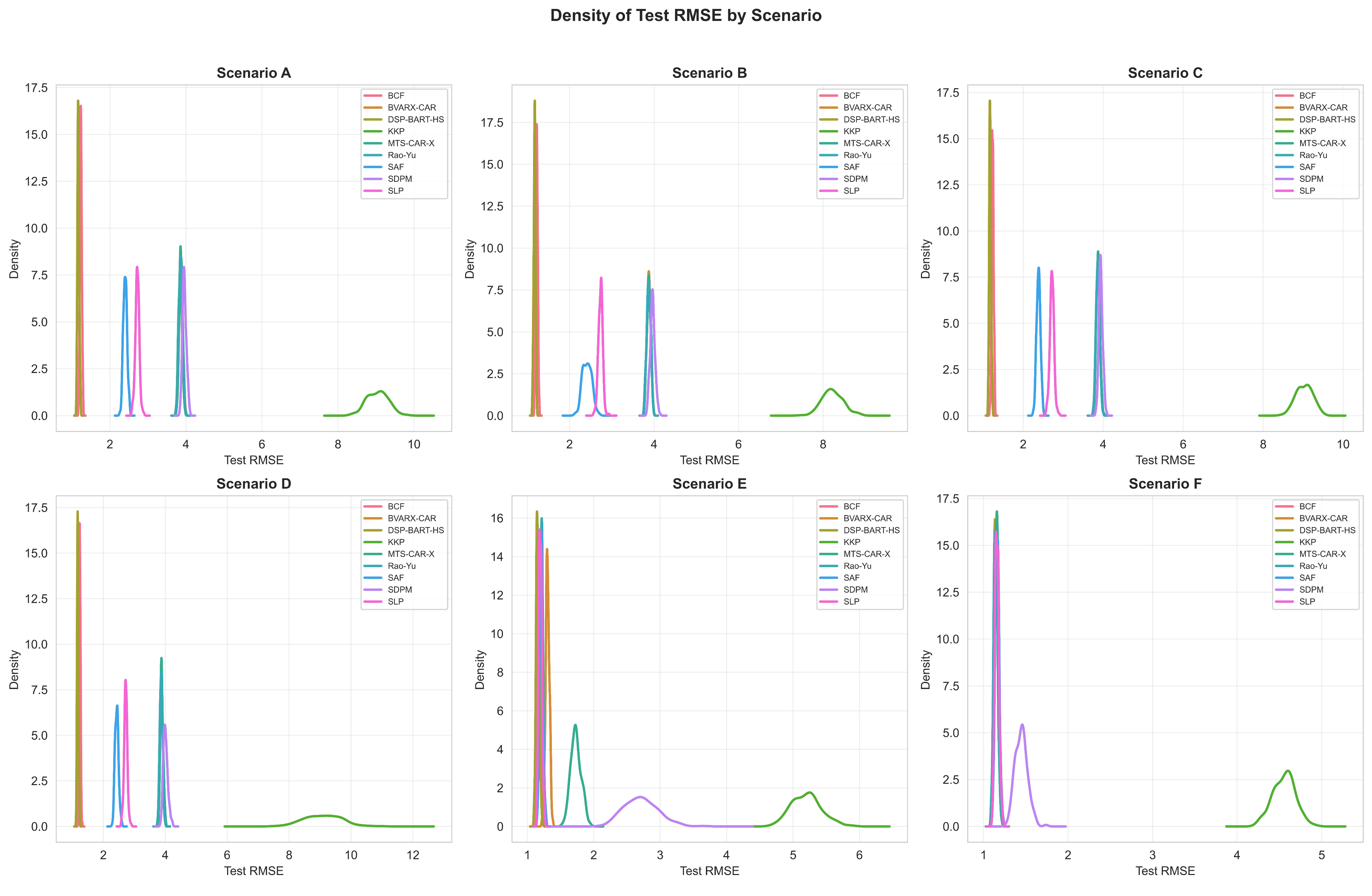}
\caption{Model Mean Test RMSE Density Across Scenarios}
\label{fig:dist}
\end{figure*}

Figure~\ref{fig:a-vs-f} places Scenarios A and F side by side on a log scale: every region-time-aggregate estimator's bar shrinks dramatically from A to F, while DSP-BART-HS's and BCF's shrink by comparatively little, since they were never relying on the $f(\bm{X})$ variance Scenario F removes. Table~\ref{tab:scenario_AF} is also summarized in Figure~\ref{fig:dist} as a model-by-scenario, replicate-level density comparison. Also, a  heatmap, a trajectory plot, and a rank plot in the Supplementary Material Section~2.5, which together show: DSP-BART-HS's and BCF's distributions are visually non-overlapping with the region-time-aggregate cluster's in Scenarios A--D (the ranking holds essentially replicate-by-replicate, not only on average); DSP-BART-HS's and Rao-Yu's distributions overlap almost entirely in Scenario F, confirming their near-exact tie is a genuine distributional overlap; KKP's distribution is visibly the widest of any model throughout; and by ordinal rank, DSP-BART-HS holds rank 1 in Scenarios A--E and rank 2 in Scenario F, the most stable ranking trajectory of any model, while SDPM's rank specifically remains second-to-last even in E and F -- the opposite of what Scenario E's design intended, and further evidence its cost is estimator-specific rather than shared.

\subsection{Robustness Simulations: Aggregated Covariate Access, Topology, and Panel Size}
\label{sec:sim-ghi-design}

Scenarios A--F share a common design choice: individual-level covariates $\bm{X}_{ijt}$ are observed by DSP-BART-HS (and BCF, and in principle SAF) but structurally unavailable to the region-time-aggregate comparators (BVARX-CAR, MTS-CAR-X, SDPM, KKP, Rao-Yu), which see only the region-period covariate block $\bm{Z}_{jt}$. This is a faithful reflection of how these comparators are actually specified and used in practice, but it leaves two questions unaddressed: whether DSP-BART-HS's advantage would persist if the aggregate comparators were given \emph{some} access to the individual-level signal rather than none, and whether the six-scenario pattern is an artifact of the fixed $5\times10$ rook-lattice topology and $J=50$, $T=10$ panel dimensions used throughout. We address both with three additional scenarios, each a targeted, single-mechanism variant of Scenario A's design so that any difference in outcome is attributable to the one manipulation under test.

\textbf{Scenario G: Aggregated-Covariate Access.} Every region-time-aggregate comparator is given an additional three columns in its contextual covariate block: the region-period mean of each of the three individual-level covariates, computed only from that region-period's own individual rows. A new comparator is introduced for this scenario alone: cell-level BART, the same underlying learner as DSP-BART-HS but supplied only this same aggregated information, with no individual-specific $\bm{X}_{ijt}$ anywhere in its input -- isolating whether any remaining DSP-BART-HS advantage reflects genuine individual-level access, or simply greater model flexibility. 

\textbf{Scenario H: Irregular Spatial Topology.} The fixed rook-lattice adjacency is replaced with a random geometric graph at the same region count (mean degree matched to 3.6, connectivity repair applied), testing whether the six-scenario pattern depends on grid regularity. 

\textbf{Scenario I: Larger Panel.} $J$ increases from 50 to 100 regions and $T$ from 10 to 20 periods, testing whether the pattern is specific to the original panel's dimensions.

Each scenario uses 200 replicates and the same paired-comparison protocol used for Scenarios A--F, with the comparator suite consisting of the eight models used throughout Scenarios A--F plus, for Scenario G only, cell-level BART. The complete technical specification (exact DGP parameters, topology construction and connectivity verification, cell-level BART's precise input specification) is given in the Supplementary Material Section~2.1; full train RMSE and SD detail for all three scenarios is given in the Supplementary Material Section~2.3.

\begin{table*}[t]
\centering
\setlength{\tabcolsep}{12pt}
\begin{threeparttable}
\caption{Mean Test RMSE by Model, Scenarios G, H, and I}
\label{tab:scenario_GHI}
\footnotesize
\begin{tabular}{lccc}
\toprule
Model & G (Agg.\ Access) & H (Topology) & I (Larger Panel) \\
\midrule

BCF & 1.2322 (0.0219) & 1.2326 (0.0231) & 1.2253 (0.0146) \\
KKP & 9.0752 (0.2735) & 9.0916 (0.2707) & 9.0715 (0.1538) \\
Rao-Yu & 3.8279 (0.0461) & 3.8683 (0.0454) & 3.8583 (0.0296) \\
SAF & 2.3997 (0.0450) & 2.4094 (0.0472) & 2.2330 (0.0308) \\
SDPM & 4.0744 (0.2825) & 4.0565 (0.1262) & 4.0046 (0.0910) \\
SLP & 2.7120 (0.0512) & 2.7122 (0.0507) & 2.7116 (0.0292) \\
BVARX-CAR & 3.8516 (0.0466) & 3.8731 (0.0474) & 3.8648 (0.0297) \\
MTS-CAR-X & 3.8122 (0.0464) & 3.8524 (0.0454) & 3.8539 (0.0295) \\
DSP-BART-HS & \textbf{1.1642 (0.0215)} & \textbf{1.1636 (0.0213)} & \textbf{1.1529 (0.0126)} \\
Cell-level BART & 4.1150 (0.0577) & -- & -- \\
\bottomrule
\end{tabular}
\begin{tablenotes}[flushleft]
\footnotesize
\item Standard deviation of test RMSE across the 200 replicates is given in parentheses.
\end{tablenotes}
\end{threeparttable}
\end{table*}

Every model's test RMSE across all three scenarios is within simulation noise of its Scenario A value, and every paired comparison against DSP-BART-HS remains significant after Holm-Bonferroni correction (9 comparators in Scenario G including cell-level BART, 8 each in Scenarios H and I; 25 tests total, Supplementary Material Section~2.6 -- the six-scenario pattern holds without qualification under aggregated-covariate access, irregular topology, and a substantially larger panel alike. The BCF margin is nearly identical across all three (mean difference $-0.068$ to $-0.072$). SDPM is the one comparator to show a consistent, repeated departure from its Scenario A baseline across all three scenarios (moving upward each time, $3.9610\to4.05$--$4.07$); we do not have a fully confirmed explanation, though a plausible contributor is that BVARX-CAR and MTS-CAR-X carry Horseshoe shrinkage on their covariate block that SDPM lacks, guarding against exactly this kind of variance inflation from weakly informative added covariates.

Cell-level BART -- given the identical aggregated information as the region-time-aggregate comparators, with no individual-level access at all -- does not close any of the gap to DSP-BART-HS in Scenario G (test RMSE 4.1150, worse than every region-time-aggregate model except KKP). Its train RMSE (3.9269) is elevated relative to BVARX-CAR's and MTS-CAR-X's, and its train-test gap (0.1881) is the largest of any comparator with both quantities available, consistent with BART's flexible tree structure being more prone to fitting individual-level noise within a region-period cell than the region-time-aggregate models' more rigid parametric forms are. This is the central test Scenario G was designed for: cell-level BART's failure to improve on the region-time-aggregate cluster, despite matching DSP-BART-HS's own underlying learner, supports genuine individual-level access as the source of DSP-BART-HS's advantage, not merely greater model flexibility.

\subsection{MCMC Convergence Diagnostics}

To assess convergence of DSP-BART-HS's own Gibbs sampler, we ran four independent chains from different starting points on a representative dataset, holding the data fixed across chains so that only the sampler's own randomness differs between them, and tracked the Gelman-Rubin $\hat{R}$ statistic \citep{gelmanrubin1992inference} and effective sample size (ESS) for ten parameters spanning the active and inactive $\beta$ coefficients, the variance hyperparameters, and a representative spatial effect. For scenario A, all ten parameters achieved $\hat{R} \leq 1.0075$ (comfortably below the stringent 1.01 threshold) and ESS well above the conventional baseline at 25{,}000 iterations (5{,}000 burn-in, thinning interval 10); full per-parameter results are reported in Supplementary Table~S10. Scenarios B - I were fit using this same validated Gibbs sampler and the same production settings and the diagnostic metrics are similar to A; no replicate across any of the scenarios showed evidence of non-convergence over the full 200-replicate run. The 25,000 iterations/chain took up to 8 minutes on a shared remote desktop computer equipped with INTEL(R) XEON(R) GOLD 6542Y 2.90 GHz  (2 processors), 512 GB RAM.

\section{The Fitted Model for Real-Data Applications}
\label{sec:realdata-model}

In the simulation study (Section~\ref{sec:sim-design}), $i$ indexes a micro-level unit: $N_j=20$ units/observations are generated per region, each with their own draw of $\bm{X}_{ijt}$ and $\alpha_i$, and $f(\bm{X})$ and $\alpha_i$ both capture true individual-level heterogeneity by construction. In both real-data applications, by contrast -- the mobility panel of Section~\ref{sec:realdata-mobility}, and the income panel of Section~\ref{sec:income} -- there is no individual-level record at all: the mobility panel is already a set of county-by-birth-cohort cell estimates (one row per $(j,t)$ combination: 45{,}560 rows against 45{,}560 unique region-period pairs), and the income panel is a set of county-by-year cell estimates, each with its own sampling standard error rather than any individual-level record. Consequently, for both real-data applications, $i$ does not exist as a separate index, $\alpha_i$ and its variance component $\sigma_\alpha^2$ are dropped entirely (not merely small or unidentified -- absent from the fitted model), and Equation~\ref{eqn:model} specializes to
\begin{equation}\label{eqn:model-realdata}
    Y_{jt} = f(\bm{X}_{jt}) + \bm{Z}_{jt}\bm{\beta} + \phi_j + \mu_{jt} + \epsilon_{jt}, \qquad \epsilon_{jt} \sim N(0, \sigma^2/w_{jt}),
\end{equation}
the model we actually fit throughout both real-data applications. Every other term retains its general-model meaning ($\bm{Z}_{jt}\bm{\beta}$ the Horseshoe-shrunk covariate block, $\phi_j$ the CAR spatial effect, $\mu_{jt}$ the FFBS dynamic component); the two changes are the disappearance of $\alpha_i$ and the introduction of the known precision weight $w_{jt}$, described below. Here $\bm{X}_{jt}$ is a small set of \emph{region-time-level}, not individual-level, covariates entering only DSP-BART-HS's own nonparametric component -- five covariates in the mobility application (\texttt{popdensity2000}, \texttt{traveltime15\_2010}, \texttt{gsmn\_math\_g3\_2013}, \texttt{singleparent\_share1990}, \texttt{frac\_coll\_plus2000}) and four in the income application (the same list without \texttt{gsmn\_math\_g3\_2013}, which has no income-application analogue). The corresponding real-data advantage, discussed in Sections~\ref{sec:realdata-mobility} and~\ref{sec:income}, is DSP-BART-HS's and BCF's ability to model $f(\bm{X}_{jt})$ nonlinearly using region-time-level covariates that most comparators (SDPM, KKP, Rao-Yu, BVARX-CAR, MTS-CAR-X) do not consume at all -- a real and useful distinction, but a different one from individual-level heterogeneity, and we no longer use that phrase to describe it.

\textbf{Known-precision likelihood weighting.} The general model equation states $\epsilon_{ijt} \sim N(0,\sigma^2)$, a single shared variance, and this is exactly what is fit in the simulation study, where the data-generating process itself is homoscedastic. Both real-data applications depart from this in one specific, disclosed way. Since there is no individual index $i$ in either real application, every cell is indexed by $(j,t)$ alone, and both the mobility percentile-rank estimates and the SAIPE income estimates come with their own sampling standard error for each $(j,t)$ cell, $\text{se}_{jt}$, reflecting the underlying sample size and measurement precision of that specific county-cohort or county-year estimate. Rather than discard this available information and treat every observation as equally precise, the likelihood is instead weighted by the known, fixed precision $w_{jt} = 1/\text{se}_{jt}^2$ -- equivalently, $\epsilon_{jt} \sim N(0, \sigma^2/w_{jt})$, matching Equation~\ref{eqn:model-realdata} above -- in every block whose full conditional depends on the likelihood: the BART component $f(\bm{X})$ (via \texttt{dbarts}'s own native weighted-likelihood support), $\bm{\beta}$'s precision-weighted regression update, and $\phi_j$'s precision-weighted CAR update. Because $w_{jt}$ is a known, fixed quantity from the source data rather than an estimated parameter, $\sigma^2$ itself is fixed at 1 for both real-data applications, with $\text{se}_{jt}^2$ alone supplying the effective per-observation variance. This is a form of \emph{known} heteroscedasticity, not an assumption-free extension to general, unknown heteroscedasticity -- see Section~\ref{sec:beyond-known-hetero}.

To avoid any ambiguity between this likelihood weighting and how performance is subsequently measured: every reported RMSE, and generalization gap throughout this paper -- for DSP-BART-HS and for every comparator -- is computed as an \emph{unweighted} mean over test (or train) rows, $\text{RMSE} = \sqrt{\frac{1}{n}\sum (Y-\hat{Y})^2}$. The known-precision weighting described above affects only how DSP-BART-HS's own likelihood is fit; it does not enter the evaluation metric for any model, so reported RMSE values are directly comparable across models on that dimension.

\section{Real Data Application: Intergenerational Economic Mobility}
\label{sec:realdata-mobility}

As in the simulation study, we apply all the models (excluding KKP) where applicable to the Opportunity Insights county-by-birth-cohort intergenerational mobility panel (45,560 county-cohort rows, 3,175 counties, birth cohorts 1978--1992), supplemented with county-level covariates from the USDA Economic Research Service and the American Community Survey. The outcome $Y_{jt}$ is Opportunity Insights' own expected income rank measure, scaled as a proportion on $[0,1]$ (range $0.080$--$0.917$, mean $0.464$), so a test RMSE of $0.0306$ corresponds to roughly $3.1$ percentile points of expected rank. We adopt a county-level train/test split for the primary comparison. Out of the 2,241 counties with a complete 15-cohort panel, 1,792 are assigned to training and the remaining 449 to test (an 80/20 split at the county level, with all cohort-years retained for every assigned county), yielding 26,880 training and 6,735 test county-cohort rows. This is a genuine out-of-sample test of spatial generalization to entirely unseen counties, not merely interpolation of missing cohort-years for counties already observed in training. BVARX-CAR and MTS-CAR-X are both withheld from this county-level comparison, but for two genuinely different reasons that we do not conflate. BVARX-CAR is excluded for a structural leakage reason: its spillover matrix $\mathbf{B}$ propagates each region's prediction using neighboring regions' \emph{realized} lagged outcomes, so a held-out county's prediction would depend on training-set information leaking in through its neighbors, which a genuine unseen-county holdout must rule out by construction. MTS-CAR-X has no such leakage risk -- its spatial term depends only on neighboring regions' current \emph{parameter} draws, not their realized outcomes, and it is mathematically well-defined under a county-level holdout -- but we tested it directly on this study's own data (Supplementary Material Section~1.4) and found it degrades substantially under a genuine unseen-region holdout, so we exclude it from this comparison on empirical rather than structural grounds. As such, we further evaluated the models under the row-level random (excluding SDPM), and last-cohort temporal splits in addition to the county-level split above, as reported in Sections~\ref{sec:county-level} and~\ref{sec:sensitivity}.

\subsection{Covariate Construction: Resolving Multi-Way Collinearity in \texorpdfstring{$\bm{Z}$}{Z}}
\label{sec:covariate-construction}

The county-level and demographic covariates available for this application ($M=22$ candidates spanning income, poverty, race/ethnicity composition, employment, and housing measures, several recorded at multiple census years) exhibit substantially more redundancy than the simulation design, which draws $\bm{Z}$ from an independent Gaussian by construction. We address this in two stages, both fit exclusively on training data and applied to test data using only training-derived parameters, so that no covariate-construction step uses information from held-out counties.

Several covariates -- \texttt{MEDHHINC\_2023}, \texttt{PCTPOVALL\_2023}, and \texttt{Metro\_2023} among others -- are recorded as of 2023, while the mobility outcome itself reflects each birth cohort's (1978--1992) income rank in adulthood, realized years to decades before 2023 depending on the cohort. These covariates therefore cannot be read as describing region conditions \emph{at the time} the outcome was determined, and any association reported between them and mobility is not a claim about a contemporaneous causal channel operating on that cohort. We use them instead under an explicit, disclosed persistence assumption: that a county's relative standing on these measures -- its poverty rate or median income \emph{relative to other counties} -- is substantially more stable over the two-to-four-decade window in question than its absolute level, so a 2023 snapshot still proxies for the county's relative socioeconomic position when each cohort was growing up, even though the specific numeric value postdates the outcome. This is an assumption about persistence of place-based relative disadvantage, not a verified fact about this specific panel, and we do not treat any coefficient on a 2023-vintage covariate as evidence of a mechanism operating in real time on the cohorts studied; all such effects should be read as reflecting durable geographic association, not a treatment applied at a specific, correctly-ordered point in each cohort's life.

\begin{sloppypar}
First, five covariates were removed as near-exact substitutes for a stronger covariate of the same underlying concept, identified by comparing each pair's univariate $R^2$ with the outcome (not their coefficients in a joint model, which are unreliable under this degree of collinearity) rather than correlation alone: \texttt{hhinc\_mean2000} ($R^2=0.0003$) in favor of \texttt{MEDHHINC\_2023} ($R^2=0.0498$); \texttt{poor\_share1990}, \texttt{poor\_share2000}, and \texttt{poor\_share2010} ($R^2=0.054$, $0.090$, $0.179$) in favor of \texttt{PCTPOVALL\_2023} ($R^2=0.182$, matching or exceeding all three); and \texttt{foreign\_share2000} in favor of \texttt{foreign\_share2010} (both negligible predictors of the outcome on their own, $R^2<0.001$). This reduced the covariate set to $M=17$.
\end{sloppypar}

\begin{sloppypar}
Second, eight of the remaining seventeen covariates exhibit a second, more diffuse form of redundancy: \texttt{share\_white2000}, \texttt{share\_black2000}, \texttt{MEDHHINC\_2023}, \texttt{share\_hisp2000}, \texttt{PCTPOVALL\_2023}, \texttt{rent\_twobed2015}, \texttt{foreign\_share2010}, and \texttt{emp2000} are each individually predictable from the other sixteen covariates combined with $R^2$ between 0.65 and 0.92 (variance inflation factors of roughly 3 to 12), despite no single pairwise correlation among them exceeding 0.62. This is joint, not pairwise, collinearity -- consistent with these eight covariates jointly describing a lower-dimensional underlying county socioeconomic profile (income, poverty, racial composition, and housing cost move together at the county level) rather than eight independent covariate margins. We resolve this via principal component analysis on this eight-covariate cluster, standardizing each covariate using training-sample means and standard deviations, extracting principal components from the training-sample covariance structure only, and projecting both training and test observations onto the resulting components. Three components are retained by the standard Kaiser criterion (eigenvalue $\geq 1$: 1.84, 1.56, and 1.05 for the first three, versus 0.68 for the fourth), jointly explaining 86.4\% of the eight-covariate cluster's variance. This reduces the maximum remaining variance-inflation-factor-equivalent $R^2$ to 0.549, comfortably below conventional collinearity thresholds (commonly $R^2$ corresponding to a VIF of 5 or 10), and yields a final covariate set of $M=12$: nine original covariates plus three orthogonal components, which we label PC1--PC3 (demographic/economic profile). Because the horseshoe prior's coefficient-level shrinkage and selection is uninformative when applied to a set of near-collinear covariates -- the sampler cannot consistently apportion a shared effect across covariates that are not separately identified by the data -- this construction is necessary for the coefficient-level inference to be meaningful, independent of its effect on predictive accuracy.
\end{sloppypar}

\subsection{County-level Split Comparison}
\label{sec:county-level}
As shown in Table~\ref{tab:realdata}, DSP-BART-HS achieves both the lowest test RMSE (0.0306) and the lowest train RMSE (0.0223) of the six models under this genuine unseen-county holdout, with SAF its closest competitor on test accuracy (0.0322). Its generalization gap (0.0083) is neither the smallest nor the largest in the panel: below BCF's (0.0094) and well below Rao-Yu's (0.0179, the largest gap in the panel by a wide margin), but above SLP's (0.0073) and SAF's (0.0046, the smallest positive gap). SDPM is the only model with a negative gap (test RMSE 0.0363 fractionally below train RMSE 0.0371). Rao-Yu's combination of the worst test RMSE (0.0416) and the largest generalization gap is worth flagging directly: despite being among the strongest comparators under the random and temporal splits below, a small-area estimator built to borrow strength across related areas evidently does not transfer that advantage to counties contributing no training data of their own, a genuine unseen-county extrapolation task the flexible individual-level models (DSP-BART-HS, SAF) handle better here. We treat out-of-sample test RMSE as the primary criterion for comparing predictive performance across models of differing flexibility, with the generalization gap reported alongside it as a secondary diagnostic.

\begin{table*}[t]
\centering
\setlength{\tabcolsep}{4pt}
\begin{threeparttable}
\caption{Mobility: Test RMSE by Model Under Three Train/Test Splits}
\label{tab:realdata}

\begin{tabular}{lccccccccc}
\toprule
 & \multicolumn{3}{c}{\textbf{Panel A. County-Level Split}} & \multicolumn{3}{c}{\textbf{Panel B. Random Row-Level Split}} & \multicolumn{3}{c}{\textbf{Panel C. Temporal Split (Last-Cohort)}} \\
\cmidrule(lr){2-4}\cmidrule(lr){5-7}\cmidrule(lr){8-10}
Model & RMSE (train) & RMSE (test) & Gen.\ Gap & RMSE (train) & RMSE (test) & Gen.\ Gap & RMSE (train) & RMSE (test) & Gen.\ Gap \\
\midrule
BCF & 0.0288 & 0.0382 & 0.0094 & 0.0308 & 0.0334 & 0.0026 & 0.0307 & 0.0336 & 0.0029 \\
Rao-Yu & 0.0236 & 0.0416 & 0.0179 & 0.0259 & 0.0310 & 0.0050 & 0.0261 & 0.0333 & 0.0073 \\
SAF & 0.0276 & 0.0322 & 0.0046 & 0.0289 & 0.0327 & 0.0037 & 0.0292 & 0.0338 & 0.0046 \\
SDPM & 0.0371 & 0.0363 & $-0.0007$ & -- & -- & -- & 0.0398 & 0.0383 & $-0.0015$ \\
SLP & 0.0276 & 0.0350 & 0.0073 & 0.0290 & 0.0327 & 0.0037 & 0.0292 & 0.0335 & 0.0043 \\
BVARX-CAR & -- & -- & -- & 0.0264 & 0.0325 & 0.0061 & 0.0283 & 0.0334 & 0.0051 \\
MTS-CAR-X & -- & -- & -- & 0.0325 & 0.0355 & 0.0030 & 0.0295 & \textbf{0.0303} & 0.0008 \\
DSP-BART-HS & 0.0223 & \textbf{0.0306} & 0.0083 & 0.0239 & \textbf{0.0307} & 0.0068 & 0.0244 & 0.0310 & 0.0066 \\
\bottomrule
\end{tabular}

\begin{tablenotes}[flushleft]
\footnotesize
\item Panel A: 1{,}792 train / 449 test counties (80/20 county-level split of the 2{,}241 counties with a complete 15-cohort panel). Panel B: 27{,}574 train / 6{,}894 test rows, 2{,}325 / 2{,}244 counties (row-level split). Panel C: train on cohorts 1978--1991, test on 1992. Gen.\ Gap is RMSE(test) $-$ RMSE(train). DSP-BART-HS's Gibbs sampler was run to full convergence under the $M=12$ covariate specification of Section~\ref{sec:covariate-construction} in all three panels. SDPM's Panel A county count (1{,}759 train, not the 1{,}792 target) excludes 33 counties that became isolated within the training subgraph once test counties were removed, a spatial-connectivity requirement of the adjacency-matrix construction used by \texttt{splm}, not a memory limitation. SDPM's Panel C run similarly restricts to the 2{,}241 counties with a complete 15-cohort panel, rather than the full $\sim$3{,}175-county set the panel-structured models (DSP-BART-HS, BCF, BVARX-CAR, MTS-CAR-X) can draw on -- a structural balanced-panel requirement of \texttt{splm}, not a subsampling limitation. Bolded figure indicates the best performing model in each panel. En dash (--) indicates the model is not evaluated under that split: BVARX-CAR is withheld from the county-level split (Panel A) for a structural leakage reason, and MTS-CAR-X on empirical grounds, both described in Section~5; SDPM's random-split run (Panel B) is infeasible under \texttt{splm}'s balanced-panel requirement.
\end{tablenotes}
\end{threeparttable}
\end{table*}

\subsection{Repeated-Split Uncertainty}
\label{sec:repeated-split-mobility}

The county-level result above reflects a single 80/20 partition. To characterize how much this result would change under a different, equally plausible partition, we refit DSP-BART-HS's county-level split $K=20$ times, each with an independently drawn 80/20 county partition (a different random seed per repeat; repeat 1 reproduces the partition used above exactly, as a validity check), at the same $200{,}000$/$50{,}000$/$3$ Gibbs settings used throughout. 

Across the 20 repeats, test RMSE has mean $0.0316$ (SD $0.0010$, 95\% CI $[0.0311, 0.0320]$, range $[0.0304, 0.0340]$), with the single-split result reported above ($0.0306$) representative of the typical partition, not a favorably selected one.

DSP-BART-HS's single-split margin over SAF, its closest county-level competitor (Table~\ref{tab:realdata}: $0.0306$ vs.\ $0.0322$), is narrow relative to DSP-BART-HS's own split-to-split variability. We therefore refit SAF under the identical 20 county partitions used for DSP-BART-HS above (same split seeds throughout), giving a genuinely paired comparison across the same 20 splits rather than two independently varying quantities. SAF's mean test RMSE across the 20 repeats is $0.0338$ (SD $0.0010$); DSP-BART-HS wins on every one of the 20 paired splits without exception (paired mean difference $-0.00219$, 95\% CI $[-0.0024,-0.0020]$, $p<10^{-15}$), including the split (repeat 12) where DSP-BART-HS's own weakest result was closest to SAF's, on which SAF's own result is also elevated ($0.0354$) rather than favorable. The margin is real and consistent across every partition tested, not a fragile artifact of any single split.

\subsection{Row-Level Random and Temporal Holdout Comparisons}
\label{sec:sensitivity}

We additionally evaluate the models under two further splits for the models that are applicable. A random row-level 80/20 split, in which individual county-cohort observations are held out rather than entire counties (allowing the same county to appear in both training and test data); and a temporal (last-period) holdout, training on cohorts 1978--1991 and testing on the final cohort, 1992, alone. Both are different tasks than the county-level split: they test interpolation of missing cohort-years or a single future cohort for counties already observed in training, not generalization to unseen counties.

Under the random row-level split in Panel B of Table~\ref{tab:realdata}, DSP-BART-HS achieves the lowest test RMSE (0.0307) of the seven models evaluated, though the margin over Rao-Yu (0.0310) and BVARX-CAR (0.0325) is modest compared to Panel A's county-level gap; SLP and SAF land close at 0.0327 each. Under the temporal holdout in Panel C, MTS-CAR-X achieves the lowest test RMSE (0.0303) and, notably, the lowest generalization gap of any model (0.0008), with DSP-BART-HS close behind (0.0310), Rao-Yu (0.0333), SLP (0.0335), BCF (0.0336), and SAF (0.0338) again trail the top cluster by a similar modest margin. We do not read either result as evidence that the field's ranking under Panel A's stricter county-level holdout is overturned: interpolating a missing cohort-year for an already-observed county, or predicting one additional cohort for counties seen in every prior period, are both structurally easier than generalizing to a county contributing zero training rows, and only Panel A's county-level result is directly comparable to SAF, SLP, and SDPM's own county-level evaluation there. We report Panels B and C as complementary robustness checks under alternative, less stringent evaluation designs (even though they are the common prediction pattern in practice), not as a substitute for Panel A's head-to-head county-level comparison.

\subsection{Cross-Model Consistency of Covariate Effects}
\label{sec:crossmodel}

Beyond predictive accuracy, a central empirical question is which county characteristics are robustly associated with intergenerational mobility, and whether that association holds across structurally different estimators rather than being an artifact of any one model's assumptions. Five of the seven models -- DSP-BART-HS, BVARX-CAR, and MTS-CAR-X (Bayesian), and Rao-Yu and SDPM (frequentist) -- report a coefficient-level effect for the same eight collinear covariates (Section~\ref{sec:covariate-construction}), back-projected from the three estimation-scale principal components onto their original, interpretable scale (formula and post-selection caveat in the Supplementary Material Section~3.5: the covariate-clustering and component-count decisions were themselves made by inspecting this study's own data, and the intervals below condition on those decisions as fixed). BCF, fully nonparametric with no linear coefficient block, does not produce a comparable quantity.

Tables~\ref{tab:c2-county}--\ref{tab:c2-temporal} report each model's back-projected posterior mean (Bayesian: DSP-BART-HS, BVARX-CAR, MTS-CAR-X) or point estimate (frequentist: Rao-Yu, SDPM) for all eight covariates, under each split for which results are available. We report both families in the same table only because their sign and significance pattern is the object of interest here, not because a posterior credible interval and a frequentist confidence interval are the same inferential quantity -- they are not, and we flag this rather than let the shared table format imply otherwise.

\begin{table*}[t]
\centering
\setlength{\tabcolsep}{10pt}
\begin{threeparttable}
\caption{Implied Covariate Effects on Original 8-Cluster Variables (County Split)}
\label{tab:c2-county}
\begin{tabular}{lccc}
\toprule
Covariate & DSP-BART-HS & Rao-Yu & SDPM \\
\midrule
share\_black2000 & -0.0108* & -0.1405* & -0.1186* \\
share\_white2000 & 0.0078* & 0.0572* & 0.0599* \\
PCTPOVALL\_2023 & -0.0043* & -0.0013* & -0.0012* \\
emp2000 & 0.0019* & 0.0339* & 0.0328* \\
rent\_twobed2015 & -0.0034* & -2.02e-05* & -2.47e-05* \\
share\_hisp2000 & 0.0011 & 0.0555* & 0.0240* \\
foreign\_share2010 & -0.0021* & 0.0132 & -0.0404* \\
MEDHHINC\_2023 & -0.0001 & 2.54e-08 & -2.32e-08 \\
\bottomrule
\end{tabular}
\begin{tablenotes}[flushleft]
\footnotesize
\item An asterisk (*) indicates that the corresponding 95\% interval does not include 0.
\end{tablenotes}
\end{threeparttable}
\end{table*}

\begin{table*}[t]
\centering
\setlength{\tabcolsep}{10pt}
\begin{threeparttable}
\caption{Implied Covariate Effects on Original 8-Cluster Variables (Random Split)}
\label{tab:c2-random}
\footnotesize
\begin{tabular}{lcccc}
\toprule
Covariate & DSP-BART-HS & BVARX-CAR & MTS-CAR-X & Rao-Yu \\
\midrule
share\_black2000 & -0.0109* & -0.0177* & -0.0779* & -0.1437* \\
share\_white2000 & 0.0078* & 0.0112* & 0.0360* & 0.0564* \\
PCTPOVALL\_2023 & -0.0047* & -0.0061* & -0.0007* & -0.0013* \\
emp2000 & 0.0023* & 0.0021* & 0.0146* & 0.0290* \\
rent\_twobed2015 & -0.0029* & -0.0054* & -1.61e-05* & -2.15e-05* \\
share\_hisp2000 & 0.0014 & 0.0042* & 0.0243* & 0.0645* \\
foreign\_share2010 & -0.0014* & -0.0014* & -0.0124* & 0.0237* \\
MEDHHINC\_2023 & 0.0005 & -0.0005 & -3.05e-08* & 9.40e-09 \\
\bottomrule
\end{tabular}
\begin{tablenotes}[flushleft]
\footnotesize
\item An asterisk (*) indicates that the corresponding 95\% interval does not include 0.
\end{tablenotes}
\end{threeparttable}
\end{table*}

\begin{table*}[t]
\centering
\setlength{\tabcolsep}{9.5pt}
\begin{threeparttable}
\caption{Implied Covariate Effects on Original 8-Cluster Variables (Temporal Split)}
\label{tab:c2-temporal}
\footnotesize
\begin{tabular}{lccccc}
\toprule
Covariate & DSP-BART-HS & BVARX-CAR & MTS-CAR-X & Rao-Yu & SDPM \\
\midrule
share\_black2000 & -0.0109* & -0.0180* & -0.0833* & -0.1453* & -0.1252* \\
share\_white2000 & 0.0079* & 0.0112* & 0.0387* & 0.0572* & 0.0607* \\
PCTPOVALL\_2023 & -0.0047* & -0.0060* & -0.0007* & -0.0013* & -0.0012* \\
emp2000 & 0.0023* & 0.0020* & 0.0150* & 0.0301* & 0.0302* \\
rent\_twobed2015 & -0.0030* & -0.0056* & -1.78e-05* & -2.16e-05* & -2.49e-05* \\
share\_hisp2000 & 0.0012 & 0.0047* & 0.0255* & 0.0645* & 0.0321* \\
foreign\_share2010 & -0.0017* & -0.0014* & -0.0159* & 0.0214* & -0.0298* \\
MEDHHINC\_2023 & 0.0004 & -0.0007 & -3.80e-08* & 1.20e-08 & -2.50e-08 \\
\bottomrule
\end{tabular}
\begin{tablenotes}[flushleft]
\footnotesize
\item An asterisk (*) indicates that the corresponding 95\% interval does not include 0.
\end{tablenotes}
\end{threeparttable}
\end{table*}

Five covariates show the same sign and statistical significance across all three model-split combinations in Tables~\ref{tab:c2-county}--\ref{tab:c2-temporal}: \texttt{share\_black2000} (negative), \texttt{share\_white2000} (positive), \texttt{PCTPOVALL\_2023} (negative), \texttt{emp2000} (positive), and \texttt{rent\_twobed2015} (negative) -- higher Black population share is associated with lower mobility and higher White share with higher mobility, holding other covariates fixed. This is consistent with the racial gap in upward mobility documented by \citet{chetty2014land}, alongside lower mobility in higher-poverty, higher-rent counties and higher mobility where historical employment was stronger. As throughout this study (Section~\ref{sec:conclusion}), "associated with" describes a regression-style coefficient, not a causal estimate: none of these models addresses confounding with unobserved county characteristics, so this is a robust predictive correlate replicated across model families, not evidence of a causal effect. That the pattern holds identically across a nonparametric Bayesian tree ensemble, two Bayesian panel-VAR specifications, and two frequentist small-area/spatial-panel estimators, under three different splits, is substantially stronger evidence than any one model's result alone.

\texttt{share\_hisp2000} is positive in all twelve combinations but reaches significance in only nine: DSP-BART-HS's own credible interval includes zero under every split, while the other four models find it significant throughout -- read as wider uncertainty for this effect, not a sign disagreement, since the point estimate is never negative under any model or split. \texttt{MEDHHINC\_2023} is null throughout, negligible in magnitude ($10^{-8}$ to $10^{-4}$) with an inconsistent sign across models and splits; only MTS-CAR-X reaches nominal significance, a consequence of an unusually tight interval around an economically negligible estimate.

\texttt{foreign\_share2010} is the one covariate where the models disagree. The negative direction found by DSP-BART-HS, BVARX-CAR, MTS-CAR-X, and SDPM is contradicted by Rao-Yu, which finds a significant \emph{positive} effect under the random and temporal splits and a positive (non-significant) point estimate under the county split. Rao-Yu's simulation performance (Section~\ref{sec:sim-design}, statistically tied with DSP-BART-HS in Scenario F) does not support setting this aside, so we treat this covariate's association with mobility as genuinely unresolved -- majority-negative but not unanimous -- rather than an established finding on four-against-one.

Tables~\ref{tab:c2-county}--\ref{tab:c2-temporal} report DSP-BART-HS's coefficient-level effects on the $\bm{Z}\bm{\beta}$ contextual covariate block; the model's own variable-importance measure on its separate $f(\bm{X})$ term is reported alongside the income application's equivalent in Supplementary Section~4. \texttt{singleparent\_share1990} is the most-split-on covariate under every split there, consistent with the income application's own ranking.

\subsection{Validation Against Established Mobility Findings}
\label{sec:validation}

Predictive accuracy and cross-model coefficient agreement establish that DSP-BART-HS behaves sensibly relative to its comparators,so we directly check its output against what is already known about U.S.\ intergenerational mobility from prior, independently-collected evidence. We check DSP-BART-HS's fitted output against three specific, well-established patterns from \citet{chetty2014land} and the associated Opportunity Atlas literature, under both the county-level holdout of Table~\ref{tab:c2-county} and the random row-level holdout of Table~\ref{tab:c2-random}.

\textbf{Racial gap in mobility.} \citet{chetty2014land} document that Black children raised in low-income families have substantially lower expected adult income rank than White children raised in comparably low-income families, controlling for other observable characteristics. DSP-BART-HS's back-projected effect for \texttt{share\_black2000} is negative and its 95\% credible interval excludes zero under both holdouts (county split: $-0.0108$, CI $[-0.0133, -0.0082]$; random split: $-0.0109$, CI $[-0.0135,-0.0083]$), with the corresponding effect for \texttt{share\_white2000} positive and significant under both (county split: $0.0078$, CI $[0.0060, 0.0095]$; random split: $0.0078$, CI $[0.0063, 0.0092]$) -- see Tables~\ref{tab:c2-county} and~\ref{tab:c2-random}. The near-identical magnitude and interval width of these effects across two holdouts that differ substantially in which counties contribute training data is itself informative: it indicates the finding is not an artifact of which particular counties happen to be held out.

\textbf{Spatial persistence of place.} A defining empirical feature of the Opportunity Atlas is that mobility is not scattered geographic noise but clusters into persistent regions of relative advantage and disadvantage. DSP-BART-HS's CAR spatial correlation parameter is estimated at $\hat\rho = 0.992$ (posterior SD 0.004, 95\% CI $[0.982, 0.999]$) under the county-level holdout and $\hat\rho = 0.993$ under the random row-level holdout, indicating strong, consistently-estimated spatial dependence in the fitted region effects $\phi_j$ under both evaluation designs. A value this close to the upper boundary of the CAR prior's support reflects strong genuine dependence rather than a prior/data conflict.

\textbf{Within-state heterogeneity.} Chetty et al.'s central geographic finding is that mobility varies substantially \emph{within} states, not merely between them -- national or state-level averages obscure most of the relevant geographic variation. Decomposing the variance of DSP-BART-HS's fitted county-level mobility values into between-state and within-state components under the county-level holdout, 41.7\% of total variance is within-state (58.3\% between-state) across all 2,241 eligible counties, and 38.3\% within-state when this decomposition is restricted to the 449 counties held out from training entirely -- confirming the pattern holds specifically for counties the model was given no training rows for, not only for counties it partially observed. Under the random row-level holdout, 41.6\% of variance is within-state, closely matching the county-level holdout's figure. This decomposition is a county-level proxy for Chetty et al.'s own finer-grained commuting-zone and tract-level analysis, not a literal reproduction of their specific statistic, but the qualitative conclusion -- that state-level averages alone would understate the geographic variation DSP-BART-HS recovers -- holds under both holdouts.

Taken together, these three checks -- run independently under two holdouts -- support treating DSP-BART-HS's real-data output as consistent with, rather than contradicting, the existing mobility literature.

\section{Real Data Application: Geographic Income Inequality}
\label{sec:income}

We apply the same comparator suite to a companion outcome: county-level geographic income inequality, using the Census Bureau's Small Area Income and Poverty Estimates (SAIPE) county-level median household income series, restricted to its 2005--2024 American Community Survey-based estimation window for the reasons described in Section 1. As in the mobility application, there is no individual-level record in this panel; the model fit throughout this section is the same real-data specialization of Equation~\ref{eqn:model} given in full in Section~\ref{sec:realdata-model} including the same known-precision likelihood weighting, with this application's own covariate specification given below. 

The outcome $Y_{jt}$ is $\log(\text{SAIPE median household income})$, not the raw dollar-denominated series; every RMSE reported for this application is therefore on the log scale, so a test RMSE of $0.0775$ corresponds to an approximate $\exp(0.0775)-1 \approx 8.1\%$ relative error in untransformed median household income. The reported standard errors ($\texttt{y\_se}\in[0.0037,0.105]$) are likewise on the log scale, consistent with a delta-method conversion of SAIPE's dollar-denominated margins of error rather than raw-dollar standard errors, which would be several orders of magnitude larger. A terminological clarification is worth making explicit here: the quantity every model in this section is fit and evaluated against, $Y_{jt}$, is a county-year income \emph{level}, not a directly-computed inequality or dispersion statistic (a Gini coefficient, a variance-across-counties measure, or similar). This application's claim to studying geographic income \emph{inequality}, introduced in Section 1, rests on what the fitted levels reveal when examined \emph{across} counties and over time -- spatial clustering in $\phi_j$, the urban-rural gap's evolution, and similar between-place patterns -- not on the outcome variable itself encoding an inequality measure. We use "income inequality" and "income level" accordingly throughout this section: any claim about inequality is a claim about the pattern of estimated levels across space, never a claim that $Y_{jt}$ itself is an inequality index. Because the outcome here is itself a median-household-income measure, two covariates from the mobility application's $M=12$ specification (Section~\ref{sec:covariate-construction}) that are themselves income- or wage-denominated -- \texttt{MEDHHINC\_2023} and \texttt{ln\_wage\_growth\_hs\_grad} -- are dropped to avoid outcome-covariate circularity, mirroring the same reasoning applied to BCF's treatment variable below. This yields a common $M=11$ contextual-covariate specification (8 direct covariates plus the same 3-component PCA construction of Section~\ref{sec:covariate-construction}, refit on the resulting 7-variable cluster with \texttt{MEDHHINC\_2023} removed), applied identically across every comparator in this section. A separate 4-variable block (\texttt{popdensity2000}, \texttt{traveltime15\_2010}, \texttt{singleparent\_share1990}, \texttt{frac\_coll\_plus2000}) enters the underlying data as $\bm{X}_{jt}$, DSP-BART-HS's own region-time-level $f(\bm{X})$ input described in Equation~\ref{eqn:model-realdata} -- not individual-level, consistent with this application having no individual index either;  these four variables are passed only to the BART component and never appear in the reported $\bm{Z}\bm{\beta}$ coefficient table. SLP and SAF, as row-level models without a separate $f(\bm{X})$ mechanism, do not consume this four-variable block either, since Rao-Yu and SDPM never have access to it -- admitting it for SLP/SAF only would give those two comparators an unearned four-covariate advantage over the rest of the suite. BCF's treatment variable is \texttt{Metro\_2023} in this application (\texttt{hhinc\_mean2000} in the mobility application), for the same circularity reason, confirmed directly against its saved output.

We report results below for all eight comparators under every split available to them. As with the mobility application, \texttt{splm}'s balanced-panel requirement for SDPM makes the random row-level split infeasible, so SDPM is reported only under the county split and temporal holdout.

\begin{table*}[t]
\centering
\setlength{\tabcolsep}{4pt}
\begin{threeparttable}
\caption{Income Inequality: Test RMSE by Model Under Three Train/Test Splits}
\label{tab:income-results}

\begin{tabular}{lccccccccc}
\toprule
 & \multicolumn{3}{c}{\textbf{Panel A. County-Level Split}} & \multicolumn{3}{c}{\textbf{Panel B. Random Row-Level Split}} & \multicolumn{3}{c}{\textbf{Panel C. Temporal Last-Period Holdout}} \\
\cmidrule(lr){2-4}\cmidrule(lr){5-7}\cmidrule(lr){8-10}
Model & RMSE (train) & RMSE (test) & Gen.\ Gap & RMSE (train) & RMSE (test) & Gen.\ Gap & RMSE (train) & RMSE (test) & Gen.\ Gap \\
\midrule
BCF & 0.1750 & 0.1899 & 0.0149 & 0.1751 & 0.1806 & 0.0055 & 0.1597 & 0.3607 & 0.2010 \\
Rao-Yu & 0.0343 & 0.1020 & 0.0677 & 0.0340 & \textbf{0.0469} & 0.0129 & 0.0340 & 0.5727 & 0.5388 \\
SAF & 0.1691 & 0.1873 & 0.0182 & 0.1676 & 0.1806 & 0.0130 & 0.1538 & 0.3591 & 0.2054 \\
SDPM & 0.1046 & 0.1041 & $-0.0005$ & -- & -- & -- & 0.0917 & 0.5803 & 0.4887 \\
SLP & 0.1703 & 0.1912 & 0.0209 & 0.1696 & 0.1790 & 0.0095 & 0.1547 & 0.3590 & 0.2044 \\
BVARX-CAR & -- & -- & -- & 0.1441 & 0.1840 & 0.0400 & 0.1532 & 0.3578 & 0.2046 \\
MTS-CAR-X & -- & -- & -- & 0.1495 & 0.1627 & 0.0131 & 0.0578 & \textbf{0.0696} & 0.0118 \\
DSP-BART-HS & 0.0277 & \textbf{0.0775} & 0.0498 & 0.0248 & 0.0557 & 0.0309 & 0.0260 & 0.1027 & 0.0767 \\
\bottomrule
\end{tabular}

\begin{tablenotes}[flushleft]
\footnotesize
\item Panel A: 47{,}720 train / 11{,}940 test rows, 2{,}386 train counties / 597 test counties (80/20 county-level split; no county overlap between train and test). Panel B: 47{,}728 train / 11{,}932 test rows, 2{,}983 / 2{,}946 counties (row-level split, counties necessarily overlapping between train and test). Panel C: 56{,}677 train / 2{,}983 test rows, all 2{,}983 counties represented in both. Gen.\ Gap is RMSE(test) $-$ RMSE(train). All eight comparators apply the identical $M=11$ contextual-covariate specification described above. DSP-BART-HS additionally observes \texttt{popdensity2000}, \texttt{traveltime15\_2010}, \texttt{singleparent\_share1990}, and \texttt{frac\_coll\_plus2000} through its own $f(\bm{X})$ term (Supplementary Table~S21), exactly as in the mobility application -- not through the $\bm{Z}\bm{\beta}$ block reported here. Bolded figure indicates the best performing model in each panel. En dash (--) indicates the model is not evaluated under that split: SDPM's random-split run is infeasible under \texttt{splm}'s balanced-panel requirement; BVARX-CAR and MTS-CAR-X are withheld from the county-level split for the same structural-leakage and empirical reasons given for the mobility application (Section~5.3).
\end{tablenotes}
\end{threeparttable}
\end{table*}

Under this genuine county holdout, DSP-BART-HS achieves both the lowest test RMSE (0.0775) and the lowest train RMSE (0.0277) of the six models, with its closest competitor on test accuracy, Rao-Yu, trailing by roughly a third (0.1020) and SDPM close behind Rao-Yu (0.1041); SAF, BCF, and SLP form a separate, substantially weaker cluster (0.1873--0.1912). The generalization gap tells a different part of the story: DSP-BART-HS's gap (0.0498) is the second-largest in the panel, and Rao-Yu's (0.0677) is the largest outright, while the three weaker-fitting models (BCF, SAF, SLP) show the smallest gaps (0.0149--0.0210), with BCF's own gap the smallest of the three despite its train RMSE being the highest - consistent with the same mechanism: the weakest training fit leaves the least room to overfit in the first place. SDPM is again the only model with a (near-zero) negative gap. This is the same qualitative pattern found in the mobility application's own county-level holdout: the models flexible enough to fit training data closely are also the models whose test performance moves the most between splits, and a small train-test gap is not on its own evidence of good generalization.

Two further results stand out. First, DSP-BART-HS's random-split test RMSE (0.0557) is about three times lower than every comparator except Rao-Yu, and closer to Rao-Yu's own 0.0469 than to the 0.1627--0.1840 cluster the other five models occupy -- consistent with the mobility application's pattern of DSP-BART-HS and Rao-Yu separating from the region-time-aggregate-only estimators, even though DSP-BART-HS does not surpass Rao-Yu outright here (unlike the formal paired significance tests established for the simulation study, this is a single train/test split with no replication, so we do not read anything into the ordering beyond what the point estimate shows). Second, MTS-CAR-X's temporal-split test RMSE (0.0696) is dramatically lower than every other model's temporal result (0.1027--0.5803) -- an order of magnitude better generalization from earlier years to the final year than any other estimator achieves, including its own random-split performance (0.1627) and DSP-BART-HS's own temporal-split performance (0.1027, the second-best in the panel but still 48\% higher than MTS-CAR-X's).

We investigate this directly with a rolling-origin design rather than reporting the single cutoff above as though it were self-validating. All eight models were refit under five different train/test cutoffs (test years corresponding to $t=16$ through $t=20$ in sequence, an expanding training window at each), rather than the single $t=20$ cutoff reported above.

\begin{table*}[t]
\centering
\setlength{\tabcolsep}{8pt}
\begin{threeparttable}
\caption{Rolling-Origin Test RMSE, Income Application, Five Origins}
\label{tab:rolling-origin-income}
\scriptsize
\begin{tabular}{lcccccccc}
\toprule
Test Year & DSP-BART-HS & MTS-CAR-X & BVARX-CAR & BCF & SAF & SLP & Rao-Yu & SDPM \\
\midrule
16 & 0.0946 & \textbf{0.0782} & 0.2335 & 0.2359 & 0.2345 & 0.2342 & 0.3854 & 0.3945 \\
17 & 0.0834 & \textbf{0.0788} & 0.2399 & 0.2424 & 0.2408 & 0.2406 & 0.4072 & 0.4159 \\
18 & 0.1088 & \textbf{0.0853} & 0.2964 & 0.2989 & 0.2974 & 0.2972 & 0.4780 & 0.4864 \\
19 & 0.0930 & \textbf{0.0658} & 0.3173 & 0.3198 & 0.3184 & 0.3182 & 0.5150 & 0.5229 \\
20 & 0.1027 & \textbf{0.0696} & 0.3578 & 0.3607 & 0.3591 & 0.3590 & 0.5727 & 0.5803 \\
\bottomrule
\end{tabular}
\end{threeparttable}
\end{table*}

MTS-CAR-X is the best-performing model at every one of the five origins, not only the single cutoff originally reported -- the pattern is real and robust, not an artifact of one particular train/test boundary. The margin over DSP-BART-HS is not narrowing at earlier origins either: it ranges from as little as 0.0046 (test year 17) to as much as 0.0331 (test year 20), with no origin where DSP-BART-HS closes the gap.

Under the random row-level split, all seven available models generalize reasonably, with generalization gaps ranging from 0.0055 (BCF) to 0.0400 (BVARX-CAR). Under the temporal holdout, every model except MTS-CAR-X shows substantially worse generalization than in the mobility application: generalization gaps range from 0.0767 (DSP-BART-HS) to 0.5388 (Rao-Yu), an order of magnitude larger than any gap observed in Table~\ref{tab:realdata}. We report this pattern directly rather than set it aside: it indicates that predicting the final year of the SAIPE series from earlier years is a substantially harder generalization problem than either the mobility application's temporal holdout or this application's own random row-level split, for every model but one, and remains a genuine open question for this companion application.

BVARX-CAR's metric B (posterior predictive uncertainty on test rows) gives a mean posterior SD on the test set of 0.0779 under the random split and 0.0362 under the temporal split, with mean 95\% credible interval width of 0.3051 (random) and 0.1418 (temporal) -- narrower under the temporal split despite that split's substantially worse point-prediction accuracy (Table~\ref{tab:income-results}), indicating BVARX-CAR is more confident, not less, in its temporal-split predictions even though those predictions are less accurate on average.

\subsection{Repeated-Split Uncertainty}
\label{sec:repeated-split-income}

As in the mobility application (Section~\ref{sec:repeated-split-mobility}), we refit DSP-BART-HS's county-level split $K=20$ times under independently drawn 80/20 county partitions. This is the first convergence check performed for the income application specifically (mobility's own diagnostics, given in the Supplementary Material Section~6, do not directly establish income's convergence properties). ESS on one repeat: $\rho$, $\tau_\phi^2$, and $\sigma_\mu^2$ in the thousands to tens of thousands, and the two tracked $\beta$ coefficients at ESS $552$ and $476$ -- both comfortably above the conventional $ESS \ge 400$threshold.

Across the 20 repeats, test RMSE has mean $0.08194$ (SD $0.00346$, 95\% CI $[0.08032, 0.08356]$, range $[0.07662, 0.08896]$). Unlike the mobility application, where the single-split result was representative of the typical repeated-split outcome, income's reported result ($0.0775$) sits noticeably below this distribution's mean -- roughly 1.3 SD more favorable than average, and below the 95\% CI's own lower bound, though still within the observed range across the 20 repeats (not an outlier). We report this directly: the county-level test RMSE reported for the income application throughout this section should be read as a plausible but somewhat favorable draw from the range of results a different 80/20 county partition would have produced, not as the typical outcome.

\subsection{Implied Covariate Effects (Partial)}
\label{sec:income-c2}

Of the eight income models, six report a back-projected coefficient-level effect with a valid confidence interval: DSP-BART-HS and Rao-Yu report all three splits; SDPM reports the county and temporal splits (its random-split run is infeasible under \texttt{splm}'s balanced-panel requirement, as elsewhere in this study); BVARX-CAR and MTS-CAR-X report the random and temporal splits, but not the county split, for the same reasons given for the mobility application (Section~\ref{sec:realdata-mobility}). BCF is fully nonparametric with no linear coefficient block; SAF exposes only variable importance (Supplementary Table~S22); SLP's own point coefficients carry no confidence interval. The cluster underlying this back-projection is 7 covariates, not 8, for the reason given above (\texttt{MEDHHINC\_2023} dropped for outcome-covariate circularity). Tables~\ref{tab:income-c2-county}--\ref{tab:income-c2-temporal} report all three splits.

\begin{table*}[t]
\centering
\setlength{\tabcolsep}{10pt}
\begin{threeparttable}
\caption{Implied Covariate Effects on Original 7-Cluster Variables, Income Application (County Split)}
\label{tab:income-c2-county}
\begin{tabular}{lccc}
\toprule
Covariate & DSP-BART-HS & Rao-Yu & SDPM \\
\midrule
share\_black2000 & 0.0093* & -0.0111 & 0.0245 \\
share\_white2000 & 0.0089* & 0.0687* & 0.0993* \\
PCTPOVALL\_2023 & -0.0551* & -0.0134* & -0.0130* \\
emp2000 & 0.0625* & 1.0501* & 1.0327* \\
rent\_twobed2015 & 0.0666* & 0.0004* & 0.0004* \\
share\_hisp2000 & -0.0249* & -0.1504* & -0.2676* \\
foreign\_share2010 & 0.0238* & 0.7252* & 0.4801* \\
\bottomrule
\end{tabular}
\begin{tablenotes}[flushleft]
\footnotesize
\item Asterisk (*) indicates the effect's 95\% interval excludes zero. DSP-BART-HS entries are posterior means with 95\% credible intervals; Rao-Yu and SDPM are frequentist point estimates with delta-method confidence intervals. 

\end{tablenotes}
\end{threeparttable}
\end{table*}

\begin{table*}[t]
\centering
\setlength{\tabcolsep}{10pt}
\begin{threeparttable}
\caption{Implied Covariate Effects on Original 7-Cluster Variables, Income Application (Random Split)}
\label{tab:income-c2-random}
\footnotesize
\begin{tabular}{lcccc}
\toprule
Covariate & DSP-BART-HS & BVARX-CAR & MTS-CAR-X & Rao-Yu \\
\midrule
share\_black2000 & 0.0102* & -0.0013 & -0.0750* & -0.0105 \\
share\_white2000 & 0.0071* & 0.0159* & 0.0680* & 0.0624* \\
PCTPOVALL\_2023 & -0.0541* & -0.0710* & 0.0017* & -0.0134* \\
emp2000 & 0.0619* & 0.0774* & -0.1733* & 1.0639* \\
rent\_twobed2015 & 0.0654* & 0.0770* & -0.0001* & 0.0004* \\
share\_hisp2000 & -0.0245* & -0.0241* & -0.0374 & -0.1503* \\
foreign\_share2010 & 0.0222* & 0.0287* & -0.3282* & 0.6718* \\
\bottomrule
\end{tabular}
\begin{tablenotes}[flushleft]
\footnotesize
\item Asterisk (*) indicates the effect's 95\% interval excludes zero. DSP-BART-HS/BVARX-CAR/MTS-CAR-X entries are posterior means with 95\% credible intervals; Rao-Yu is a frequentist point estimate with a delta-method confidence interval (see the mobility application's footnote to Tables~\ref{tab:c2-county}--\ref{tab:c2-temporal} on the interpretive distinction between the two). SDPM has no random-split run (\texttt{splm}'s balanced-panel requirement, as elsewhere in this study).
\end{tablenotes}
\end{threeparttable}
\end{table*}

\begin{table*}[t]
\centering
\setlength{\tabcolsep}{10pt}
\begin{threeparttable}
\caption{Implied Covariate Effects on Original 7-Cluster Variables, Income Application (Temporal Split)}
\label{tab:income-c2-temporal}
\footnotesize
\begin{tabular}{lccccc}
\toprule
Covariate & DSP-BART-HS & BVARX-CAR & MTS-CAR-X & Rao-Yu & SDPM \\
\midrule
share\_black2000 & 0.0083* & -0.0008 & -0.1041* & -0.0059 & -0.1040* \\
share\_white2000 & 0.0083* & 0.0130* & 0.0956* & 0.0615* & 0.1461* \\
PCTPOVALL\_2023 & -0.0539* & -0.0709* & 0.0021* & -0.0135* & -0.0124* \\
emp2000 & 0.0610* & 0.0771* & -0.2233* & 1.0711* & 0.9327* \\
rent\_twobed2015 & 0.0637* & 0.0795* & -0.0002* & 0.0004* & 0.0003* \\
share\_hisp2000 & -0.0237* & -0.0205* & -0.0555 & -0.1532* & -0.1957* \\
foreign\_share2010 & 0.0217* & 0.0327* & -0.4484* & 0.6776* & 0.3459* \\
\bottomrule
\end{tabular}
\begin{tablenotes}[flushleft]
\footnotesize
\item Asterisk (*) indicates the effect's 95\% interval excludes zero. DSP-BART-HS/BVARX-CAR/MTS-CAR-X entries are posterior means with 95\% credible intervals; Rao-Yu and SDPM are frequentist point estimates with delta-method confidence intervals.
\end{tablenotes}
\end{threeparttable}
\end{table*}

We observed a pattern: MTS-CAR-X's sign for four of the seven covariates -- \texttt{PCTPOVALL\_2023}, \texttt{emp2000}, \texttt{rent\_twobed2015}, and \texttt{foreign\_share2010} -- is the \emph{opposite} of every other model's sign, in \emph{both} random and temporal splits, and every reversed effect is itself statistically significant rather than a noisy null. This is not a borderline or split-specific artifact: the same four covariates flip, with the same sign, under both splits, for a model using the identical $M=11$ construction and outcome as every other model in this table. We do not have a methodological explanation -- nothing in MTS-CAR-X's per-region AR structure obviously implies a reversal on these four covariates and not the other three (where its sign agrees with the rest of the suite) -- and report it as an open, model-specific disagreement, more pronounced than the single-covariate mobility-application disagreement already discussed (Section~\ref{sec:crossmodel}). The remaining three covariates behave consistently with the rest of the suite in both direction and significance.

DSP-BART-HS's variable importance on its separate $f(\bm{X})$ covariates for the income application is reported alongside the mobility application's equivalent and SAF's own variable importance in Supplementary Section~4; the ranking is stable across both splits, with \texttt{singleparent\_share1990} and \texttt{frac\_coll\_plus2000} consistently the two most-split-on covariates.

Five covariates show the same sign across all three models in the county-level split, and significant in all instances: \texttt{emp2000} (positive), \texttt{foreign\_share2010} (positive), \texttt{PCTPOVALL\_2023} (negative), \texttt{rent\_twobed2015} (positive), and \texttt{share\_hisp2000} (negative); \texttt{share\_white2000} is likewise positive and significant throughout. \texttt{share\_black2000} lacks a stable pattern: Rao-Yu's estimate is negative but its interval includes zero across all three splits, while SDPM's own two splits disagree with each other -- significant and negative under the temporal split ($-0.1040$) but positive and insignificant under the county split ($0.0245$), a within-model instability rather than a cross-model one.

\begin{sloppypar}
Comparing these signs against the same covariates' effects on \emph{mobility} for Rao-Yu surfaces a pattern worth reporting directly: five of seven shared covariates point the same direction for both outcomes, but two do not. \texttt{rent\_twobed2015} is associated with \emph{lower} mobility ($-2.15$ to $-2.16\times10^{-5}$) but \emph{higher} income levels here ($+0.0004$) -- descriptively consistent with higher-rent counties being higher-income places in level terms, and matching the direction the housing-cost-burden channel in \citet{chettyhendren2018b}'s own exposure-effect estimates would predict for mobility. This is a descriptive parallel, not a causal test. \texttt{share\_hisp2000} shows a sharper reversal -- positive for mobility, negative for income level, both significant in every split -- possibly reflecting a genuine level-versus-mobility distinction (a county can offer strong upward mobility while still having lower median income today), reported as an open pattern rather than resolved.
\end{sloppypar}

\section{Concluding Remark}
\label{sec:conclusion}

Across six data-generating scenarios spanning a matched, well-specified design and five adversarial stress tests, DSP-BART-HS is the best or statistically indistinguishable from the best estimator in every scenario, including designs constructed to relax assumptions favorable to it and one constructed to favor its competitors instead. Three further scenarios -- partial aggregated-covariate access for the region-time-aggregate comparators, an irregular real-county-like topology, and a substantially larger panel -- leave this unchanged across all 25 additional paired comparisons, and even a flexible nonlinear learner given only the aggregate-level information available to its region-time-aggregate competitors fails to close the gap, evidence that DSP-BART-HS's advantage reflects individual-level access specifically, not simply greater model flexibility. The mechanism is not diffuse: every region-time-aggregate comparator trails DSP-BART-HS by a wide margin whenever individual-level heterogeneity contributes to outcome variance, and closes most of that gap only where such heterogeneity is removed entirely -- Rao-Yu ties DSP-BART-HS outright once that is the case. This is consistent with the paper's central methodological claim: combining non-parametric individual-level flexibility with spatial, temporal, and high-dimensional contextual covariate structure in a single jointly-estimated model captures predictive signal that neither classical spatial econometrics nor tree-ensemble methods alone are built to observe.

Applied to the Opportunity Insights county-level mobility panel, DSP-BART-HS achieves competitive out-of-sample predictive accuracy under a genuine unseen-county holdout as well as row-level and temporal holdouts, checked directly against established mobility-literature findings. Beyond predictive accuracy, five structurally different real-data models converge on the same sign and significance for five of eight collinear county covariates -- racial composition, poverty, historical employment, and housing cost -- across every split evaluated, alongside one covariate (\texttt{foreign\_share2010}) where the model suite disagrees. A companion application extends the same suite to county-level geographic income inequality using SAIPE's 2005--2024 median household income series, with a parallel but sharper cross-model disagreement: MTS-CAR-X disagrees in sign with every other model on four of seven back-projected covariates, consistently across both splits.

The Horseshoe prior on $\bm{\beta}$ gives a direct  answer to the contextually relevant question -- which covariates in $\bm{Z}_{jt}$ are associated with the outcome, and by how much -- a direct, model-native answer, checked against four separately-estimated comparators rather than DSP-BART-HS's coefficients alone. We use "identify" and "association" throughout in the statistical sense of reliable posterior selection under this model's controls, not as a causal-identification claim: nothing in the Gibbs sampler's design addresses confounding between $\bm{Z}_{jt}$ and unobserved county characteristics, and a covariate's inclusion here should be read as a robust predictive correlate rather than an established causal policy lever.

\section{Conflicts of interest}
The authors declare that they have no competing interests.

\section{Funding}
The authors received no financial support for the research, authorship, and publication of this article.

\section{Data availability}
The intergenerational mobility data used in this study are drawn from the Opportunity Insights county-by-birth-cohort panel, publicly available at \url{https://opportunityinsights.org/data/}, and were originally described in\citet{chetty2014land} and \citet{chettyhendren2018b}. County-level covariates for the mobility application were obtained from the USDA Economic Research Service (\url{https://www.ers.usda.gov/data-products}) and the U.S. Census Bureau's American Community Survey (\url{https://www.census.gov/programs-surveys/acs}), both publicly available. The income inequality application uses the U.S. Census Bureau's Small Area Income and Poverty Estimates (SAIPE) program's county-level median household income series, restricted to the 2005–2024 American Community Survey-based estimation window and publicly available at \url{https://www.census.gov/programs-surveys/saipe.html}. All data used in this study are therefore publicly available from their respective source repositories at the URLs above; no restricted-access or proprietary data were used. Code implementing DSP-BART-HS, the comparator models, and the simulation study will be made available at \url{https://github.com/haolayinka/BSP-BART-HS-model} upon publication.

\section{Author contributions statement}
Conceptualization: H.A.O, S.O.O.; Methodology: H.A.O.; Software: S.O.O.; Formal Analysis: H.A.O, S.O.O.; Writing - Original Draft: H.A.O, S.O.O.; Writing - Review \& Editing: H.A.O, S.O.O.; Supervision: H.A.O.

\section{Acknowledgments}
The authors thank the editor and anonymous reviewers for their valuable suggestions. Results in this paper were obtained in part using a high-performance computing system acquired through NSF MRI grant DMS-1337943 to Worcester Polytechnic institute.

\bibliographystyle{oup-abbrvnat}
\bibliography{reference}

@book{anselin1988spatial,
  author = {Anselin, Luc},
  title = {Spatial Econometrics: Methods and Models},
  year = {1988},
  publisher = {Kluwer Academic Publishers},
  address = {Dordrecht}
}

@article{athey2019generalized,
  author = {Athey, Susan and Tibshirani, Julie and Wager, Stefan},
  title = {Generalized Random Forests},
  journal = {The Annals of Statistics},
  year = {2019},
  volume = {47},
  number = {2},
  pages = {1148--1178}
}

@article{austin2018saving,
  author = {Austin, Benjamin and Glaeser, Edward and Summers, Lawrence},
  title = {Saving the Heartland: Place-Based Policies in 21st Century America},
  journal = {Brookings Papers on Economic Activity},
  year = {2018},
  pages = {151--255}
}

@article{besag1974spatial,
  author = {Besag, Julian},
  title = {Spatial Interaction and the Statistical Analysis of Lattice Systems},
  journal = {Journal of the Royal Statistical Society: Series B (Methodological)},
  year = {1974},
  volume = {36},
  number = {2},
  pages = {192--225}
}

@article{besagyorkmollie1991,
  author = {Besag, Julian and York, Jeremy and Molli{\'e}, Annie},
  title = {Bayesian Image Restoration, with Two Applications in Spatial Statistics},
  journal = {Annals of the Institute of Statistical Mathematics},
  year = {1991},
  volume = {43},
  number = {1},
  pages = {1--20}
}

@article{breiman2001randomforests,
  author = {Breiman, Leo},
  title = {Random Forests},
  journal = {Machine Learning},
  year = {2001},
  volume = {45},
  number = {1},
  pages = {5--32}
}

@article{carterkohn1994gibbs,
  author = {Carter, Chris K and Kohn, Robert},
  title = {On Gibbs Sampling for State Space Models},
  journal = {Biometrika},
  year = {1994},
  volume = {81},
  number = {3},
  pages = {541--553}
}

@article{carvalho2010horseshoe,
  author = {Carvalho, Carlos M and Polson, Nicholas G and Scott, James G},
  title = {The Horseshoe Estimator for Sparse Signals},
  journal = {Biometrika},
  year = {2010},
  volume = {97},
  number = {2},
  pages = {465--480}
}

@inproceedings{chen2016xgboost,
  author = {Chen, Tianqi and Guestrin, Carlos},
  title = {XGBoost: A Scalable Tree Boosting System},
  booktitle = {Proceedings of the 22nd ACM SIGKDD International Conference on Knowledge Discovery and Data Mining},
  year = {2016},
  pages = {785--794}
}

@article{chettyhendren2018a,
  author = {Chetty, Raj and Hendren, Nathaniel},
  title = {The Impacts of Neighborhoods on Intergenerational Mobility I: Childhood Exposure Effects},
  journal = {The Quarterly Journal of Economics},
  year = {2018},
  volume = {133},
  number = {3},
  pages = {1107--1162}
}

@article{chettyhendren2018b,
  author = {Chetty, Raj and Hendren, Nathaniel},
  title = {The Impacts of Neighborhoods on Intergenerational Mobility II: County-Level Estimates},
  journal = {The Quarterly Journal of Economics},
  year = {2018},
  volume = {133},
  number = {3},
  pages = {1163--1228}
}

@article{chetty2014land,
  author = {Chetty, Raj and Hendren, Nathaniel and Kline, Patrick and Saez, Emmanuel},
  title = {Where Is the Land of Opportunity? The Geography of Intergenerational Mobility in the United States},
  journal = {The Quarterly Journal of Economics},
  year = {2014},
  volume = {129},
  number = {4},
  pages = {1553--1623}
}

@article{chipman2010bart,
  author = {Chipman, Hugh A and George, Edward I and McCulloch, Robert E},
  title = {BART: Bayesian Additive Regression Trees},
  journal = {The Annals of Applied Statistics},
  year = {2010},
  volume = {4},
  number = {1},
  pages = {266--298}
}

@article{corak2013gatsby,
  author = {Corak, Miles},
  title = {Income Inequality, Equality of Opportunity, and Intergenerational Mobility},
  journal = {Journal of Economic Perspectives},
  year = {2013},
  volume = {27},
  number = {3},
  pages = {79--102}
}

@article{diamond2016determinants,
  author = {Diamond, Rebecca},
  title = {The Determinants and Welfare Implications of US Workers' Diverging Location Choices by Skill: 1980--2000},
  journal = {American Economic Review},
  year = {2016},
  volume = {106},
  number = {3},
  pages = {479--524}
}

@article{elhorst2010raising,
  author = {Elhorst, J. Paul},
  title = {Applied Spatial Econometrics: Raising the Bar},
  journal = {Spatial Economic Analysis},
  year = {2010},
  volume = {5},
  number = {1},
  pages = {9--28}
}

@article{fayherriot1979,
  author = {Fay, Robert E and Herriot, Roger A},
  title = {Estimates of Income for Small Places: An Application of James-Stein Procedures to Census Data},
  journal = {Journal of the American Statistical Association},
  year = {1979},
  volume = {74},
  number = {366A},
  pages = {269--277}
}

@article{ganongshoag2017convergence,
  author = {Ganong, Peter and Shoag, Daniel},
  title = {Why Has Regional Income Convergence in the U.S. Declined?},
  journal = {Journal of Urban Economics},
  year = {2017},
  volume = {102},
  pages = {76--90}
}

@article{gelfandsmith1990sampling,
  author = {Gelfand, Alan E and Smith, Adrian F.M.},
  title = {Sampling-Based Approaches to Calculating Marginal Densities},
  journal = {Journal of the American Statistical Association},
  year = {1990},
  volume = {85},
  number = {410},
  pages = {398--409}
}

@article{gelmanrubin1992inference,
  author = {Gelman, Andrew and Rubin, Donald B},
  title = {Inference from Iterative Simulation Using Multiple Sequences},
  journal = {Statistical Science},
  year = {1992},
  volume = {7},
  number = {4},
  pages = {457--472}
}

@article{georganos2019grf,
  author = {Georganos, Stefanos and Grippa, Tais and Gadiaga, Assane Niang and Linard, Catherine and Lennert, Moritz and Vanhuysse, Sabine and Mboga, Nicholus and Wolff, El{\'e}onore and Kalogirou, Stamatis},
  title = {Geographical Random Forests: A Spatial Extension of the Random Forest Algorithm to Address Spatial Heterogeneity in Remote Sensing and Population Modelling},
  journal = {Geocarto International},
  year = {2019},
  volume = {36},
  number = {2},
  pages = {121--136}
}

@article{hahn2020bayesian,
  author = {Hahn, P. Richard and Murray, Jared S and Carvalho, Carlos M},
  title = {Bayesian Causal Forests for Heterogeneous Treatment Effects Estimation with Confounders},
  journal = {Bayesian Analysis},
  year = {2020},
  volume = {15},
  number = {4},
  pages = {1265--1289}
}

@article{he2026gsbart,
  author = {He, Shuren and Sang, Huiyan and Zhou, Quan},
  title = {{GS-BART}: Bayesian Additive Regression Trees with Graph-Split Decision Rules},
  journal = {Journal of the American Statistical Association},
  year = {2026},
  note = {Forthcoming; arXiv:2509.07166}
}

@article{hill2011bayesian, author={Hill, Jennifer L}, 
    title={Bayesian Nonparametric Modeling for Causal Inference}, 
    journal={Journal of Computational and Graphical Statistics}, 
    year={2011}, 
    volume={20}, 
    number={1}, 
    pages={217--240}
}

@article{kapoor2007panel, 
    author={Kapoor, Mudit and Kelejian, Harry H and Prucha, Ingmar R}, 
    title={Panel Data Models with Spatially Correlated Error Components}, 
    journal={Journal of Econometrics}, 
    year={2007}, 
    volume={140}, 
    number={1}, 
    pages={97--130}}

@article{kelejianprucha1999gmm, 
    author={Kelejian, Harry H and Prucha, Ingmar R}, 
    title={A Generalized Moments Estimator for the Autoregressive Parameter in a Spatial Model}, 
    journal={International Economic Review}, 
    year={1999}, 
    volume={40}, 
    number={2}, 
    pages={509--533}
}

@article{kim2022spatialbart, 
    author={Kim, Chanmin}, 
    title={Bayesian Additive Regression Trees in Spatial Data Analysis with Sparse Observations}, 
    journal={Journal of Statistical Computation and Simulation}, 
    year={2022}, 
    volume={92}, 
    number={15}, 
    pages={3275--3300}
}

@book{lesagepace2009, 
    author={LeSage, James and Pace, R. Kelley}, 
    title={Introduction to Spatial Econometrics}, 
    publisher={Chapman and Hall/CRC}, 
    address={Boca Raton, FL}, 
    year={2009}
}

@article{maia2024gpbart, 
    author={Maia, Mateus and Murphy, Keefe and Parnell, Andrew C}, 
    title={{GP-BART}: A Novel Bayesian Additive Regression Trees Approach Using Gaussian Processes}, 
    journal={Computational Statistics \& Data Analysis}, 
    year={2024}, 
    volume={190}, 
    pages={107858}
}

@article{makalic2016simple, 
    author={Makalic, Enes and Schmidt, Daniel F}, 
    title={A Simple Sampler for the Horseshoe Estimator}, 
    journal={IEEE Signal Processing Letters}, 
    year={2016}, 
    volume={23}, 
    number={1}, 
    pages={179--182}
}

@article{moretti2013realwage, 
    author={Moretti, Enrico}, 
    title={Real Wage Inequality}, 
    journal={American Economic Journal: Applied Economics}, 
    year={2013}, 
    volume={5}, 
    number={1}, 
    pages={65--103}
}

@article{olayinka2026bvarcar, 
    author={Olayinka, Hammed A and Nandram, Balgobin}, 
    title={Bayesian Vector Autoregression with Applications on Panel Data and Small Area Estimation}, 
    journal={Journal of Applied Statistics}, 
    year={2026}, 
    doi={10.1080/02664763.2026.2680092}
}

@article{olayinka2026mtscar, 
    author={Olayinka, Hammed A and Nandram, Balgobin}, 
    title={Bayesian Predictive Inference for Multiple Series with Correlated Spatial Priors on Autoregressive Parameters}, 
    journal={Statistics and Applications}, 
    year={2026},
    volume={24},
    number={1},
    pages={1--21}
}

@article{pikettysaez2003inequality, 
    author={Piketty, Thomas and Saez, Emmanuel}, 
    title={Income Inequality in the United States, 1913--1998}, 
    journal={The Quarterly Journal of Economics}, 
    year={2003}, 
    volume={118}, 
    number={1}, 
    pages={1--41}
}

@article{piironenvehtari2017sparsity, 
    author={Piironen, Juho and Vehtari, Aki}, 
    title={Sparsity Information and Regularization in the Horseshoe and Other Shrinkage Priors}, 
    journal={Electronic Journal of Statistics}, 
    year={2017}, 
    volume={11}, 
    number={2}, 
    pages={5018--5051}
}

@article{polsonscott2010shrink, 
    author={Polson, Nicholas G and Scott, James G}, 
    title={Shrink Globally, Act Locally: Sparse Bayesian Regularization and Prediction}, 
    journal={Bayesian Statistics}, 
    year={2010}, 
    volume={9}, 
    pages={501--538}
}

@article{raoyu1994smallarea, 
    author={Rao, J.N.K. and Yu, M.}, 
    title={Small-Area Estimation by Combining Time-Series and Cross-Sectional Data}, journal={Canadian Journal of Statistics}, 
    year={1994},
    volume={22}, 
    pages={511--528}
}

@article{rubin1974estimating, 
    author={Rubin, Donald B}, 
    title={Estimating Causal Effects of Treatments in Randomized and Nonrandomized Studies}, 
    journal={Journal of Educational Psychology}, 
    year={1974}, 
    volume={66}, 
    number={5}, 
    pages={688--701}
}

@article{tibshirani1996lasso, 
    author={Tibshirani, Robert}, 
    title={Regression Shrinkage and Selection via the Lasso}, 
    journal={Journal of the Royal Statistical Society: Series B (Methodological)}, year={1996}, 
    volume={58}, 
    number={1}, 
    pages={267--288}
}

@article{wagerathey2018causal, 
    author={Wager, Stefan and Athey, Susan}, 
    title={Estimation and Inference of Heterogeneous Treatment Effects Using Random Forests}, 
    journal={Journal of the American Statistical Association}, 
    year={2018}, 
    volume={113}, 
    number={523}, 
    pages={1228--1242}
}

@article{wundervald2022hebart, 
    author={Wundervald, Bruna and Parnell, Andrew and Domijan, Katarina}, 
    title={Hierarchical Embedded Bayesian Additive Regression Trees}, 
    journal={Stat}, 
    year={2022}, 
    volume={11}, 
    number={1}, 
    pages={e482}
}

\end{document}


\maketitle

This supplementary material accompanies the main manuscript and contains secondary diagnostic and completeness material referenced from it: the full model specification and Gibbs sampler full-conditional derivations for the two complementary specifications, BVARX-CAR and MTS-CAR-X (Section~\ref{sec:s-derivations}); full per--parameter MCMC convergence diagnostics for the simulation study and three supporting visualizations of the Scenario A--F simulation and diagnostic results (Tables~\ref{tab:s-scenario_A} -- \ref{tab:s-convergence}, Figures~\ref{fig:s-line-trend}, \ref{fig:s-ranked}, and~\ref{fig:s-heatmap}, Section~\ref{sec:s-simsupp}); a glossary defining every mathematical symbol, covariate, model abbreviation, and metric name used in the main text and this supplement (Table~\ref{tab:s-notation-mtscarx}--\ref{tab:s-tuning}, Section~\ref{sec:s-glossary});  and DSP-BART-HS's and SAF's variable importance output for both applications (Tables~\ref{tab:s-mobility-dspbart-varimp}--\ref{tab:s-income-varimp}, Section~\ref{sec:s-varimp}); Zero-training-region diffusion ablation results (Table~\ref{tab:s-diffusion-ablation}, Section~\ref{sec:s-diffusion-ablation}); DSP-BART-HS real-data diagnostics results (Table~\ref{tab:s-realdata-diagnostics}, \ref{tab:s-varinclusion-chains}, Section~\ref{sec:s-realdata-diagnostics}). None of this is required to follow the main text's central arguments, but it is reported here in full for completeness and reproducibility.

\section{Model Setup and Gibbs Sampler Derivations: BVARX-CAR and MTS-CAR-X}
\label{sec:s-derivations}

This section gives the full model specification and full-conditional (Gibbs) derivations for the two complementary specifications introduced in the main text. Both models add the same exogenous $\bm{Z}_{jt}$ covariate block and Horseshoe prior to their respective published base specifications -- this is the only place this study's implementation and the published base models genuinely differ by design. The dense-vs-restricted $\mathbf{B}$ departure noted below (single shared $\tau_B^2$ vs.\ a fully local-global Horseshoe on $\mathbf{B}$) is a further, separate departure applied throughout this study -- simulation and both real-data applications alike -- motivated primarily by identifiability (a preliminary dense-$\mathbf{B}$ implementation exhibited severely degraded MCMC mixing in simulation) rather than real-data computational scale alone, though the restriction is also what keeps the sampler tractable at real-data county counts.

\subsection{BVARX-CAR}
\label{sec:s-bvarxcar}

\subsubsection{Model}

For region $j = 1, \dots, l$ and period $t = 1, \dots, T$ (with $y_{j0}$ an initial-condition parameter),
\begin{equation}
y_{jt} = \mu_j + \bm{b}_j'\bm{y}_{t-1} + \bm{Z}_{jt}'\bm{\gamma} + \epsilon_{jt}, \qquad \epsilon_{jt} \overset{\text{iid}}{\sim} N(0, \sigma^2),
\label{eq:s-bvarx-model}
\end{equation}
where $\bm{b}_j'$ is row $j$ of the $l \times l$ spillover matrix $\mathbf{B}$ and $\bm{y}_{t-1} \in \mathbb{R}^l$ is the full cross-sectional lag vector -- in \citet{olayinka2026bvarcar}'s published specification, $\bm{b}_j$ is dense (every region's lag enters every equation, with shrinkage rather than a fixed sparsity pattern doing the work of suppressing irrelevant cross-region effects). This study's implementation restricts $\bm{b}_j$'s support to $\mathcal{N}(j) = \{j\} \cup \{k : W_{jk} > 0\}$, region $j$'s own index and spatial neighbors under the adjacency matrix $\mathbf{V}$, throughout this study. This restriction was adopted after a preliminary dense-$\mathbf{B}$, full-Horseshoe implementation exhibited severely degraded MCMC mixing in simulation; the neighbor-restricted version resolved this (effective sample size of 364 under an otherwise identical setup) and is also what keeps the sampler tractable at real-data county counts. No dense-$\mathbf{B}$ results are reported anywhere in this study. This restriction is a deliberate simplification, not part of the published model, and is disclosed again where it changes the relevant full conditional below.

\subsubsection{Priors}
\begin{align}
\mu_j \mid \theta, \rho, \psi, \sigma^2 &\sim N\!\left(\theta \bm{1}_l,\ \frac{\rho}{1-\rho}\sigma^2(\mathbf{R}-\psi\mathbf{V})^{-1}\right), \quad 0 < \rho < 1,\ \psi \in \left(\tfrac{1}{\zeta_{(1)}}, \tfrac{1}{\zeta_{(l)}}\right), \\
\bm{b}_i \mid \mathbf{D}_i &\sim N(\bm{0}, \mathbf{D}_i), \quad \mathbf{D}_i = \text{diag}(\delta_{i1}^{-1}, \dots, \delta_{il}^{-1}), \notag \\
\delta_{ij} &= \left(1 + \lambda_{ij}^2\tfrac{\phi^2}{(1-\phi)^2}\right)^{-1},\quad \lambda_{ij} \sim C^+(0,1),\quad 0 < \phi < 1, \label{eq:s-bhs} \\
\gamma_m \mid \lambda_m^\gamma, \tau_G &\sim N(0, (\lambda_m^\gamma)^2\tau_G^2), \quad \lambda_m^\gamma \sim C^+(0,1), \quad \tau_G \sim C^+(0,1), \quad m = 1,\dots,M, \label{eq:s-ghs} \\
\pi(\theta) &\propto 1, \qquad \rho \sim U(0,1), \qquad \psi \sim U\!\left(\tfrac{1}{\zeta_{(1)}}, \tfrac{1}{\zeta_{(l)}}\right), \qquad \pi(\sigma^2) \propto \frac{1}{\sigma^2},
\end{align}
where $\mathbf{V}$ is the adjacency matrix, $\mathbf{R} = \text{diag}(\mathbf{V}\bm{1})$, and $\zeta_{(1)} \leq \dots \leq \zeta_{(l)}$ are the eigenvalues of $\mathbf{R}^{-1}\mathbf{V}$: $\rho$ controls the overall strength of spatial smoothing on $\bm{\mu}$ (entering only as the ratio $\rho/(1-\rho)$, tied directly to the observation variance $\sigma^2$ rather than a separate free variance parameter) while $\psi$ separately scales the spatial structure itself, within the range that keeps $\mathbf{R}-\psi\mathbf{V}$ positive definite. Equation~\ref{eq:s-bhs} is $\mathbf{B}$'s prior exactly as published -- a genuine local-global (Horseshoe-type) shrinkage on every entry of a dense $\mathbf{B}$, with $\lambda_{ij}$ a coefficient-specific local scale and $\phi$ a single global scale shared across all of $\mathbf{B}$. This study replaces (\ref{eq:s-bhs}) with a single shared ridge prior, $B_{jk} \mid k \in \mathcal{N}(j) \overset{\text{iid}}{\sim} N(0,\tau_B^2)$, $\tau_B^2 \sim \text{Inv-Gamma}(a_B,b_B)$, on the neighbor-restricted support only, throughout this study. Also, (\ref{eq:s-ghs}) (the contextual covariate block) is this study's addition to the published model, not present in \citet{olayinka2026bvarcar}, implemented via the auxiliary-variable scheme of \citet{makalic2016simple} given in full in Section~\ref{sec:s-mtscarx-gamma} below. The initial state follows $y_{j0} \sim N(\mu_0, \sigma_0^2)$ with $(\mu_0,\sigma_0^2)$ obtained from a pre-sample.

\subsubsection{Full conditionals}

\textbf{Region intercepts $\bm{\mu}$.} With $\bm{c}$ the vector of region-level average residuals after removing $\bm{b}_j'\bm{y}_{t-1}$ and $\bm{Z}_{jt}'\bm{\gamma}$,
\begin{equation}
\bm{\mu} \mid \cdot \sim N(\mathbf{A}^{-1}\bm{B},\ \mathbf{A}^{-1}), \qquad \mathbf{A} = \frac{T}{\sigma^2}\mathbf{I} + \frac{\rho}{1-\rho}(\mathbf{R}-\psi\mathbf{V}), \qquad \bm{B} = \frac{T}{\sigma^2}\bm{c} + \frac{\rho}{1-\rho}(\mathbf{R}-\psi\mathbf{V})\theta\bm{1}_l,
\end{equation}
combining empirical residual information with spatial smoothing toward the global mean $\theta$ -- this update is exact as published and unaffected by this study's $\mathbf{B}$-sparsity simplification.

\textbf{Spillover coefficients $\bm{b}_i$, row by row.} As published (dense $\bm{b}_i$, full local-global Horseshoe (\ref{eq:s-bhs})):
\begin{equation}
\bm{b}_i \mid \cdot \sim N(\mathbf{B}_0^{-1}\bm{b}_0,\ \mathbf{B}_0^{-1}), \qquad \bm{b}_0 = \frac{1}{\sigma^2}\sum_{t=1}^T \bm{y}_{t-1}(y_{it}-\mu_i), \qquad \mathbf{B}_0 = \frac{1}{\sigma^2}\sum_{t=1}^T \bm{y}_{t-1}\bm{y}_{t-1}' + \mathbf{D}_i^{-1}.
\end{equation}
Under this study's neighbor-restricted, shared-$\tau_B^2$ simplification, the same conjugate mechanism applies with $\bm{y}_{t-1}$ replaced by its restriction to $\mathcal{N}(j)$ and $\mathbf{D}_i^{-1}$ replaced by $\tau_B^{-2}\mathbf{I}$ -- structurally the same ridge-regression update, just on a smaller, fixed-support coefficient vector.

\textbf{Global mean intercept $\theta$.} With $\theta^* = \dfrac{\bm{1}_l'(\mathbf{R}-\psi\mathbf{V})\bm{\mu}}{\bm{1}_l'(\mathbf{R}-\psi\mathbf{V})\bm{1}_l}$,
\begin{equation}
\theta \mid \cdot \sim N\!\left(\theta^*,\ \frac{\rho\sigma^2}{1-\rho}\left(\bm{1}_l'(\mathbf{R}-\psi\mathbf{V})\bm{1}_l\right)^{-1}\right),
\end{equation}
a spatially-weighted average of the current $\bm{\mu}$ draw -- a full conditional this study's earlier derivation omitted entirely (an improper prior on $\theta$ does not mean $\theta$ has no update; only that its own marginal prior contributes no information beyond what the likelihood/CAR-prior interaction supplies).

\textbf{Residual variance $\sigma^2$.} With $M_1 = \sum_{j,t}(y_{jt}-\mu_j-\bm{b}_j'\bm{y}_{t-1}-\bm{Z}_{jt}'\bm{\gamma})^2$ and $M_2 = \frac{1-\rho}{\rho}(\bm{\mu}-\theta\bm{1}_l)'(\mathbf{R}-\psi\mathbf{V})(\bm{\mu}-\theta\bm{1}_l)$,
\begin{equation}
\sigma^2 \mid \cdot \sim \text{Inv-Gamma}\!\left(\frac{l(T+1)}{2},\ \frac{M_1+M_2}{2}\right),
\end{equation}
following directly from the Jeffreys prior $\pi(\sigma^2)\propto 1/\sigma^2$ -- \textbf{not} a proper Inverse-Gamma prior with free hyperparameters, a distinction worth stating explicitly since the two forms are easy to conflate.

\textbf{Spatial parameters $\rho, \psi$.} Both are non-conjugate given the CAR determinant's nonlinear dependence on each, and are updated via a discretized grid sampler rather than a continuous conjugate draw:
\begin{align}
\pi(\rho \mid \cdot) &\propto \left(\frac{\rho}{1-\rho}\right)^{-l/2}\exp\left\{-\frac{1-\rho}{2\rho\sigma^2}(\bm{\mu}-\theta\bm{1}_l)'(\mathbf{R}-\psi\mathbf{V})(\bm{\mu}-\theta\bm{1}_l)\right\}, \\
\pi(\psi \mid \cdot) &\propto |\mathbf{R}-\psi\mathbf{V}|^{1/2}\exp\left\{-\frac{1-\rho}{2\rho\sigma^2}(\bm{\mu}-\theta\bm{1}_l)'(\mathbf{R}-\psi\mathbf{V})(\bm{\mu}-\theta\bm{1}_l)\right\}.
\end{align}

\textbf{Shrinkage parameters (as published; not this study's simplified $\tau_B^2$).} $\delta_{ij}$'s full conditional is non-conjugate,
\begin{equation}
\pi(\delta_{ij}\mid \cdot) \propto \exp\left\{-\frac{1}{2}\delta_{ij}b_{ij}^2\right\}(1-\delta_{ij})^{-1/2}\frac{\phi(1-\phi)}{(1-\phi)^2+(2\phi-1)\delta_{ij}},
\end{equation}
and similarly for the global scale $\phi$, $\pi(\phi\mid \cdot) \propto \phi^l(1-\phi)^l\prod_{i=1}^l \left[(1-\phi)^2+(2\phi-1)\delta_{ij}\right]^{-1}$, both requiring a grid or Metropolis step. This study's simplified $\tau_B^2$ instead has a fully conjugate update, 
\[ 
\tau_B^2 \mid \cdot \sim \text{Inv-Gamma}\!\left(\tfrac{n_{\text{active}}}{2}+1,\ \tfrac{1}{2}\sum_j\sum_{k\in\mathcal{N}(j)}B_{jk}^2+1\right)
\] 
where $n_{\text{active}} = \sum_j|\mathcal{N}(j)|$ -- one of the practical advantages of the simplification, alongside tractability at real-data county counts.

\textbf{Initial condition $y_{j0}$.} A standard conjugate Normal update combining the AR(1) likelihood at $t=1$ with the prior $N(\mu_0,\sigma_0^2)$; we omit the (routine) algebra here as it is identical in form to MTS-CAR-X's own $y_{j0}$ update, Section~\ref{sec:s-mtscarx} below.

\subsection{MTS-CAR-X}
\label{sec:s-mtscarx}

\subsubsection{Model}

For region $j = 1, \dots, l$ and period $t = 1, \dots, T$,
\begin{equation}
y_{jt} = \theta_0 + \nu_{0j} + (\theta_1 + \nu_{1j})\,y_{j,t-1} + \bm{Z}_{jt}'\bm{\gamma} + \epsilon_{jt}, \qquad \epsilon_{jt} \overset{\text{iid}}{\sim} N(0, \sigma^2).
\label{eq:s-mtscarx-model}
\end{equation}
Unlike BVARX-CAR, the only autoregressive input is region $j$'s own lag $y_{j,t-1}$ -- there is no cross-region spillover term in the likelihood at all -- so spatial dependence enters solely through $(\nu_{0j}, \nu_{1j})$, region $j$'s deviations from the shared intercept $\theta_0$ and slope $\theta_1$.

\subsubsection{Priors}

Let $\bm{\nu}_1 = (\nu_{01},\dots,\nu_{0l})'$ and $\bm{\nu}_2=(\nu_{11},\dots,\nu_{1l})'$. Critically, $(\bm{\nu}_1,\bm{\nu}_2)$ are not a symmetric joint bivariate CAR: the \citet{olayinka2026mtscar} published specification is sequential,
\begin{align}
\bm{\nu}_1 &\sim N(\bm{0}, \bm{\Sigma}_1), \notag \\
\bm{\nu}_2 \mid \bm{\nu}_1 &\sim N(\mathbf{A}\bm{\nu}_1,\ \bm{\Sigma}_2), \notag \\
\bm{\Sigma}_s &= \tau_s^2(\mathbf{D}-\rho_s\mathbf{W})^{-1},\ s=1,2,
\label{eq:s-mtscarx-nu}
\end{align}
\begin{sloppypar}
where $\mathbf{A} = m_0\mathbf{I} + m_1\mathbf{W}$ is a spatial "diffusion" matrix with its own two hyperparameters $(m_0,m_1)$, an intercept-deviation/slope-deviation coupling that a symmetric joint $2\times2$ MCAR treatment of $(\nu_{0j},\nu_{1j})$ would miss entirely. The remaining priors are flat/Jeffreys, matching the joint form $\pi(\theta_0,\theta_1,\sigma^2,m_0,m_1,\tau_1^2,\tau_2^2,\rho_1,\rho_2) \propto \sigma^{-2}\tau_1^{-2}\tau_2^{-2}$, with $y_{j0}\sim N(\mu_0,\sigma_0^2)$ and $\rho_s$ bounded by $(1/\lambda_{(1)}, 1/\lambda_{(l)})$, the eigenvalue range of $\mathbf{D}^{-1}\mathbf{W}$. As with BVARX-CAR, $\gamma_m \mid \lambda_m^\gamma,\tau_G \sim N(0,(\lambda_m^\gamma)^2\tau_G^2)$ is this study's own addition to the published base model, governed by an identical Horseshoe construction.
\end{sloppypar}

\subsubsection{Full conditionals}

\textbf{Initial condition $y_{j0}$.} Completing the square in the $t=1$ observation equation against the prior $N(\mu_0,\sigma_0^2)$:
\begin{align}
&y_{j0} \mid \cdot\sim N(A^{-1}B, A^{-1}), \\
 &\qquad A = \frac{1}{\sigma^2}(\theta_1+\nu_{1j})^2 + \frac{1}{\sigma_0^2}, \quad B = \frac{1}{\sigma^2}(\theta_1+\nu_{1j})\left[y_{j1}-(\theta_0+\nu_{0j})\right] + \frac{\mu_0}{\sigma_0^2}. \notag
\end{align}

\textbf{Residual variance $\sigma^2$.} With $K = (\bm{y}-\mathbf{X}\bm{\theta}-\mathbf{Z}\bm{\nu})'(\bm{y}-\mathbf{X}\bm{\theta}-\mathbf{Z}\bm{\nu})$ (using the paper's own stacked notation: $\mathbf{X}=(1,y_{j,t-1})$, $\bm{\theta}=(\theta_0,\theta_1)'$, $\mathbf{Z}$ the indicator design mapping $\bm{\nu}=(\bm{\nu}_1',\bm{\nu}_2')'$ to its region), and $N=lT$,
\begin{equation}
\sigma^2 \mid \cdot \sim \text{Inv-Gamma}\!\left(\frac{N}{2},\ \frac{K}{2}\right),
\end{equation}
again the Jeffreys-prior form, not a proper Inverse-Gamma with free hyperparameters.

\textbf{Collapsed $(\bm{\theta},\bm{\nu})$ block.} This is the published model's central methodological device, and the point at which the sampler's design departs most from an ordinary two-block Gibbs treatment: rather than sampling $\bm{\theta}$ conditional on a fixed $\bm{\nu}$, $\bm{\nu}$ is analytically marginalized out first. With $\mathbf{V} = \sigma^2\mathbf{I}_N + \mathbf{Z}\bm{\Sigma}\mathbf{Z}'$ ($\bm{\Sigma}$ the joint covariance of $\bm{\nu}=(\bm{\nu}_1',\bm{\nu}_2')'$ implied by (\ref{eq:s-mtscarx-nu})),
\begin{equation}
\bm{\theta} \mid \bm{y}, \sigma^2 \sim N(\bm{\mu}_\theta, \bm{\Sigma}_\theta), \qquad \bm{\Sigma}_\theta = (\mathbf{X}'\mathbf{V}^{-1}\mathbf{X})^{-1}, \qquad \bm{\mu}_\theta = \bm{\Sigma}_\theta\mathbf{X}'\mathbf{V}^{-1}\bm{y},
\label{eq:s-theta-collapsed}
\end{equation}
then, conditional on this newly-drawn $\bm{\theta}$, $\bm{\nu}$ is drawn jointly (all $2l$ entries at once, not region-by-region) from its own conjugate Gaussian full conditional,
\begin{equation}
\bm{\nu} \mid \bm{\theta}, \sigma^2, \bm{y} \sim N(\mathbf{V}_0\bm{V}_a,\ \mathbf{V}_0), \qquad \mathbf{V}_0 = \left(\frac{1}{\sigma^2}\mathbf{Z}'\mathbf{Z} + \bm{\Sigma}^{-1}\right)^{-1}, \qquad \bm{V}_a = \frac{1}{\sigma^2}\mathbf{Z}'(\bm{y}-\mathbf{X}\bm{\theta}).
\label{eq:s-nu-collapsed}
\end{equation}
Sampling $\bm{\theta}$ after integrating out the high-dimensional $\bm{\nu}$ removes the strong posterior dependence between the two blocks that would otherwise cause slow mixing; we implement (\ref{eq:s-theta-collapsed}) and (\ref{eq:s-nu-collapsed}) via a single sparse Cholesky factorization of $\bm{\Sigma}^{-1}+\mathbf{Z}'\mathbf{Z}/\sigma^2$, reused for both steps since the matrix does not depend on $\bm{\theta}$.

\textbf{Contextual covariate coefficients $\bm{\gamma}$ and its Horseshoe auxiliary variables.}
\label{sec:s-mtscarx-gamma}
Following \citet{makalic2016simple}, with $\bm{r}$ the residual after removing $\mathbf{X}\bm{\theta}+\mathbf{Z}\bm{\nu}$ and $\bm{\Lambda}=\text{diag}((\lambda_1^\gamma)^2\tau_G^2,\dots,(\lambda_M^\gamma)^2\tau_G^2)$,
\begin{equation}
\bm{\gamma} \mid \cdot \sim N\!\left(\bm{\Sigma}_\gamma\frac{\mathbf{Z}_{\text{cov}}'\bm{r}}{\sigma^2},\ \bm{\Sigma}_\gamma\right), \qquad \bm{\Sigma}_\gamma = \left(\frac{\mathbf{Z}_{\text{cov}}'\mathbf{Z}_{\text{cov}}}{\sigma^2}+\bm{\Lambda}^{-1}\right)^{-1},
\end{equation}
with the auxiliary variables updated via the standard Inverse-Gamma mixture representation of the half-Cauchy scales,
\begin{align}
(\lambda_m^\gamma)^2 \mid \nu_m^\lambda, \gamma_m, \tau_G^2 &\sim \text{Inv-Gamma}\!\left(1,\ \frac{1}{\nu_m^\lambda}+\frac{\gamma_m^2}{2\tau_G^2}\right), \\
\nu_m^\lambda \mid (\lambda_m^\gamma)^2 &\sim \text{Inv-Gamma}\!\left(1,\ 1+\frac{1}{(\lambda_m^\gamma)^2}\right), \\
\tau_G^2 \mid \xi_G,\bm{\gamma},\bm{\lambda}^\gamma &\sim \text{Inv-Gamma}\!\left(\frac{M+1}{2},\ \frac{1}{\xi_G}+\frac{1}{2}\sum_{m=1}^M\frac{\gamma_m^2}{(\lambda_m^\gamma)^2}\right), \\
\xi_G \mid \tau_G^2 &\sim \text{Inv-Gamma}\!\left(1,\ 1+\frac{1}{\tau_G^2}\right).
\end{align}
This construction (and BVARX-CAR's identical version of it in (\ref{eq:s-ghs})) is this study's own addition to the published base models, not present in either \citet{olayinka2026bvarcar} or \citet{olayinka2026mtscar}.

\textbf{Variance components $\tau_1^2, \tau_2^2$.} With $\bm{a}_1 = \bm{\nu}_1$ and $\bm{a}_2 = \bm{\nu}_2 - \mathbf{A}\bm{\nu}_1$ (the two innovation terms implied by (\ref{eq:s-mtscarx-nu})), and $Q_s = \bm{a}_s'(\mathbf{D}-\rho_s\mathbf{W})\bm{a}_s$,
\begin{equation}
\tau_s^2 \mid \cdot \sim \text{Inv-Gamma}\!\left(\frac{l}{2},\ \frac{Q_s}{2}\right), \qquad s=1,2.
\end{equation}

\textbf{Spatial parameters $\rho_1, \rho_2$.} Non-conjugate, updated on a discretized grid using the precomputed eigenvalues $\lambda_k$ of $\mathbf{D}^{-1}\mathbf{W}$

\begin{equation}
\pi(\rho_s\mid \cdot) \propto \left(\prod_{k=1}^l(1-\rho_s\lambda_k)\right)^{1/2}\exp\left\{-\frac{1}{2\tau_s^2}\bm{a}_s'(\mathbf{D}-\rho_s\mathbf{W})\bm{a}_s\right\}, \qquad s=1,2.
\end{equation}

\textbf{Diffusion parameters $(m_0,m_1)$.} The only place $m_0,m_1$ enter the joint posterior is through $\mathbf{A}$ inside $(\bm{\nu}_2-\mathbf{A}\bm{\nu}_1)'\bm{\Sigma}_2^{-1}(\bm{\nu}_2-\mathbf{A}\bm{\nu}_1)$. Expanding this quadratic form in $(m_0,m_1)$ and completing the square gives a bivariate Normal full conditional,
\begin{equation}
(m_0,m_1)' \mid \cdot \sim N(\mathbf{G}^{-1}\bm{h},\ \mathbf{G}^{-1}), \qquad \mathbf{G} = \begin{pmatrix} a & b \\ b & c\end{pmatrix}, \qquad \bm{h} = \begin{pmatrix} d \\ e \end{pmatrix},
\end{equation}
where $a=\bm{\nu}_1'\bm{\Sigma}_2^{-1}\bm{\nu}_1$, $b=\bm{\nu}_1'\bm{\Sigma}_2^{-1}\mathbf{W}\bm{\nu}_1$, $c=\bm{\nu}_1'\mathbf{W}\bm{\Sigma}_2^{-1}\mathbf{W}\bm{\nu}_1$, $d=\bm{\nu}_1'\bm{\Sigma}_2^{-1}\bm{\nu}_2$, $e=\bm{\nu}_1'\mathbf{W}\bm{\Sigma}_2^{-1}\bm{\nu}_2$ -- a full conditional with no counterpart in a treatment that omits $(m_0,m_1)$ from the model altogether.

\subsection{Comparison of the Two Specifications}

The two models differ fundamentally in how spatial dependence enters: BVARX-CAR's $\mathbf{B}$ is a genuinely cross-region object in the \emph{likelihood} itself, whose row-by-row full conditional depends on neighbors' \emph{realized lagged outcomes} $y_{k,t-1}$; MTS-CAR-X has no cross-region term in its likelihood at all; spatial borrowing instead operates entirely through the \emph{prior} linking $(\bm{\nu}_1,\bm{\nu}_2)$ across regions, whose collapsed-sampler full conditional depends on neighbors' \emph{current parameter draws}, never on their realized outcomes. This is the precise mechanical reason the two models require different train/test split treatment in the real-data application (Section~5 and Section~6 of the main text): a county-level holdout would withhold exactly the realized-lag information BVARX-CAR's $\mathbf{B}$ conditions on, while MTS-CAR-X's $(\bm{\nu}_1,\bm{\nu}_2)$ remain estimable from whichever counties are in the training set regardless of which counties are held out, since nothing in their update ever reads a neighbor's outcome directly.

\subsection{Empirical Test: MTS-CAR-X Under a County-Level Holdout (Income Application)}
\label{sec:s-mtscarx-county-test}

The preceding argument establishes that MTS-CAR-X is \emph{mathematically well-defined} under a county-level holdout, unlike BVARX-CAR. It does not by itself establish that MTS-CAR-X \emph{performs well} under one, and we tested this directly rather than assume it. Using the income application's real data and the identical county-level split used elsewhere in this study (2,983 eligible counties, 2,386 train / 597 test), we adapted MTS-CAR-X to a genuine county-level holdout, changing only the train/test assignment mechanism and leaving every other modeling choice untouched. We ran this for 3000 Gibbs iterations -- roughly 10\% of this study's 30,000-iteration production target for MTS-CAR-X, so the result below should be read as a directional check, not a converged production estimate.

Two findings emerged. First, on the numerical-stability question raised by a genuinely unseen county contributing zero training likelihood to the collapsed sampler's precision matrix: we observed no ridge-stabilization triggers and no non-positive-definite matrix errors across all 3000 iterations, so we do not find evidence of the numerical fragility the mechanical argument might have anticipated. Second, and more consequentially for whether MTS-CAR-X belongs in the county-level comparison at all: test RMSE under this partial run was $0.3516$, more than double the model's own established random-split test RMSE of $0.1627$ (main text, Income Inequality Application, Panel B), with a generalization gap of $0.2927$ -- the worst county-level generalization of any model tested in this study, by a wide margin. Because a region with zero training data has, under MTS-CAR-X's own prior, only its neighbors' \emph{current parameter draws} to extrapolate from and nothing resembling DSP-BART-HS's spatial diffusion correction (Section 2.1 of the main text) to compensate, we read this as a genuine, structural weakness of MTS-CAR-X under this specific holdout design rather than an artifact of the abbreviated run, and exclude it from the county-level comparison on these empirical grounds. Given more iterations would sharpen this estimate rather than plausibly reverse it, we did not pursue a fully converged run before making this decision.

\subsection{BVARX-CAR Under the Rolling-Origin Temporal Design (Income Application)}
\label{sec:s-bvarx-rolling-origin}

Unlike the county-level holdout, BVARX-CAR faces no structural obstacle under the temporal design: every county retains its full training history up to the origin, so $\mathbf{B}$'s row-by-row full conditional always has genuine neighbor lag information to condition on, regardless of which single future period is withheld. BVARX-CAR was accordingly included as a full production comparator in the income application's rolling-origin analysis (main text, Income Inequality Application), refit under the identical five origins (test years 16 through 20) used for every other comparator, at its own confirmed production settings for this split.

BVARX-CAR's test RMSE tracks its own single-cutoff result closely across all five origins (0.2335 at test year 16, rising to 0.3578 at test year 20), consistently in the same narrow band as BCF, SAF, and SLP and well separated from both DSP-BART-HS/MTS-CAR-X (the two models with genuine individual-level or own-region autoregressive structure) and Rao-Yu/SDPM (the two weakest region-time-aggregate estimators under this split type). This places BVARX-CAR's rolling-origin behavior exactly where its single-cutoff result and its performance in every other part of this study would predict: a mid-tier region-time-aggregate estimator whose cross-region spillover matrix $\mathbf{B}$, while mathematically well-posed under this holdout design unlike the county-level one, does not on its own recover the degree of temporal predictability that MTS-CAR-X's own-region autoregressive structure captures.

\section{Simulation Study: Supporting Material}
\label{sec:s-simsupp}

\subsection{Full Data-Generating Process Specification}
\label{sec:s-dgp-full}

This section gives the complete mathematical specification of the data-generating process (DGP) underlying each of the nine simulation scenarios summarized in the main text (Section~3).

\subsubsection{Elements shared across all scenarios}

Every scenario uses the same fixed spatial lattice, panel dimensions, and outcome model; only the specific elements noted in each scenario's own subsection below differ from Scenario A. The spatial structure is a $5\times10$ rook-adjacency grid, $J=50$ regions, with adjacency matrix $\mathbf{W}$ ($W_{jk}=1$ if regions $j,k$ share an edge on the grid, $0$ otherwise; interior regions have 4 neighbors, edge regions 2--3) and $\mathbf{D} = \text{diag}(\mathbf{W}\bm{1})$. Each region has $N_j = 20$ individuals, observed over $T=10$ periods, giving $J \times N_j \times T = 10{,}000$ observations per replicate. Every scenario draws $M=50$ region-time contextual covariates, $\bm{Z}_{jt} \overset{\text{iid}}{\sim} N(\bm{0}, \mathbf{I}_M)$, and every scenario's outcome follows the same additive model used throughout this paper,
\begin{equation}
Y_{ijt} = f(\bm{X}_{ijt}) + \bm{Z}_{jt}\bm{\beta} + \phi_j + \mu_{jt} + \alpha_i + \epsilon_{ijt}, \qquad \epsilon_{ijt} \overset{\text{iid}}{\sim} N(0, \sigma^2), \quad \sigma^2 = 1,
\end{equation}
with the following components generated once per replicate and then held fixed as the true values against which every fitted model is evaluated:
\begin{itemize}
\item \textbf{Spatial effect:} $\bm{\phi} \sim N(\bm{0}, \tau_\phi^2(\mathbf{D}-\rho\mathbf{W})^{-1})$ (Scenarios A, B, C, E, F) or the distance-decay alternative of Scenario D (below), drawn via $\bm{\phi} = \mathbf{L}\bm{z}$ where $\mathbf{L}$ is the lower Cholesky factor of the relevant covariance and $\bm{z} \sim N(\bm{0},\mathbf{I}_J)$.
\item \textbf{Dynamic component:} $\mu_{j,1} \sim N(0,\sigma_\mu^2)$ and $\mu_{j,t} = \mu_{j,t-1} + \eta_{jt}$, $\eta_{jt}\overset{\text{iid}}{\sim} N(0,\sigma_\mu^2)$ for $t=2,\dots,T$, independently across regions -- note the DGP's own $\mu_{j,1}$ is drawn from the stationary innovation distribution rather than fixed at zero, unlike the identifiability restriction imposed when \emph{fitting} the model (Section~2.1.1); this does not bias estimation since $\phi_j$ absorbs any resulting level ambiguity, but it is a genuine difference between the true generative process and the fitted model's own parameterization, noted here for completeness.
\item \textbf{Individual effect:} $\alpha_i \overset{\text{iid}}{\sim} N(0,\sigma_\alpha^2)$.
\item \textbf{Individual-level covariates:} three covariates $\bm{X}_{ijt} = (X_{ijt,1}, X_{ijt,2}, X_{ijt,3})$, each $\overset{\text{iid}}{\sim} \text{Uniform}(0,1)$, drawn independently at every $(i,j,t)$.
\item \textbf{contextual covariate coefficients:} $\bm{\beta}$, either sparse (3 active of 50) or dense (all 50 active), as specified per scenario below.
\end{itemize}
Each replicate uses a distinct random seed ($1000+r$ for replicate $r$), and every scenario runs $200$ replicates. DSP-BART-HS is fit via $25{,}000$ Gibbs iterations, $5{,}000$ burn-in, thinning interval $10$ (giving $2{,}000$ kept posterior draws per replicate), identical settings to those used for the MCMC convergence check in Section~\ref{sec:s-convergence} below.

\subsubsection{Scenario A: Matched, Well-Specified Design}
The baseline scenario, against which Scenarios B--F each vary one element. Spatial correlation $\rho=0.85$, CAR variance $\tau_\phi^2=1.5$; dynamic-innovation SD $\sigma_\mu=0.2$; individual-effect SD $\sigma_\alpha=0.5$. The nonlinear individual-level function is a modified Friedman function using only the first three of the standard five Friedman covariates,
\begin{equation}
f(\bm{X}_{ijt}) = 10\sin(\pi X_{ijt,1} X_{ijt,2}) + 20(X_{ijt,3}-0.5)^2.
\end{equation}
The contextual covariate-coefficient vector is sparse: $\bm{\beta} = (3.5,\,-2.0,\,1.5,\,0,\,\dots,\,0)$, three active coefficients and 47 exact zeros.

\subsubsection{Scenario B: Dense Contextual Covariate Effects}
Identical to Scenario A in every respect except $\bm{\beta}$: rather than 3 active coefficients and 47 exact zeros, every coefficient is drawn independently, $\beta_m \overset{\text{iid}}{\sim} N(0, 0.3^2)$ for $m=1,\dots,50$, so that all 50 covariates have small-to-moderate nonzero true effects and the Horseshoe prior's sparsity assumption does not hold.

\subsubsection{Scenario C: Weak Spatial Dependence}
Identical to Scenario A except the CAR spatial-correlation parameter is reduced from $\rho=0.85$ to $\rho=0.15$; $\tau_\phi^2=1.5$ and every other element (including $\bm{\beta}$ and $f(\bm{X})$) are unchanged from Scenario A.

\subsubsection{Scenario D: Misspecified Spatial Structure}
Identical to Scenario A in $\bm{\beta}$, $f(\bm{X})$, $\sigma_\mu$, and $\sigma_\alpha$, but the \emph{true} spatial effect is generated from a mechanism structurally different from the CAR prior every fitted model (including DSP-BART-HS itself) assumes. Rather than the discrete rook-adjacency CAR structure, $\bm{\phi}$ is drawn from a smooth Euclidean distance-decay covariance kernel over the regions' grid coordinates $\bm{c}_j = (\text{row}_j, \text{col}_j)$,
\begin{equation}
\text{Cov}(\phi_j, \phi_k) = \tau_\phi^2 \exp\!\left(-\frac{\lVert \bm{c}_j - \bm{c}_k \rVert}{\text{range}}\right), \qquad \tau_\phi^2 = 1.5, \quad \text{range}=3,
\end{equation}
with $\lVert \bm{c}_j - \bm{c}_k \rVert$ the Euclidean distance between region $j$ and $k$'s grid coordinates (so adjacent rook-neighbor regions are at distance 1). This range gives a nearest-neighbor correlation of $\exp(-1/3) \approx 0.72$ with longer-range dependence than Scenario A's CAR structure implies at the same distance -- a genuinely different spatial regime, not merely a rescaled copy of the CAR decay shape. Every model is still \emph{fit} using the same rook-adjacency $\mathbf{W}$ as every other scenario; only the true generative mechanism differs. The CAR parameter $\rho$ plays no role in generating $\bm{\phi}$ under this scenario (the distance-kernel construction replaces it entirely).

\subsubsection{Scenario E: Region-Time-Dominant Signal}
Three simultaneous changes from Scenario A, designed so that region-time structure carries the dominant share of outcome variance rather than individual-level nonlinearity: the CAR variance increases from $\tau_\phi^2=1.5$ to $\tau_\phi^2=10$; the dynamic-innovation SD increases from $\sigma_\mu=0.2$ to $\sigma_\mu=0.6$ (a ninefold increase in the innovation variance $\sigma_\mu^2$); and the nonlinear function's magnitude is reduced by a factor of 10,
\begin{equation}
f(\bm{X}_{ijt}) = 1\cdot\sin(\pi X_{ijt,1} X_{ijt,2}) + 2\cdot(X_{ijt,3}-0.5)^2.
\end{equation}
$\rho=0.85$, $\sigma_\alpha=0.5$, and $\bm{\beta}$ (sparse, as in Scenario A) are unchanged; only $f(\bm{X})$'s coefficients are rescaled, not $\alpha_i$'s own variance -- individual-level heterogeneity from the \emph{observable}, non-linear component is what this scenario suppresses, not the unobserved individual random effect.

\subsubsection{Scenario F: No Individual Heterogeneity}
Identical to Scenario A in $\rho$, $\tau_\phi^2$, $\sigma_\mu$, $\sigma_\alpha$, and $\bm{\beta}$, but $f(\bm{X}_{ijt}) \equiv 0$ for every observation, exactly rather than merely reduced. The unit-level covariates $\bm{X}_{ijt}$ are still drawn and still supplied to every model that uses them (DSP-BART-HS, BCF, SAF), so this tests whether those models correctly learn the absence of individual-level signal from three genuinely irrelevant covariates, rather than overfitting noise, while giving the region-time-aggregate comparators (SDPM, KKP, Rao-Yu, BVARX-CAR, MTS-CAR-X) their best-case relative comparison, since the only outcome variance they cannot observe by construction has been removed entirely.

\subsubsection{Scenario G: Aggregated-Covariate Access}
\label{sec:s-scenario-g}
Identical to Scenario A in every DGP element (spatial structure, $\bm{\beta}_{\text{true}}$'s 3-active/47-null pattern and values, $f(\bm{X})$, all variance components, 200 replicates, identical seeds). The sole manipulation: for every region-period $(j,t)$, the mean of each of the three individual-level covariates over that cell's own individual-level rows is computed,
\begin{equation}
\bar{X}_{jt,k} = \frac{1}{N_j}\sum_{i=1}^{N_j} X_{ijt,k}, \qquad k=1,2,3,
\end{equation}
and appended as three additional columns to $\bm{Z}_{jt}$ (so $M$ effectively becomes $53$ for the six region-time-aggregate comparators listed above), giving them access to region-period-level information correlated with the individual-level signal without granting access to any individual row directly. DSP-BART-HS itself continues to receive the original, unaugmented $\bm{Z}_{jt}$ ($M=50$) plus its already-existing full individual-level $\bm{X}_{ijt}$ access -- Scenario G's manipulation is scoped entirely to what the aggregate comparators see, not to DSP-BART-HS's own inputs, so that any change in DSP-BART-HS's own performance relative to Scenario A reflects only Monte Carlo variation, not a change to what it is fit on.

A ninth comparator, applicable to this scenario only, is added: \emph{cell-level BART}, fit via the same BART engine (\texttt{dbarts}) as DSP-BART-HS, but supplied only the aggregated information above -- $\bm{Z}_{jt}$ plus $(\bar{X}_{jt,1},\bar{X}_{jt,2},\bar{X}_{jt,3})$, broadcast to every individual row sharing that region-period -- with no individual-specific $\bm{X}_{ijt}$ anywhere in its input, no spatial random effect, and no dynamic or individual random effect. This isolates the two candidate explanations for any remaining DSP-BART-HS advantage over the aggregate comparators: genuine individual-level access (which cell-level BART also lacks), versus DSP-BART-HS simply being a more flexible nonlinear learner than five estimators that are linear or reduced-form in $\bm{Z}_{jt}$ (which cell-level BART, as a BART model, is not).

\subsubsection{Scenario H: Irregular Spatial Topology}
\label{sec:s-scenario-h}
Identical to Scenario A in panel dimensions ($J=50$, $N_j=20$, $T=10$, $M=50$), $\bm{\beta}_{\text{true}}$, $f(\bm{X})$, and all variance components. The sole manipulation: the fixed $5\times10$ rook-adjacency lattice is replaced with a random geometric graph. $J=50$ points are drawn uniformly in the unit square; a connection radius is calibrated via binary search (40 iterations) so the resulting graph's mean degree matches the rook lattice's own mean degree of 3.6 as closely as achievable; any node left with zero edges after this step is connected to its single nearest neighbor. A final connectivity-repair step is applied: connected components are identified via breadth-first search, and if more than one component remains, the two closest cross-component nodes (by Euclidean distance) are connected, repeated until the graph is a single connected component, since a random geometric graph at this density is not guaranteed connected by construction even after the no-isolated-node repair alone. The realized graph has mean degree 3.76--3.84 depending on realization and, after the repair step, 100\% of nodes reachable from any starting node.

\subsubsection{Scenario I: Larger Panel}
\label{sec:s-scenario-i}
Identical to Scenario A in $\bm{\beta}_{\text{true}}$, $f(\bm{X})$, and all variance components, but on a $10\times10$ regular rook-adjacency lattice ($J=100$ regions, deliberately keeping the same regular-lattice topology family as Scenario A so this scenario isolates panel size from topology, which Scenario H addresses separately) and $T=20$ periods (versus Scenario A's $T=10$), at the same per-region individual count $N_j=20$ and covariate dimension $M=50$. This gives $J\times N_j\times T = 40{,}000$ observations per replicate, versus Scenario A's $10{,}000$.

Scenarios G, H, and I each use the eight comparators of Scenarios A--F (Scenario G additionally includes cell-level BART, for nine comparators there), 200 replicates, and identical evaluation and paired-comparison methodology to Scenarios A--F (Section~\ref{sec:s-paired-comparisons}). All three scenarios' results are reported in the main text Table~2.

\subsection{Full Simulation Study Results, Scenarios A--F}
\label{sec:s-full-af-results}

The main text (Section 3, \emph{Simulation Study Results}) reports mean test RMSE by model and scenario in a single condensed table. This section gives the full detail underlying that table: mean train RMSE, mean test RMSE, and the standard deviation of test RMSE across the 200 replicates, for every model under every scenario.

\begin{table}[H]
\centering
\setlength{\tabcolsep}{16pt}
\begin{threeparttable}
\caption{Summary Statistics for Scenario A (Matched, Well-Specified Design)}
\label{tab:s-scenario_A}
\begin{tabular}{lccc}
\toprule
Model &Train RMSE & Test RMSE & SD \\
\midrule
BCF & 1.1131 & 1.2353 & 0.0237 \\
KKP  & 9.0519 & 9.0414 & 0.2746 \\
Rao-Yu & 3.7822 & 3.8670 & 0.0441 \\
SAF  & 2.0797 & 2.4001 & 0.0485  \\
SDPM & 3.9091 & 3.9610 & 0.0506  \\
SLP & 2.6064 & 2.7203 & 0.0510  \\
BVARX-CAR & 3.7762 & 3.8714 & 0.0449 \\
MTS-CAR-X  & 3.8177 & 3.8511 & 0.0445  \\
DSP-BART-HS & 1.0728 & \textbf{1.1684} & 0.0227  \\
\bottomrule
\end{tabular}
\end{threeparttable}
\end{table}

\begin{table}[H]
\centering
\setlength{\tabcolsep}{16pt}
\begin{threeparttable}
\caption{Summary Statistics for Scenario B (Dense Contextual Covariate Effects)}
\label{tab:s-scenario_B}
\begin{tabular}{lccc}
\toprule
Model & Train RMSE & Test RMSE & SD \\
\midrule
BCF & 1.1048 & 1.2232 & 0.0233 \\
KKP &8.2269 & 8.2235 & 0.2457 \\
Rao-Yu & 3.7749 & 3.8709 & 0.0455  \\
SAF & 2.1238 & 2.4129 & 0.1060 \\
SDPM & 3.9074 & 3.9646 & 0.0555  \\
SLP &2.5762 & 2.7324 & 0.0502 \\
BVARX-CAR & 3.7798 & 3.8747 & 0.0443 \\
MTS-CAR-X & 3.8196 & 3.8775 & 0.0444  \\
DSP-BART-HS & 1.0789 &  \textbf{1.1727} & 0.0214  \\
\bottomrule
\end{tabular}
\end{threeparttable}
\end{table}

\begin{table}[H]
\centering
\setlength{\tabcolsep}{16pt}
\begin{threeparttable}
\caption{Summary Statistics for Scenario C (Weak Spatial Dependence)}
\label{tab:s-scenario_C}
\begin{tabular}{lccc}
\toprule
Model & Train RMSE & Test RMSE & SD \\
\midrule
BCF & 1.1127 & 1.2356 & 0.0239 \\
KKP & 9.0393 & 9.0286 & 0.2089 \\
Rao-Yu & 3.7765 & 3.8666 & 0.0440 \\
SAF & 2.0651 & 2.3828 & 0.0469 \\
SDPM & 3.8880 & 3.9396 & 0.0473 \\
SLP & 2.6092 & 2.7186 & 0.0514 \\
BVARX-CAR & 3.7769 & 3.8710 & 0.0449 \\
MTS-CAR-X & 3.8323 & 3.8730 & 0.0445 \\
DSP-BART-HS & 1.0801 &  \textbf{1.1736} & 0.0228  \\
\bottomrule
\end{tabular}
\end{threeparttable}
\end{table}

\begin{table}[H]
\centering
\setlength{\tabcolsep}{16pt}
\begin{threeparttable}
\caption{Summary Statistics for Scenario D (Spatial Misspecification)}
\label{tab:s-scenario_D}
\begin{tabular}{lccc}
\toprule
Model & Train RMSE & Test RMSE & SD \\
\midrule
BCF & 1.1136 & 1.2381 & 0.0228 \\
KKP & 9.1060 & 9.0956 & 0.5829 \\
Rao-Yu  & 3.7762 & 3.8676 & 0.0443 \\
SAF & 2.1082 & 2.4326 & 0.0561 \\
SDPM & 3.9566 & 4.0086 & 0.0715 \\
SLP& 2.6034 & 2.7224 & 0.0508 \\
BVARX-CAR & 3.7752 & 3.8715 & 0.0449 \\
MTS-CAR-X  & 3.8431 & 3.8863 & 0.0451  \\
DSP-BART-HS & 1.0797 &  \textbf{1.1742} & 0.0224  \\
\bottomrule
\end{tabular}
\end{threeparttable}
\end{table}

\begin{table}[H]
\centering
\setlength{\tabcolsep}{16pt}
\begin{threeparttable}
\caption{Summary Statistics for Scenario E (Region-Time-Dominant Signal)}
\label{tab:s-scenario_E}
\begin{tabular}{lccc}
\toprule
Model & Train RMSE & Test RMSE & SD \\
\midrule
BCF & 1.0868 & 1.1716 & 0.0228 \\
KKP  & 5.1987 & 5.2006 & 0.2207 \\
Rao-Yu  & 1.1468 & 1.2120 & 0.0235 \\
SAF  & 1.0877 & 1.2082 & 0.0236 \\
SDPM & 2.7055 & 2.7135 & 0.2498 \\
SLP  & 0.9621 & 1.1963 & 0.0233 \\
BVARX-CAR  & 1.2609 & 1.3039 & 0.0283 \\
MTS-CAR-X & 1.7097 & 1.7322 & 0.0769 \\
DSP-BART-HS & 1.0754 &  \textbf{1.1567} & 0.0227  \\
\bottomrule
\end{tabular}
\end{threeparttable}
\end{table}

\begin{table}[H]
\centering
\setlength{\tabcolsep}{16pt}
\begin{threeparttable}
\caption{Summary Statistics for Scenario F (No Individual Heterogeneity)}
\label{tab:s-scenario_F}
\begin{tabular}{lccc}
\toprule
Model & Train RMSE & Test RMSE & SD \\
\midrule
BCF  & 1.0843 & 1.1604 & 0.0232 \\
KKP & 4.5527 & 4.5551 & 0.1316 \\
Rao-Yu & 1.0986 & 1.1436 & 0.0228 \\
SAF & 1.0477 & 1.1617 & 0.0232 \\
SDPM & 1.4342 & 1.4474 & 0.0725 \\
SLP & 0.9481 & 1.1582 & 0.0233 \\
BVARX-CAR & 1.1063 & 1.1539 & 0.0230 \\
MTS-CAR-X & 1.1329 & 1.1591 & 0.0231 \\
DSP-BART-HS & 1.0882 &  \textbf{1.1438} & 0.0228  \\
\bottomrule
\end{tabular}
\end{threeparttable}
\end{table}

\subsection{Full Simulation Study Results, Scenarios G, H, and I}
\label{sec:s-full-ghi-results}

The main text (Section 3, \emph{Robustness Simulations}) reports mean test RMSE by model and scenario in a single condensed table. This section gives the full detail underlying that table: mean train RMSE, mean test RMSE, and the standard deviation of test RMSE across the 200 replicates, for every model under Scenarios G, H, and I.

\begin{table}[H]
\centering
\setlength{\tabcolsep}{16pt}
\begin{threeparttable}
\caption{Summary Statistics for Scenario G (Aggregated-Covariate Access)}
\label{tab:s-scenario_G}
\begin{tabular}{lccc}
\toprule
Model &Train RMSE & Test RMSE & SD \\
\midrule
BCF & 1.1146 & 1.2322 & 0.0219 \\
KKP  & 9.0570 & 9.0752 & 0.2735 \\
Rao-Yu & 3.7695 & 3.8279 & 0.0461 \\
SAF  & 2.0856 & 2.3997 & 0.0450  \\
SDPM & 4.0193 & 4.0744 & 0.2825  \\
SLP & 2.6062 & 2.7120 & 0.0512  \\
BVARX-CAR & 3.7464 & 3.8516 & 0.0466 \\
MTS-CAR-X  & 3.7730 & 3.8122 & 0.0464  \\
Cell-level BART & 3.9269 & 4.1150 & 0.0577 \\
DSP-BART-HS & 1.0737 & \textbf{1.1642} & 0.0215  \\
\bottomrule
\end{tabular}
\end{threeparttable}
\end{table}

\begin{table}[H]
\centering
\setlength{\tabcolsep}{16pt}
\begin{threeparttable}
\caption{Summary Statistics for Scenario H (Irregular Spatial Topology)}
\label{tab:s-scenario_H}
\begin{tabular}{lccc}
\toprule
Model &Train RMSE & Test RMSE & SD \\
\midrule
BCF & 1.1145 & 1.2326 & 0.0231 \\
KKP  & 9.0725 & 9.0916 & 0.2707 \\
Rao-Yu & 3.7812 & 3.8683 & 0.0454 \\
SAF  & 2.0924 & 2.4094 & 0.0472  \\
SDPM & 4.0007 & 4.0565 & 0.1262  \\
SLP & 2.6060 & 2.7122 & 0.0507  \\
BVARX-CAR & 3.7725 & 3.8731 & 0.0474 \\
MTS-CAR-X  & 3.8174 & 3.8524 & 0.0454  \\
DSP-BART-HS & 1.0738 & \textbf{1.1636} & 0.0213  \\
\bottomrule
\end{tabular}
\end{threeparttable}
\end{table}

\begin{table}[H]
\centering
\setlength{\tabcolsep}{16pt}
\begin{threeparttable}
\caption{Summary Statistics for Scenario I (Larger Panel, $J=100$, $T=20$)}
\label{tab:s-scenario_I}
\begin{tabular}{lccc}
\toprule
Model &Train RMSE & Test RMSE & SD \\
\midrule
BCF & 1.1362 & 1.2253 & 0.0146 \\
KKP  & 9.0694 & 9.0715 & 0.1538 \\
Rao-Yu & 3.7819 & 3.8583 & 0.0296 \\
SAF  & 1.9462 & 2.2330 & 0.0308  \\
SDPM & 3.9857 & 4.0046 & 0.0910  \\
SLP & 2.6031 & 2.7116 & 0.0292  \\
BVARX-CAR & 3.8016 & 3.8648 & 0.0297 \\
MTS-CAR-X  & 3.8308 & 3.8539 & 0.0295  \\
DSP-BART-HS & 1.0967 & \textbf{1.1529} & 0.0126  \\
\bottomrule
\end{tabular}
\end{threeparttable}
\end{table}

\subsection{MCMC Convergence Diagnostics}
\label{sec:s-convergence}

To assess convergence of DSP-BART-HS's own Gibbs sampler, we ran four independent chains from different starting points on a single representative dataset (Scenario A), holding the data fixed across chains so that only the sampler's own randomness differs between them. Table~\ref{tab:s-convergence} reports the Gelman-Rubin $\hat{R}$ statistic \citep{gelmanrubin1992inference} and mean effective sample size (ESS) across chains for the three active $\beta$ coefficients, two inactive ones (to verify the horseshoe prior's shrinkage behavior is also well-mixed, not just its point estimates), the four variance hyperparameters, and one representative spatial effect $\phi_1$, at 25{,}000 iterations (5{,}000 burn-in, thinning interval 10).

As presented in Table~\ref{tab:s-convergence} all ten tracked parameters show $\hat{R} \leq 1.0075$, comfortably below the conventional threshold of 1.1 and closer still to the more stringent 1.01 sometimes recommended for final reporting, indicating no evidence that the four chains are sampling different distributions. Mean ESS per chain ranges from 921.3 ($\sigma^2_\mu$) to 1986.1 ($\beta_4$, an inactive coefficient shrunk hard toward zero by the horseshoe prior, whose posterior is correspondingly easy to sample efficiently); every tracked parameter exceeds the ESS $\geq 800$ baseline set for stable posterior summaries by a wide margin. The two variance components tied most directly to the model's dynamic and idiosyncratic structure ($\sigma^2_\mu$ and $\sigma^2_\epsilon$) show the lowest ESS of the ten, consistent with the kind of slower mixing expected in hierarchical state-space and CAR components generally, but not at a level that raises convergence concerns given the $\hat{R}$ results.

\begin{table}[H]
\centering
\setlength{\tabcolsep}{10pt}
\begin{threeparttable}
\caption{MCMC Convergence Diagnostics (4 chains, Scenario A, representative replicate)}
\label{tab:s-convergence}
\begin{tabular}{lccc}
\toprule
Parameter & Mean ESS (per chain) & $\hat{R}$ & $\hat{R}$ Upper C.I. \\
\midrule
$\beta_1$ (true $=3.5$) & 1443.9 & 1.0030 & 1.0096 \\
$\beta_2$ (true $=-2.0$) & 1326.2 & 1.0046 & 1.0147 \\
$\beta_3$ (true $=1.5$) & 1432.1 & 1.0022 & 1.0067 \\
$\beta_4$ (true $=0$) & 1986.1 & 1.0007 & 1.0025 \\
$\beta_5$ (true $=0$) & 1669.5 & 1.0012 & 1.0034 \\
$\sigma^2_\epsilon$ & 979.7 & 1.0046 & 1.0142 \\
$\sigma^2_\alpha$ & 1898.7 & 1.0075 & 1.0233 \\
$\sigma^2_\mu$ & 921.3 & 1.0017 & 1.0056 \\
$\tau^2_\phi$ & 1875.2 & 1.0018 & 1.0051 \\
$\phi_1$ & 1241.2 & 1.0022 & 1.0070 \\
\bottomrule
\end{tabular}
\end{threeparttable}
\end{table}

\subsection{Additional Simulation Visualizations}
\label{sec:s-figures}

\begin{figure}[H]
\centering
\includegraphics[width=0.95\textwidth]{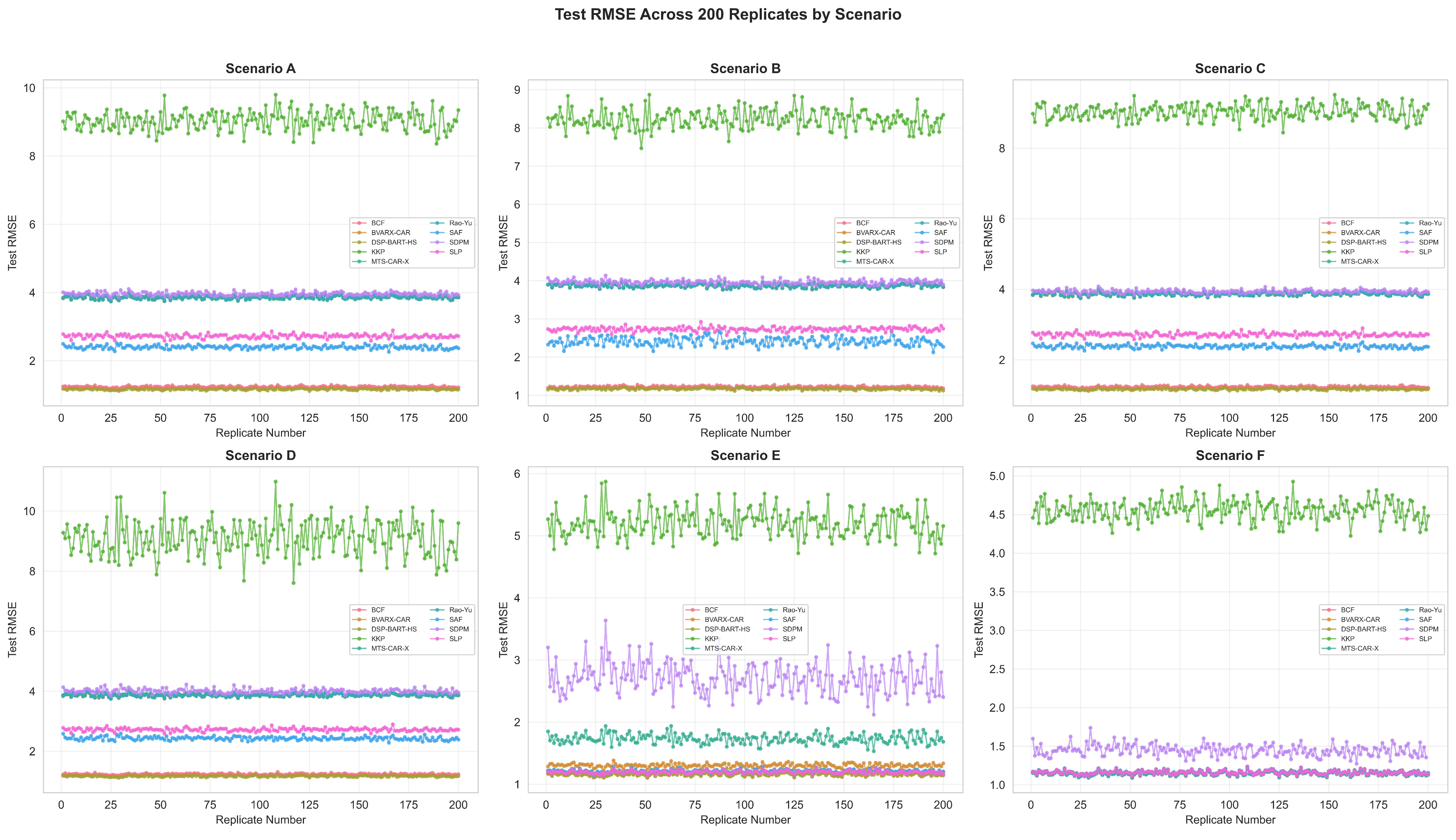}
\caption{Mean test RMSE by model across Scenarios A--F, one line per model, connecting the Test RMSE column of the main text's Scenario A--F tables in sequence.}
\label{fig:s-line-trend}
\end{figure}

Figure~\ref{fig:s-line-trend} traces each model's test RMSE across the six scenarios in sequence. DSP-BART-HS's and BCF's lines stay close to flat across Scenarios A--D, dip further in Scenario E, and converge with the rest of the field only in Scenario F -- the flattest trajectories of the nine models, reflecting their comparative insensitivity to the four adversarial manipulations (B--D) that leave individual-level heterogeneity intact. The five region-time-aggregate estimators' lines, by contrast, all show a pronounced drop between Scenario A/B/C/D and Scenario F, with BVARX-CAR's and MTS-CAR-X's lines showing the steepest single-step decline of the nine models, the visual counterpart to the roughly 70\% reduction in their test RMSE documented in the main text's discussion of Scenario F.

\begin{figure}[H]
\centering
\includegraphics[width=0.95\textwidth]{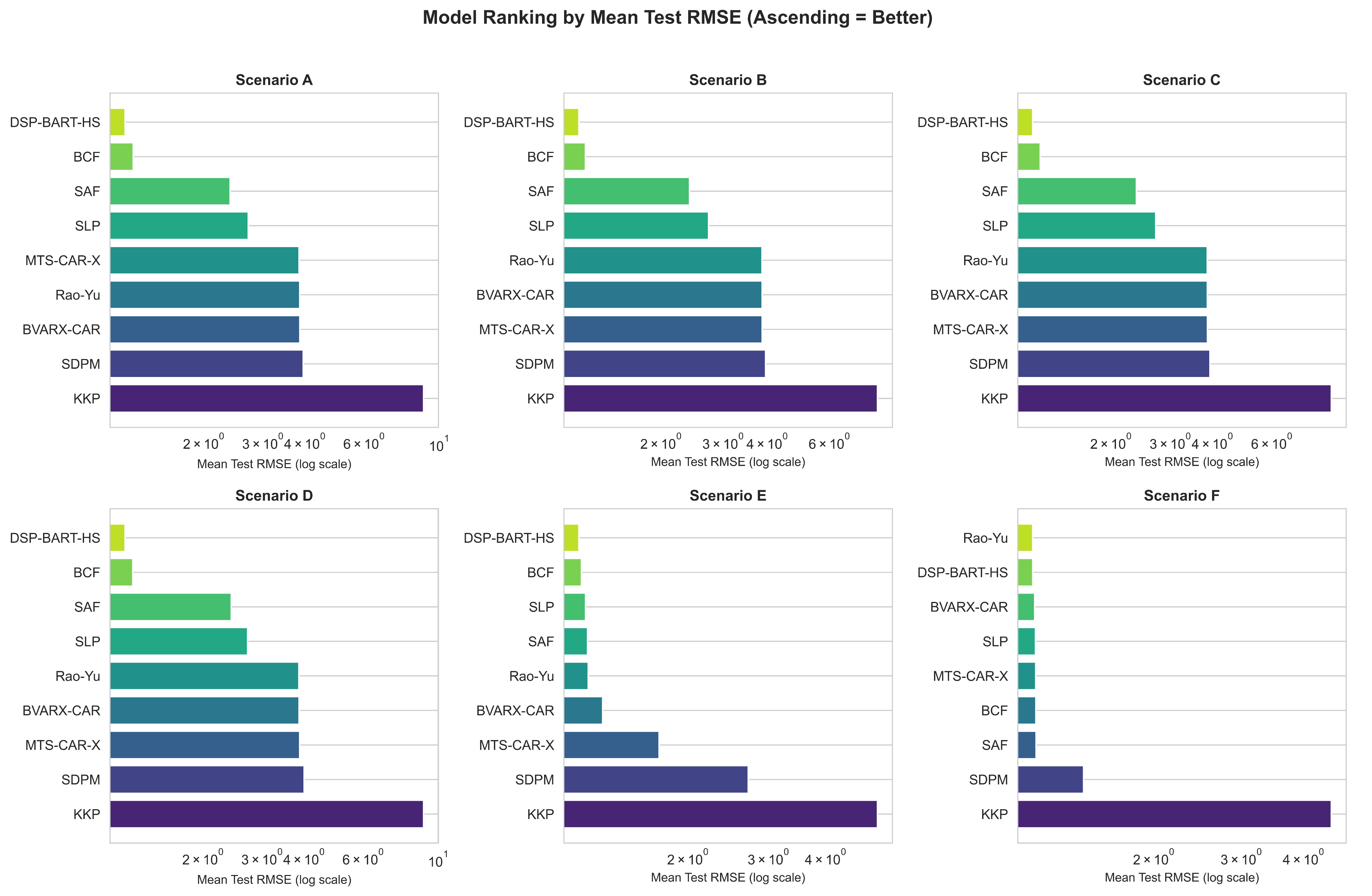}
\caption{Model rank by test RMSE within each of the six scenarios.}
\label{fig:s-ranked}
\end{figure}

Figure~\ref{fig:s-ranked} reports each model's ordinal rank, rather than its raw RMSE, within each scenario. DSP-BART-HS holds rank 1 in Scenarios A--E and rank 2 in Scenario F (behind Rao-Yu, by a margin of 0.0002), the most stable ranking trajectory of any model. KKP holds the last-place rank in every scenario without exception. The rank view makes a pattern visible that raw RMSE values alone can obscure: SDPM's rank remains second-to-last across Scenarios A--F, indicating its relative standing degrades even in the two scenarios constructed to favor region-time-aggregate estimators generally -- the opposite of what Scenario E's design intended, and consistent with SDPM's spatial-lag reduced form carrying a cost that persists, or worsens, even when individual heterogeneity is reduced or removed. This pattern is discussed further in the main text, which reports it independently from the two scenarios' own tabulated results.

\begin{figure}[H]
\centering
\includegraphics[width=0.90\textwidth]{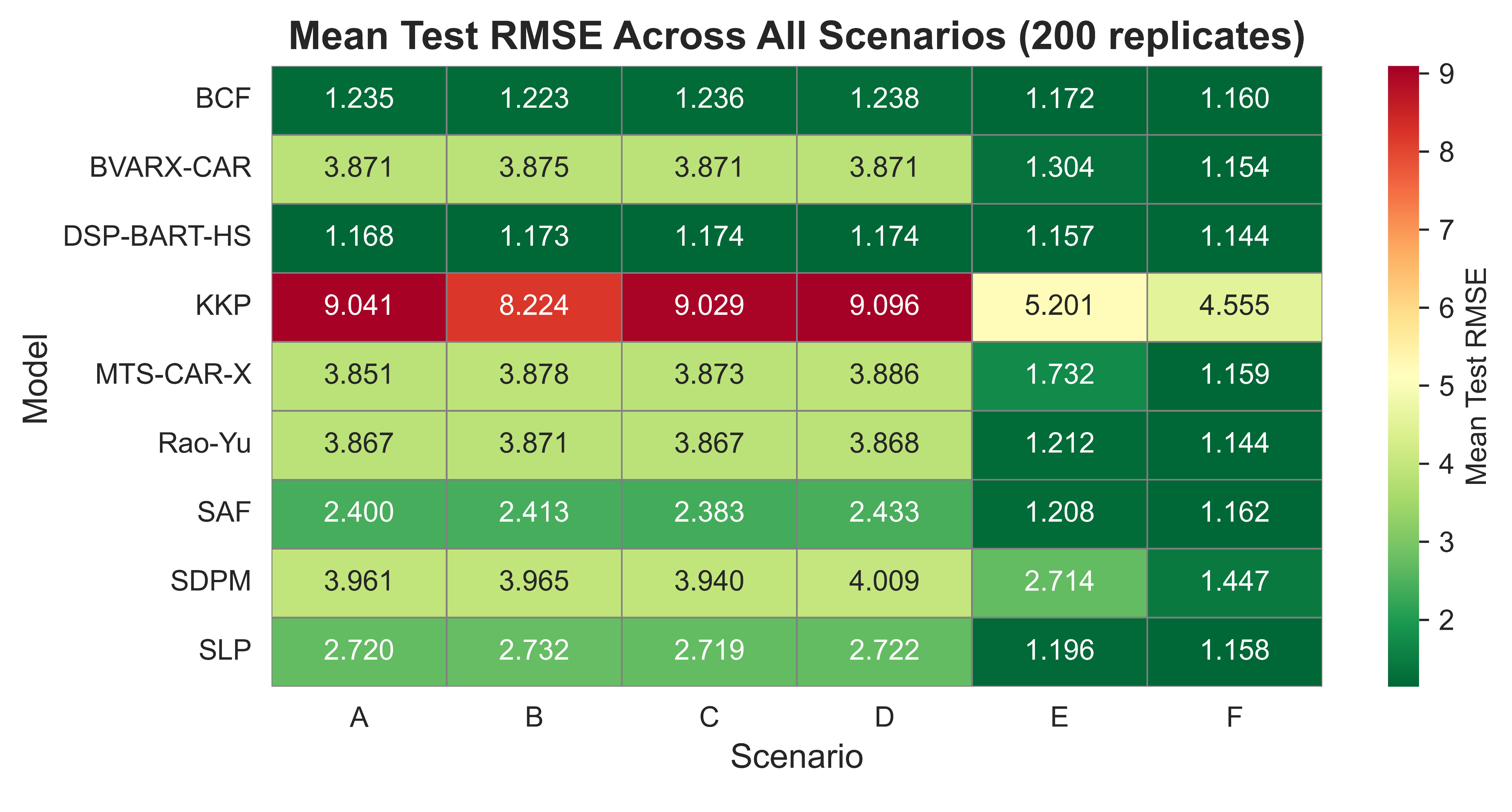}
\caption{Mean test RMSE across all nine estimators and six scenarios, arranged as a model-by-scenario grid. Color reflects $\log_{10}$(mean test RMSE); exact values are annotated in each cell.}
\label{fig:s-heatmap}
\end{figure}

Figure~\ref{fig:s-heatmap} summarizes the main text's Scenario A--F tables in a single grid, the most compact view of the full simulation study in this paper. Two column-wise patterns are visible that are not obvious from any single table in isolation: DSP-BART-HS's row is the most uniformly shaded of the nine, indicating its test RMSE is the least sensitive to which scenario is in effect, while KKP's row is both the darkest overall and the most variable column-to-column, consistent with it being both the weakest and the least stable estimator in the suite. The region-time-aggregate cluster's rows (SDPM, KKP, Rao-Yu, BVARX-CAR, MTS-CAR-X) visibly lighten in the Scenario F column relative to every other column, the grid-level signature of the individual-heterogeneity mechanism that motivates this study's simulation design.

\subsection{Expanded Paired Comparisons Against All Comparators}
\label{sec:s-paired-comparisons}

In the main text, Section~3.2 (\emph{Simulation Study Results}), Scenario F, reports a formal paired comparison between DSP-BART-HS and Rao-Yu alone. Here we extend this to every comparator (BCF, SAF, SLP, Rao-Yu, BVARX-CAR, MTS-CAR-X, SDPM, and KKP) across all six scenarios -- $8 \times 6 = 48$ paired tests in total -- using the same 200-replicate, common-random-numbers design underlying every scenario table in this paper (each replicate's synthetic dataset is shared across all nine models, so the comparison is genuinely paired, not an independent-samples comparison of two separate Monte Carlo runs). For each test, we report the paired mean difference in test RMSE (DSP-BART-HS minus comparator; negative favors DSP-BART-HS), a 95\% confidence interval, and a paired $t$-test $p$-value, with Holm-Bonferroni correction applied across the full family of 48 tests to control the family-wise error rate under multiple comparisons.

\textbf{Result.} 47 of 48 paired comparisons remain significant after Holm-Bonferroni correction, every one favoring DSP-BART-HS (negative mean difference). The single exception is Scenario F versus Rao-Yu (mean difference $0.0002$, 95\% CI $[-0.0002, 0.0006]$, $p=0.38$) -- the same comparison already reported as statistically indistinguishable in the main text (Section 3, \emph{Simulation Study}), now confirmed against the full 200-replicate paired sample rather than summary statistics alone, and the only cell in the entire $8\times6$ family where DSP-BART-HS's advantage is not distinguishable from zero after correction. Full results (all 48 rows: scenario, comparator, $n$, mean difference, 95\% CI, $t$-statistic, raw $p$-value, Holm-adjusted decision) are provided as a supplementary CSV file accompanying this manuscript.

We extend the same paired-comparison protocol to the three robustness scenarios (main text, \emph{Robustness Simulations: Aggregated Covariate Access, Topology, and Panel Size}): every comparator against DSP-BART-HS, 200 replicates, common random numbers, paired $t$-test, Holm-Bonferroni correction. Scenario G includes nine comparators (BCF, SAF, SLP, Rao-Yu, BVARX-CAR, MTS-CAR-X, SDPM, KKP, and cell-level BART); Scenarios H and I include the same eight as Scenarios A--F (cell-level BART does not apply outside Scenario G). This gives $9+8+8=25$ paired tests, corrected as their own family separate from the $8\times6=48$ tests above, since Scenarios G, H, and I test a different question (robustness of the six-scenario pattern to design choices held fixed throughout Scenarios A--F) than the original adversarial-design family.

\textbf{Result.} All 25 paired comparisons remain significant after Holm-Bonferroni correction, every one favoring DSP-BART-HS. This includes BCF, DSP-BART-HS's closest competitor throughout this study, in every one of the three scenarios (Scenario G: mean difference $-0.068$, $p<10^{-157}$; Scenario H: $-0.069$, $p<10^{-148}$; Scenario I: $-0.072$, $p<10^{-192}$) and Scenario G's cell-level BART baseline (mean difference $-2.951$, $p$ below double-precision floating-point representation, effectively zero), the comparator built specifically to test whether DSP-BART-HS's advantage reflects genuine individual-level access rather than model flexibility alone. Full results (all 25 rows, same column format as above) are provided as a supplementary CSV file accompanying this manuscript.

\section{Glossary of Notation, Covariates, and Abbreviations}
\label{sec:s-glossary}

This section defines every mathematical symbol, real-world covariate, model abbreviation, and metric name used in the main manuscript and this supplement, so that a reader can look up any variable code name encountered in either document without searching back through the text where it was first introduced.

\subsection{Mathematical Notation}

\begin{table}[H]
\centering
\setlength{\tabcolsep}{6pt}
\begin{threeparttable}
\caption{General and DSP-BART-HS Notation}
\label{tab:s-notation-general}
\small
\begin{tabular}{@{}p{1.7in}p{4.1in}@{}}
\toprule
Symbol & Meaning \\
\midrule
$i$, $N_j$ & Individual index within a region; number of individuals in region $j$ (mobility application only) \\
$j$, $J$ (or $l$) & Region (county) index; total number of regions \\
$t$, $T$ & Time period (birth cohort or year) index; total number of periods \\
$Y_{ijt}$ & Outcome: income percentile rank (mobility) or county median household income (income application) \\
$\bm{X}_{ijt}$ & Individual-level covariates entering DSP-BART-HS's nonparametric $f(\cdot)$ term \\
$\bm{Z}_{jt}$, $M$ & Region-time contextual covariate vector; its dimension (the "$M$" in $M=12$/$M=11$) \\
$f(\bm{X}_{ijt})$ & DSP-BART-HS's BART component: a sum of $K$ regression trees over individual-level covariates \\
$\bm{\beta}$, $\bm{\gamma}$ & Contextual covariate-block coefficients on $\bm{Z}_{jt}$: $\bm{\beta}$ for DSP-BART-HS, $\bm{\gamma}$ for BVARX-CAR/MTS-CAR-X \\
$\lambda_m$, $\tau$ & DSP-BART-HS's own Horseshoe local (per-covariate, $m=1,\dots,M$) and global shrinkage scales on $\bm{\beta}$, distinct from BVARX-CAR/MTS-CAR-X's own Horseshoe scales on $\bm{\gamma}$ (Tables~\ref{tab:s-notation-bvarx}--\ref{tab:s-notation-mtscarx}), which carry a "G" subscript ($\tau_G$) specifically to distinguish them from this $\tau$ \\
$\phi_j$ & DSP-BART-HS's region-level spatial random effect (CAR prior) \\
$\mu_{jt}$ & DSP-BART-HS's region-specific dynamic (random-walk) macroeconomic shock. For a region with zero training observations at every period (arising only under the county-level holdout), $\mu_{j,\cdot}$ is set via iterative diffusion from neighboring regions' own fitted values rather than left at the prior mean of zero -- see the main text's derivation of $\bm{\eta}_j$ and $\sigma_\mu$. \\
$\alpha_i$ & DSP-BART-HS's individual random effect \\
$\sigma^2$, $\sigma_\mu^2$, $\sigma_\alpha^2$ & Residual variance; variance of $\mu_{jt}$'s random-walk innovations; variance of $\alpha_i$ \\
$\tau_\phi^2$, $\rho$ & CAR prior variance and spatial-correlation parameter for $\phi_j$ \\
$\mathbf{W}$, $\mathbf{D}$ & Spatial adjacency matrix; diagonal matrix of row sums of $\mathbf{W}$ ($\mathbf{D}=\text{diag}(\mathbf{W}\bm{1})$) \\
\bottomrule
\end{tabular}
\end{threeparttable}
\end{table}

\begin{table}[H]
\centering
\setlength{\tabcolsep}{6pt}
\begin{threeparttable}
\caption{BVARX-CAR Notation}
\label{tab:s-notation-bvarx}
\begin{tabular}{@{}p{1.7in}p{4.1in}@{}}
\toprule
Symbol & Meaning \\
\midrule
$\mu_j$, $\theta$ & Region-level intercept; the global mean intercept $\bm{\mu}$ is smoothed toward \\
$\mathbf{B}$, $b_{ij}$ & Spillover matrix; its $(i,j)$ entry (row $i$'s response to region $j$'s lagged outcome) \\
$\rho$, $\psi$ & Spatial-smoothing weight and spatial-scale parameter for $\bm{\mu}$'s CAR prior \\
$\mathbf{V}$, $\mathbf{R}$ & Adjacency matrix and its row-sum diagonal -- the same underlying spatial adjacency matrix as the main text's and DSP-BART-HS's $\mathbf{W}$ (and $l$ the same region count as $J$), retained here as $\mathbf{V}/l$ solely to match \citet{olayinka2026bvarcar}'s own published notation \\
$\lambda_{ij}$, $\phi$ (shrinkage) & Local and global shrinkage scales in $\mathbf{B}$'s published Horseshoe-type prior \\
$\lambda_m^\gamma$, $\tau_G$ & Local and global Horseshoe shrinkage scales on the contextual covariate-block coefficients $\bm{\gamma}$ (this study's own addition, Equation~\ref{eq:s-ghs}) -- the "G" subscript on $\tau_G$ distinguishes it from $\phi$ (the unrelated global scale in $\mathbf{B}$'s own published Horseshoe prior, row above) and from DSP-BART-HS's own $\tau$ (Table~\ref{tab:s-notation-general}) \\
$\delta_{ij}$ & Per-entry shrinkage factor induced by $(\lambda_{ij},\phi)$ in $\mathbf{B}$'s published prior \\
$\tau_B^2$ & This study's single shared ridge variance replacing the published per-entry Horseshoe on $\mathbf{B}$ \\
$\zeta_{(1)},\dots,\zeta_{(l)}$ & Eigenvalues of $\mathbf{R}^{-1}\mathbf{V}$, bounding $\psi$'s admissible range \\
\bottomrule
\end{tabular}
\end{threeparttable}
\end{table}

\begin{table}[H]
\centering
\setlength{\tabcolsep}{6pt}
\begin{threeparttable}
\caption{MTS-CAR-X Notation}
\label{tab:s-notation-mtscarx}
\begin{tabular}{@{}p{1.7in}p{4.1in}@{}}
\toprule
Symbol & Meaning \\
\midrule
$\theta_0$, $\theta_1$ & Shared (region-invariant) intercept and AR(1) slope \\
$\nu_{0j}$, $\nu_{1j}$ & Region $j$'s CAR-linked deviations from $\theta_0$, $\theta_1$ \\
$\mathbf{A}$, $m_0$, $m_1$ & Spatial "diffusion" matrix linking $\bm{\nu}_2$ to $\bm{\nu}_1$ ($\mathbf{A}=m_0\mathbf{I}+m_1\mathbf{W}$); its two hyperparameters \\
$\tau_1^2$, $\tau_2^2$ & CAR prior variances for $\bm{\nu}_1$ and $\bm{\nu}_2\mid\bm{\nu}_1$ \\
$\rho_1$, $\rho_2$ & CAR spatial-correlation parameters for $\bm{\nu}_1$ and $\bm{\nu}_2\mid\bm{\nu}_1$ \\
$\lambda_m^\gamma$, $\tau_G$ & Local and global Horseshoe shrinkage scales on the contextual covariate-block coefficients $\bm{\gamma}$ (this study's own addition), governed by an identical construction to BVARX-CAR's own $\tau_G$ (Table~\ref{tab:s-notation-bvarx}) -- the same symbol is used deliberately, since both models add an identical contextual covariate block \\
$\lambda_{(1)},\dots,\lambda_{(l)}$ & Eigenvalues of $\mathbf{D}^{-1}\mathbf{W}$, bounding $\rho_1,\rho_2$'s admissible range \\
\bottomrule
\end{tabular}
\end{threeparttable}
\end{table}

\subsection{Covariates}

\begin{table}[H]
\centering
\setlength{\tabcolsep}{6pt}
\begin{threeparttable}
\caption{PCA Components (Replacing the 7/8-Variable Collinear Cluster)}
\label{tab:s-covariates-pc}
\small
\begin{tabular}{@{}p{1.7in}p{4.1in}@{}}
\toprule
Covariate & Description \\
\midrule
\texttt{PC1\_demog\_econ} & First principal component of the demographic/economic collinear cluster \\
\texttt{PC2\_demog\_econ} & Second principal component of the same cluster \\
\texttt{PC3\_demog\_econ} & Third principal component of the same cluster \\
\bottomrule
\end{tabular}
\begin{tablenotes}[flushleft]
\footnotesize
\item The cluster is 8 variables for the mobility application (including \texttt{MEDHHINC\_2023}) and 7 for the income application (excluding it); the three principal components are re-fit separately for each application and each train/test split (train-only, projected onto test).
\end{tablenotes}
\end{threeparttable}
\end{table}

\begin{table}[H]
\centering
\setlength{\tabcolsep}{6pt}
\begin{threeparttable}
\caption{Region-Time Contextual Covariates ($\bm{Z}_{jt}$ Block)}
\label{tab:s-covariates-z}
\small
\begin{tabular}{@{}p{1.7in}p{4.1in}@{}}
\toprule
Covariate & Description \\
\midrule
\texttt{foreign\_share1990} & Share of county population that is foreign-born, 1990 \\
\texttt{foreign\_share2000} & Share of county population that is foreign-born, 2000 (mobility only; dropped, near-duplicate) \\
\texttt{foreign\_share2010} & Share of county population that is foreign-born, 2010 \\
\texttt{share\_black2000.x} & Share of county population that is Black, 2000 \\
\texttt{share\_white2000.x} & Share of county population that is White, 2000 \\
\texttt{share\_hisp2000} & Share of county population that is Hispanic, 2000 \\
\texttt{share\_asian2000} & Share of county population that is Asian, 2000 \\
\texttt{emp2000} & County employment rate, 2000 \\
\texttt{ln\_wage\_growth\_hs\_grad} & Log wage growth for high-school graduates (mobility only; dropped for income, wage-denominated) \\
\texttt{mail\_return\_rate2010} & 2010 Census mail-back response rate (a standard Census-response/social-capital proxy) \\
\texttt{job\_density\_2013} & Jobs per unit land area, 2013 \\
\texttt{ann\_avg\_job\_\allowbreak growth\_2004\_2013} & Average annual job growth rate, 2004--2013 \\
\texttt{Unemployment\_rate\_2023} & County unemployment rate, 2023 \\
\texttt{MEDHHINC\_2023} & County median household income, 2023 (mobility only; dropped for income, outcome-circular) \\
\texttt{PCTPOVALL\_2023} & County poverty rate (all ages), 2023 \\
\texttt{R\_NET\_MIG\_2023} & County net migration rate, 2023 \\
\texttt{rent\_twobed2015} & Median two-bedroom rent, 2015 \\
\texttt{mean\_commutetime2000} & Mean commute time, 2000 \\
\texttt{hhinc\_mean2000} & Mean household income, 2000 (dropped as a near-duplicate of \texttt{MEDHHINC\_2023}) \\
\texttt{poor\_share1990/\allowbreak 2000/2010} & Poverty share, 1990/2000/2010 (dropped as near-duplicates of \texttt{PCTPOVALL\_\allowbreak 2023}) \\
\bottomrule
\end{tabular}
\begin{tablenotes}[flushleft]
\footnotesize
\item The mobility application's $M=12$ specification uses all rows above except the three marked "dropped" for mobility; the income application's $M=11$ specification additionally drops \texttt{MEDHHINC\_2023} and \texttt{ln\_wage\_growth\_hs\_grad} (Section 6 of the main text).
\end{tablenotes}
\end{threeparttable}
\end{table}

\begin{table}[H]
\centering
\setlength{\tabcolsep}{6pt}
\begin{threeparttable}
\caption{DSP-BART-HS's Individual-Level $f(\bm{X})$ Covariates and Other Model-Specific Variables}
\label{tab:s-covariates-x}
\small
\begin{tabular}{@{}p{1.7in}p{4.1in}@{}}
\toprule
Covariate & Description \\
\midrule
\texttt{singleparent\_share1990} & Share of children in single-parent households, 1990 \\
\texttt{popdensity2000} & Population density, 2000 \\
\texttt{frac\_coll\_plus2000} & Share of adults with a college degree or higher, 2000 \\
\texttt{traveltime15\_2010} & Share of workers with a commute under 15 minutes, 2010 \\
\texttt{gsmn\_math\_g3\_2013} & Average 3rd-grade standardized math test score, 2013 (mobility application only; a childhood-cohort measure with no income-application analogue) \\
\texttt{region\_idx} & Numeric county index used by SAF (Table~\ref{tab:s-income-varimp}) as a raw predictive feature, not a substantive covariate \\
\texttt{Metro\_2023} & Metropolitan-area indicator; BCF's treatment variable in the income application (\texttt{hhinc\_mean2000} in the mobility application) \\
\bottomrule
\end{tabular}
\begin{tablenotes}[flushleft]
\footnotesize
\item The mobility application's $f(\bm{X})$ block is all five rows above except \texttt{region\_idx}/\texttt{Metro\_2023}; the income application's is the same four covariates minus \texttt{gsmn\_math\_g3\_2013} (see the main text's DSP-BART-HS variable-importance tables for the mobility and income applications, respectively).
\end{tablenotes}
\end{threeparttable}
\end{table}

\subsection{Metrics}

\begin{table}[H]
\centering
\setlength{\tabcolsep}{6pt}
\begin{threeparttable}
\caption{Real-Data Evaluation Metrics}
\label{tab:s-metrics}
\small
\begin{tabular}{@{}p{1.7in}p{4.1in}@{}}
\toprule
Metric & Description \\
\midrule
Metric A & RMSE, MAE, bias, and generalization gap \\
Metric B & Posterior predictive SD and 95\% credible interval width on test rows -- Bayesian models only \\
Metric C & Posterior mean, 95\% CI, and significance of each $\bm{Z}_{jt}$ coefficient \\
Metric C2 & Back-projected implied effect on each original clustered covariate (Table~\ref{tab:s-covariates-z}), reconstructed from the PCA components' coefficients \\
Metric D & BART/random-forest variable importance (split frequency or \%IncMSE) \\
Metric E & Multi-chain $\hat{R}$/ESS convergence diagnostics \\
Gen.\ Gap & RMSE(test) $-$ RMSE(train) \\
\bottomrule
\end{tabular}
\end{threeparttable}
\end{table}

\subsection{Model and Method Abbreviations}

\begin{table}[H]
\centering
\setlength{\tabcolsep}{6pt}
\begin{threeparttable}
\caption{Comparator Suite}
\label{tab:s-model-abbrev}
\small
\begin{tabular}{@{}p{1.7in}p{4.1in}@{}}
\toprule
Abbreviation & Full name / description \\
\midrule
DSP-BART-HS & Dynamic Spatial BART with Horseshoe shrinkage -- this study's proposed model \\
BVARX-CAR & Bayesian VAR with exogenous covariates and a CAR-smoothed intercept \citep{olayinka2026bvarcar} \\
MTS-CAR-X & Multiple Time Series with CAR-correlated priors and exogenous covariates \citep{olayinka2026mtscar} \\
SDPM & Spatial Durbin Panel Model \\
SLP & Spatial Lasso Panel \\
SAF & Spatially-Adjusted (random) Forest \\
BCF & Bayesian Causal Forests \\
KKP & Kapoor-Kelejian-Prucha spatial-error-component estimator \\
Rao-Yu & Rao-Yu small-area time-series/cross-sectional model \\
CAR & Conditional Autoregressive (prior) \\
GL / Horseshoe & Global-local shrinkage prior of the Horseshoe family \\
PCA & Principal Component Analysis \\ 
\bottomrule
\end{tabular}
\end{threeparttable}
\end{table}

\subsection{Back-Projection Formula and Post-Selection Caveat (Metric C2)}
\label{sec:s-backprojection}

For a covariate $k$ in the $p$-variable collinear cluster ($p=8$ for mobility, $p=7$ for income), let $s_k$ be its training-sample standard deviation and let $\mathbf{L} \in \mathbb{R}^{p \times 3}$ be the loading matrix of the three retained principal components (columns), estimated once on training-sample standardized covariates and held fixed thereafter (main text, \emph{Covariate Construction: Resolving Multi-Way Collinearity in $\bm{Z}$}). Writing $z_k = (x_k-\bar{x}_k)/s_k$ for covariate $k$'s standardized value and $\text{PC}_r = \sum_{k=1}^p L_{kr} z_k$ for $r=1,2,3$, the model's estimated linear effect on the three components, $\hat\gamma_{\text{PC}_1},\hat\gamma_{\text{PC}_2},\hat\gamma_{\text{PC}_3}$ (posterior means for DSP-BART-HS/BVARX-CAR/MTS-CAR-X, point estimates for Rao-Yu/SDPM), implies a linear effect on each original covariate by direct substitution and collecting terms:
\begin{equation}
\hat\gamma_{\text{PC}_1}\text{PC}_1+\hat\gamma_{\text{PC}_2}\text{PC}_2+\hat\gamma_{\text{PC}_3}\text{PC}_3 = \sum_{k=1}^p \left(\frac{\sum_{r=1}^3 L_{kr}\hat\gamma_{\text{PC}_r}}{s_k}\right)x_k + \text{const.}, \qquad \text{implied effect on } x_k := \frac{\sum_{r=1}^3 L_{kr}\hat\gamma_{\text{PC}_r}}{s_k}.
\label{eq:s-backproject}
\end{equation}
Uncertainty is propagated by applying (\ref{eq:s-backproject}) -- a fixed linear transformation, since $\mathbf{L}$ and $s_k$ are treated as known constants at this step -- to every posterior draw of $(\gamma_{\text{PC}_1},\gamma_{\text{PC}_2},\gamma_{\text{PC}_3})$ for the three Bayesian models, yielding an exact posterior distribution (and 95\% credible interval) for each covariate's implied effect; for the two frequentist models, the same linear map is applied via the standard delta method to $(\hat\gamma_{\text{PC}_1},\hat\gamma_{\text{PC}_2},\hat\gamma_{\text{PC}_3})$'s estimated sampling covariance.

\textbf{Post-selection caveat.} Both the choice of which $p$ covariates form the collinear cluster (main text, \emph{Covariate Construction: Resolving Multi-Way Collinearity in $\bm{Z}$}, VIF/$R^2$-based diagnosis) and the choice to retain exactly three components (the Kaiser eigenvalue-$\geq 1$ criterion) are themselves decisions made by inspecting this study's own training data, prior to and separately from the coefficient-estimation step that produces $\hat\gamma_{\text{PC}_r}$. Equation~\ref{eq:s-backproject} and its associated intervals treat $\mathbf{L}$, $s_k$, the cluster membership, and the retained-component count as fixed and known, propagating only the sampling or posterior uncertainty in $\hat\gamma_{\text{PC}_r}$ itself. This is a standard conditional-on-the-selected-model treatment, not a selection-adjusted one: it does not account for the uncertainty in, or data-dependence of, the preliminary clustering and dimension-reduction decisions, in the sense formalized by the post-selection inference literature (e.g., a different training sample could plausibly select a different cluster membership or component count, a source of variability not reflected in the reported intervals). We do not attempt a selection-adjusted correction (for instance, via sample splitting or simultaneous post-selection confidence regions) here; the back-projected effects and their intervals should be read as conditional on this study's specific, data-driven covariate-clustering choice, not as accounting for the additional uncertainty that choice itself carries.

\subsection{Comparator Tuning Settings: Real-Data Applications}
\label{sec:s-tuning}

\begin{table}[H]
\centering
\setlength{\tabcolsep}{6pt}
\begin{threeparttable}
\caption{Comparator Tuning Settings, Real-Data Applications}
\label{tab:s-tuning}
\small
\begin{tabular}{@{}p{1.0in}p{1.9in}p{1.9in}@{}}
\toprule
Model & Mobility & Income \\
\midrule
DSP-BART-HS & Gibbs: 200{,}000 iter, 50{,}000 burn-in, thin 3; BART: 200 trees (\texttt{dbarts} default, not overridden) & Same (200{,}000 / 50{,}000 / 3; 200 trees) \\
BVARX-CAR & Gibbs: 11{,}000 iter, 1{,}000 burn-in, thin 5 & Gibbs: 25{,}000 iter, 5{,}000 burn-in, thin 5 \\
MTS-CAR-X & Gibbs: 30{,}000 iter, 5{,}000 burn-in, thin 5 & Same (30{,}000 / 5{,}000 / 5) \\
BCF & \texttt{nburn}=2{,}000, \texttt{nsim}=5{,}000 & Same (2{,}000 / 5{,}000) \\
SLP & LASSO (\texttt{glmnet}, $\alpha=1$); penalty selected via \texttt{cv.glmnet}, \texttt{lambda.min} & Same \\
SAF & Random forest: \texttt{ntree}=300, \texttt{mtry} at \texttt{randomForest} default (not overridden) & Same (\texttt{ntree}=300, default \texttt{mtry}) \\
Rao-Yu & Single \texttt{optim()} fit; no MCMC or iterative tuning applicable & Same \\
SDPM & \texttt{splm::spml()}, closed-form/ML panel fit; no MCMC or iterative tuning applicable & Same \\
\bottomrule
\end{tabular}
\begin{tablenotes}[flushleft]
\footnotesize
\item BVARX-CAR's mobility and income iteration counts differ intentionally (not a typo or oversight): each was set independently based on that application's own convergence behavior. KKP is applied only in the simulation study (Section~\ref{sec:s-simsupp}), not in either real-data application, and is therefore omitted from this table.
\end{tablenotes}
\end{threeparttable}
\end{table}

Table~\ref{tab:s-tuning} reports every comparator's tuning settings for both real-data applications (mobility and income). Settings are reported only where a model has a tunable quantity of the relevant kind: Rao-Yu and SDPM are fit via a single non-iterative optimization (\texttt{optim()} and \texttt{splm::spml()} respectively), so no MCMC iteration/burn-in/thinning applies to either, and SLP's LASSO penalty is selected by cross-validation rather than fixed a priori.

\section{Real Data Applications: Variable Importance}
\label{sec:s-varimp}

\subsection{DSP-BART-HS Variable Importance, Mobility Application}
\label{sec:s-mobility-varimp}

Beyond its coefficient-level effects on the $\bm{Z}\bm{\beta}$ contextual covariate block (reported in the main text), DSP-BART-HS produces a variable-importance measure of its own on the covariates entering its separate $f(\bm{X})$ term: \texttt{dbarts}'s per-iteration split-count frequency, aggregated across all trees and kept posterior draws. Table~\ref{tab:s-mobility-dspbart-varimp} reports it for all three mobility-application splits.

\begin{sloppypar}
\texttt{singleparent\_share1990} is the most-split-on covariate under every split (29--33\% of splits), and \texttt{popdensity2000} is second. \texttt{frac\_coll\_plus2000} and \texttt{gsmn\_math\_g3\_2013} are effectively tied for last under the county split (14.76\% each, differing only in the fifth decimal place), rather than the clear gap between them seen under the random and temporal splits; given three different orderings across three splits for these lower-ranked covariates, we read their exact relative order as sensitive to which counties happen to be held out rather than a stable feature of the model.
\end{sloppypar}

\begin{table}[H]
\centering
\setlength{\tabcolsep}{6pt}
\begin{threeparttable}
\caption{DSP-BART-HS Variable Importance ($f(\bm{X})$ Covariates), Mobility Application}
\label{tab:s-mobility-dspbart-varimp}
\small
\begin{tabular}{lccc}
\toprule
Covariate & Proportion (County) & Proportion (Random) & Proportion (Temporal) \\
\midrule
\texttt{singleparent\_share1990} & 0.3315 & 0.2961 & 0.3111 \\
\texttt{popdensity2000} & 0.2052 & 0.2262 & 0.2723 \\
\texttt{frac\_coll\_plus2000} & 0.1476 & 0.1757 & 0.1392 \\
\texttt{traveltime15\_2010} & 0.1680 & 0.1643 & 0.1638 \\
\texttt{gsmn\_math\_g3\_2013} & 0.1476 & 0.1377 & 0.1135 \\
\bottomrule
\end{tabular}
\begin{tablenotes}[flushleft]
\footnotesize
\item Proportion is each covariate's share of total BART splits across all trees and kept iterations (\texttt{dbarts}'s \texttt{varcount}), so the three proportions sum to 1 within each split. This mobility-application $f(\bm{X})$ block is five covariates, not the four used in the income application (Table~\ref{tab:s-income-dspbart-varimp}): \texttt{gsmn\_math\_g3\_2013} (a childhood-cohort third-grade math test score measure) is specific to the individual/cohort-level framing of the mobility outcome and has no analogue in the income application's county-year panel.
\end{tablenotes}
\end{threeparttable}
\end{table}

\subsection{DSP-BART-HS Variable Importance, Income-inequality Application}
\label{sec:s-income-dspbart-varimp}

\begin{table}[H]
\centering
\setlength{\tabcolsep}{6pt}
\begin{threeparttable}
\caption{DSP-BART-HS Variable Importance ($f(\bm{X})$ Covariates), Income Application}
\label{tab:s-income-dspbart-varimp}
\small
\setlength{\tabcolsep}{6pt}
\begin{tabular}{lccc}
\toprule
Covariate & Proportion (County) & Proportion (Random) & Proportion (Temporal) \\
\midrule
\texttt{singleparent\_share1990} & 0.2931 & 0.3035 & 0.2964 \\
\texttt{frac\_coll\_plus2000} & 0.2855 & 0.2880 & 0.2906 \\
\texttt{traveltime15\_2010} & 0.2455 & 0.2348 & 0.2445 \\
\texttt{popdensity2000} & 0.1759 & 0.1737 & 0.1685 \\
\bottomrule
\end{tabular}
\begin{tablenotes}[flushleft]
\footnotesize
\item Proportion is each covariate's share of total BART splits across all trees and kept iterations (\texttt{dbarts}'s \texttt{varcount}), so the four proportions sum to 1 within each split. This covers only the $f(\bm{X})$ covariates, distinct from and not comparable to the $\bm{Z}\bm{\beta}$ coefficients reported in the main text, or to SAF's variable importance below (which covers the $\bm{Z}$ block instead).
\end{tablenotes}
\end{threeparttable}
\end{table}

\begin{sloppypar}
The ranking is identical and the proportions closely stable across all three splits: \texttt{singleparent\_share1990} and \texttt{frac\_coll\_plus2000} are consistently the two most-split-on covariates (roughly 29--30\% of splits each), \texttt{traveltime15\_2010} third (23--25\%), and \texttt{popdensity2000} last (16--18\%). We read this as a genuinely stable importance ordering for DSP-BART-HS's individual-level component rather than a split-specific artifact, though we note these four covariates were not selected by any theoretical argument about their relevance to income specifically -- they are the same $f(\bm{X})$ block used throughout this study for whichever covariates are not absorbed into the $\bm{Z}\bm{\beta}$ covariate block -- so this ranking is a property of what DSP-BART-HS actually did with the covariates it was given, not a claim about which county characteristics matter most for income in general.
\end{sloppypar}

\subsection{SAF Variable Importance, Income-Inequality Application}
\label{sec:s-income-varimp}

SAF reports variable importance rather than a signed coefficient, so it is not directly comparable to the main text's implied-covariate-effect tables for the income application, or to DSP-BART-HS's own variable importance above (which covers a disjoint set of covariates -- SAF's importance is computed on the $\bm{Z}$ contextual covariate block, DSP-BART-HS's on its separate $f(\bm{X})$ block); it is reported here for completeness. Values are computed under the corrected $M=11$ covariate specification described in the main text.

\begin{table}[H]
\centering
\setlength{\tabcolsep}{6pt}
\begin{threeparttable}
\caption{SAF Variable Importance, Income Application (Top 5)}
\label{tab:s-income-varimp}
\small
\begin{tabular}{@{}lp{4.8in}@{}}
\toprule
Split & Top covariates by \%IncMSE \\
\midrule
Random & \texttt{PC1\_demog\_econ}, \texttt{PC2\_demog\_econ}, \texttt{region\_idx}, \texttt{R\_NET\_MIG\_2023}, \texttt{mean\_commutetime2000} \\
Temporal & \texttt{PC1\_demog\_econ}, \texttt{R\_NET\_MIG\_2023}, \texttt{PC2\_demog\_econ}, \texttt{mean\_commutetime2000}, \texttt{region\_idx} \\
\bottomrule
\end{tabular}
\end{threeparttable}
\end{table}

\texttt{PC1\_demog\_econ} (the affluence-index component of the covariate PCA construction) is the single most important covariate for SAF under both splits, and \texttt{PC2\_demog\_econ} is in the top three under both as well, indicating the same PCA-derived demographic/economic composite that dominates DSP-BART-HS's and the other Bayesian models' coefficient tables also dominates SAF's split selection, despite the two being entirely different kinds of quantity (a coefficient effect versus a variable-importance score). \texttt{region\_idx} itself ranks third under the random split and fifth under the temporal split, indicating SAF's random forest partially recovers region-level structure directly from the county index rather than through the covariates alone, unsurprising given the model has no explicit spatial term the way the CAR-based comparators do.

\section{Zero-Training-Region Diffusion: Ablation}
\label{sec:s-diffusion-ablation}

The main text (\emph{Model}) describes the iterative diffusion step that replaces the uninformative $\mu=0$ FFBS fallback for regions with zero training observations under a county-level holdout. We verify its effect directly by refitting both county-split applications (mobility and income) with the diffusion step toggled off -- zero-training regions held at the $\mu=0$ fallback throughout.

\begin{table}[H]
\centering
\setlength{\tabcolsep}{6pt}
\begin{threeparttable}
\caption{Zero-Training-Region Diffusion: Ablation Results}
\label{tab:s-diffusion-ablation}
\small
\begin{tabular}{@{}lcccc@{}}
\toprule
Application & Diffusion & RMSE (train) & RMSE (test) & Bias (test) \\
\midrule
Mobility & On & 0.0223 & 0.0306 & 0.00035 \\
Mobility & Off & 0.0226 & 0.0334 & $-$0.00470 \\
Income & On & 0.0277 & 0.0775 & 0.00235 \\
Income & Off & 0.0299 & 0.2813 & 0.21231 \\
\bottomrule
\end{tabular}
\begin{tablenotes}[flushleft]
\footnotesize
\item Both applications: county-level holdout, 20\% of counties held out with zero training observations. Mobility: $n_{\text{train}}=26{,}880$, $n_{\text{test}}=6{,}735$ (1{,}792/449 counties). Income: $n_{\text{train}}=47{,}720$, $n_{\text{test}}=11{,}940$ (2{,}386/597 counties).
\end{tablenotes}
\end{threeparttable}
\end{table}

The effect is directionally consistent and, for income, an order of magnitude larger than for mobility: disabling diffusion increases mobility's test RMSE by 9\% (0.0306 $\to$ 0.0334) but increases income's test RMSE by 263\% (0.0775 $\to$ 0.2813), driven almost entirely by test bias exploding from 0.00235 to 0.21231 -- roughly a 24\% multiplicative bias on the untransformed dollar scale (main text, \emph{Real Data Application: Geographic Income Inequality}, log-outcome interpretation). We do not have a fully confirmed explanation for why income is affected so much more severely than mobility under an identical mechanism and directionally identical (20\%) holdout fraction; a plausible contributor is that a larger absolute number of zero-training counties (597 vs.\ 449) increases the chance of extended chains of adjacent zero-training regions, on which the $\mu=0$ fallback's failure to propagate any signal compounds across more hops before reaching a data-having region, but we have not verified this specific mechanism directly against the two applications' adjacency structures and note it as a plausible rather than confirmed explanation. In both applications, refitting the random row-level and temporal-holdout splits (where every region retains training data by construction) with diffusion toggled off reproduces results identical to the diffusion-on run to numerical precision, confirming the step is a no-op outside the county-level holdout, exactly as claimed in the main text.

\section{DSP-BART-HS Real-Data MCMC Diagnostics}
\label{sec:s-realdata-diagnostics}

Section~\ref{sec:s-simsupp} reports multi-chain convergence diagnostics for the simulation study. Here we report the analogous check for a real-data fit -- the mobility county-level split, DSP-BART-HS -- since convergence evidence confined to simulation alone would leave the real-data application's own MCMC behavior unverified. Four independent chains were run at the production Gibbs settings ($200{,}000$/$50{,}000$/$3$), differing only in RNG seed; the train/test split itself, and every other input, is identical across chains.

\begin{table}[H]
\centering
\setlength{\tabcolsep}{6pt}
\begin{threeparttable}
\caption{Real-Data Convergence Diagnostics: Mobility County-Split, DSP-BART-HS, 4 Chains}
\label{tab:s-realdata-diagnostics}
\small
\begin{tabular}{@{}lccc@{}}
\toprule
Parameter & $\hat{R}$ & $\hat{R}$ (97.5\% CI) & Total ESS (4 chains) \\
\midrule
$\rho$ & 1.008 & 1.022 & 4{,}520 \\
$\tau_\phi^2$ & 1.000 & 1.001 & 21{,}915 \\
$\sigma_\mu^2$ & 1.000 & 1.000 & 150{,}666 \\
$\phi_j$  & 1.001 & 1.002 & 6{,}566 \\
$\beta_{\text{PC1\_demog\_econ}}$ & 1.024 & 1.071 & 569 \\
$\beta_{\text{PC2\_demog\_econ}}$ & 1.006 & 1.012 & 562 \\
$\beta_{\text{PC3\_demog\_econ}}$ & 1.016 & 1.044 & 572 \\
$\beta_{\text{foreign\_share1990}}$ & 1.009 & 1.027 & 1{,}253 \\
\bottomrule
\end{tabular}
\begin{tablenotes}[flushleft]
\footnotesize
\item $\hat{R}$ via \citet{gelmanrubin1992inference} (Gelman-Rubin potential scale reduction), matching the citation and methodology already used for the simulation study's own diagnostics (Section~\ref{sec:s-simsupp}); All eight tracked parameters have point-estimate $\hat{R} < 1.03$ and upper-CI $\hat{R} < 1.05$, indicating they are under the conventional $\hat{R}<1.1$ convergence threshold.
\end{tablenotes}
\end{threeparttable}
\end{table}

\textbf{BART tree-mixing proxy.} \texttt{dbarts} exposes no native tree-structure convergence statistic, so standard scalar $\hat{R}$/ESS cannot directly diagnose ensemble mixing. As a descriptive proxy, we compare each chain's mean per-iteration variable-inclusion proportions (split-count share) across the five individual-level covariates in $f(\bm{X})$:

\begin{table}[H]
\centering
\setlength{\tabcolsep}{6pt}
\begin{threeparttable}
\caption{BART Variable-Inclusion Proportions by Chain}
\label{tab:s-varinclusion-chains}
\small
\begin{tabular}{@{}lcccc@{}}
\toprule
Covariate & Chain 1 & Chain 2 & Chain 3 & Chain 4 \\
\midrule
\texttt{popdensity2000} & 0.2037 & 0.2074 & 0.2129 & 0.2202 \\
\texttt{traveltime15\_2010} & 0.1956 & 0.1853 & 0.1846 & 0.1826 \\
\texttt{gsmn\_math\_g3\_2013} & 0.1345 & 0.1349 & 0.1396 & 0.1381 \\
\texttt{singleparent\_share1990} & 0.2889 & 0.2802 & 0.2724 & 0.2824 \\
\texttt{frac\_coll\_plus2000} & 0.1774 & 0.1922 & 0.1905 & 0.1767 \\
\bottomrule
\end{tabular}
\begin{tablenotes}[flushleft]
\footnotesize
\item Maximum cross-chain range for any covariate: 0.0165 (\texttt{popdensity2000}). All four chains agree closely on both the rank ordering and approximate magnitude of every covariate's inclusion share -- evidence against any one chain settling into a qualitatively different region of tree space, though this proxy cannot rule out finer-grained mixing differences within a shared inclusion pattern, and we do not present it as a formal ensemble-convergence guarantee.
\end{tablenotes}
\end{threeparttable}
\end{table}

\textbf{Posterior predictive check.} Each chain's per-iteration mean predicted test-set $Y$ was compared against the observed test-set mean ($0.4608$).

\begin{table}[H]
\centering
\setlength{\tabcolsep}{6pt}
\begin{threeparttable}
\caption{Posterior Predictive Check: Mean Test-Set $Y$}
\label{tab:s-ppc-chains}
\small
\begin{tabular}{@{}lcccc@{}}
\toprule
Chain & Posterior Mean & 95\% CI Lower & 95\% CI Upper & Covers Observed \\
\midrule
1 & 0.4604 & 0.4173 & 0.5030 & Yes \\
2 & 0.4605 & 0.4178 & 0.5036 & Yes \\
3 & 0.4602 & 0.4180 & 0.5026 & Yes \\
4 & 0.4604 & 0.4177 & 0.5025 & Yes \\
\bottomrule
\end{tabular}
\end{threeparttable}
\end{table}

The observed test-set mean falls within every chain's 95\% posterior predictive interval, with near-identical posterior means and intervals across chains -- consistent with the R-hat/ESS evidence above that the four chains have converged to the same posterior.

\bibliographystyle{abbrvnat}
\bibliography{reference}